%% file: ms.tex
\documentclass[11pt,a4paper]{article}
\pdfoutput=1
\usepackage{jheppub}

\usepackage[numbers,sort&compress]{natbib}
\usepackage{subcaption}
\usepackage{slashed}
\usepackage{bbm} 

\renewcommand*{\le}{\left}
\newcommand*{\ri}{\right}

\renewcommand*{\a}{\alpha}
\renewcommand*{\b}{\beta}
\newcommand*{\g}{\gamma}
\renewcommand*{\d}{\delta}
\newcommand*{\e}{\epsilon}
\newcommand*{\ve}{\varepsilon}
\newcommand*{\h}{\eta}

\renewcommand*{\k}{\kappa}
\renewcommand*{\l}{\lambda}
\newcommand*{\m}{\mu}
\newcommand*{\n}{\nu}
\renewcommand*{\r}{\rho}
\newcommand*{\s}{\sigma}
\renewcommand*{\t}{\tau}
\newcommand*{\f}{\phi}

\newcommand*{\y}{\psi}
\newcommand*{\w}{\omega}
\newcommand*{\G}{\Gamma}
\newcommand*{\Q}{\Theta}

\newcommand*{\cA}{\mathcal{A}}
\newcommand*{\cB}{\mathcal{B}}
\newcommand*{\cC}{\mathcal{C}}
\newcommand*{\cF}{\mathcal{F}}
\newcommand*{\cG}{\mathcal{G}}
\newcommand*{\cL}{\mathcal{L}}
\newcommand*{\cM}{\mathcal{M}}

\newcommand*{\cO}{\mathcal{O}}
\newcommand*{\cZ}{\mathcal{Z}}

\newcommand*{\gn}{G_\mathrm{N}}     
\newcommand*{\diff}{\mathrm{d}}
\newcommand*{\p}{\partial}
\renewcommand*{\Re}{\operatorname{Re}} 
\renewcommand*{\Im}{\operatorname{Im}} 
\newcommand*{\id}{\mathbbm{1}} 
\newcommand*{\tr}{\operatorname{tr}} 

\newcommand*{\ket}[1]{|#1\rangle}
\newcommand*{\bra}[1]{\langle#1|}

\newcommand*{\vev}[1]{\langle#1\rangle}

\newcommand{\ads}[1][]{\ifmmode \mathrm{AdS}_{#1} \else AdS\(_{#1}\)\fi}

\title{Thermodynamics and transport of holographic nodal line semimetals}
\author{Ronnie Rodgers,}
\author{Enea Mauri,}
\author{Umut G\"ursoy,}
\author{Henk~T.\,C.~Stoof}

\affiliation{Institute for Theoretical Physics, Utrecht University, \\Princetonplein 5, 3584 CC Utrecht, the Netherlands}

\emailAdd{r.j.rodgers@uu.nl}
\emailAdd{e.mauri@uu.nl}
\emailAdd{u.gursoy@uu.nl}
\emailAdd{h.t.c.stoof@uu.nl}

\abstract{We study various thermodynamic and transport properties of a holographic model of a nodal line semimetal (NLSM) at finite temperature, including the quantum phase transition to a topologically trivial phase, with Dirac semimetal-like conductivity. At zero temperature, composite fermion spectral functions obtained from holography are known to exhibit multiple Fermi surfaces. Similarly, for the holographic NLSM we observe multiple nodal lines instead of just one. We show, however, that as the temperature is raised these nodal lines broaden and disappear into the continuum one by one, so there is a finite range of temperatures for which there is only a single nodal line visible in the spectrum. We compute several transport coefficients in the holographic NLSM as a function of temperature, namely the charge and thermal conductivities, and the shear viscosities. By adding a new non-linear coupling to the model we are able to control the low frequency limit of the electrical conductivity in the direction orthogonal to the plane of the nodal line, allowing us to better match the conductivity of real NLSMs. The boundary quantum field theory is anisotropic and therefore has explicitly broken Lorentz invariance, which leads to a stress tensor that is not symmetric. This has important consequences for the energy and momentum transport: the thermal conductivity at vanishing charge density is not simply fixed by a Ward identity, and there are a much larger number of independent shear viscosities than in a Lorentz-invariant system.}

\begin{document}
    
\maketitle

\input{intro}

\input{holographic_model}

\input{fermions}

\input{transport}

\input{discussion}

\section*{Acknowledgements}
We thank Stijn Claerhoudt for collaboration at an early stage of this project. This work is supported by the Stichting voor Fundamenteel Onderzoek der Materie (FOM) and is part of the D-ITP consortium, a program of the Netherlands Organization for Scientific Research (NWO) that is funded by the Dutch Ministry of Education, Culture and Science (OCW).

\appendix

\input{holo_rg}

\input{two_point_fn_wi}

\input{coefficients}

\bibliographystyle{JHEP}
\bibliography{nodal_loops_ref}

\end{document}

%% file: intro.tex
\section{Introduction}

Nodal line semimetals (NLSMs) are a recently discovered class of materials, in which two electronic bands intersect along a closed curve in momentum space at or near the Fermi energy. As reviewed below, this intersection is protected by the non-trivial topology of the electronic band structure, combined with the discrete symmetries of the system. Examples of materials with band structures containing nodal lines are Ca\(_3\)P\(_2\)~\cite{doi:10.1063/1.4926545,PhysRevB.93.205132}, PbTaSe\(_2\)~\cite{Bian_2016}, and ZrSiS~\cite{Schoop_2016,PhysRevB.93.201104}. As with other recently discovered semimetals, such as the Dirac and Weyl semimetals, these topological materials are of particular interest because of their potential for energy-efficient electronics, (pseudo)spintronics devices, and quantum information processing. This is not only because of the topologically protected and often dissipationless edge states that these materials possess, but also due to the sometimes anomalous bulk properties of these semimetals~\cite{doi:10.1146/annurev-conmatphys-031016-025458,RevModPhys.82.3045,PhysRevLett.107.127205,PhysRevB.83.205101}.

A simple model exhibiting a nodal line is a free fermion \(\y\) of mass \(M\) coupled to an external antisymmetric tensor field \(b_{\m\n}\) and having the Lagrangian density~\cite{PhysRevB.84.235126}
\begin{equation} \label{eq:toy_model_lagrangian}
    \cL =i \bar{\y} \le( \g^\m \p_\m -  M + i b_{\m\n} \g^{\m\n} \ri) \y,
\end{equation}
where \(\g^{\m\n} = \frac{1}{2}[\g^\m, \g^\n]\), and throughout this paper we use units such that \(\hbar = c = 1\). Greek indices such as \(\m\) and \(\n\) label all four spacetime coordinates, while Latin indices such as \(i\) and \(j\) will correspond only to the three spatial coordinates. Our convention for the Dirac matrices is \(\{\g^\m, \g^\n\} = 2 \h^{\m\n}\), where \(\h_{\m\n}\) is the Minkowski metric in mostly-plus signature, and we take \((\g^0)^\dagger = - \g^0\) and \((\g^i)^\dagger = \g^i\).

\begin{figure}
    \begin{subfigure}{0.5\textwidth}
        \includegraphics[width=\textwidth,height=5cm,keepaspectratio]{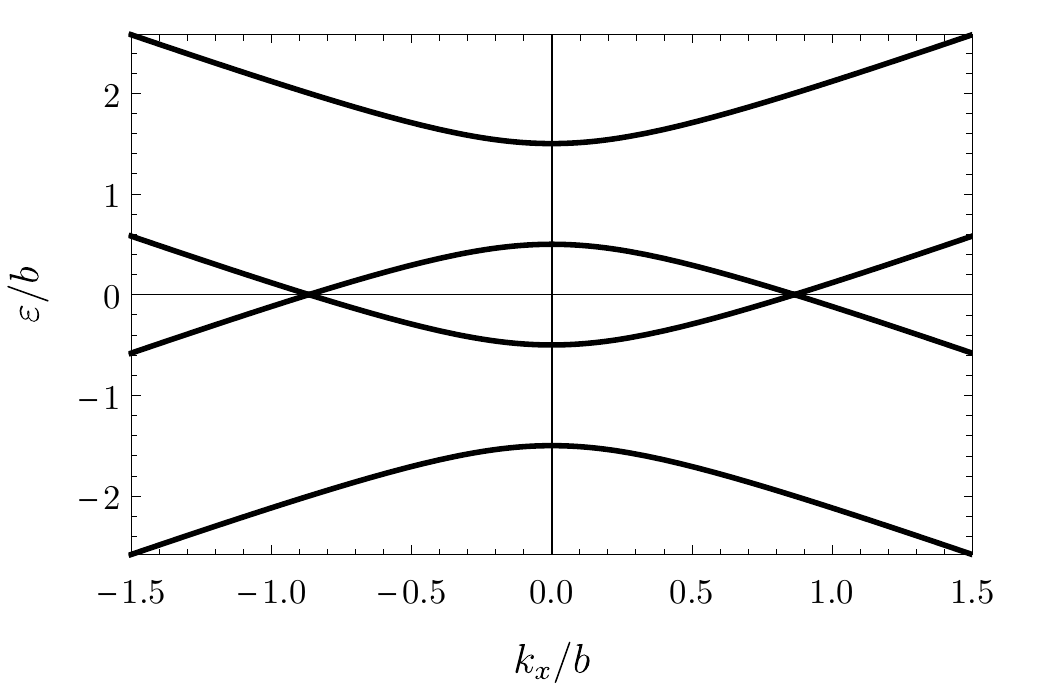}
        \caption{}
        \label{fig:example_loop}
    \end{subfigure}
    \begin{subfigure}{0.5\textwidth}
        \includegraphics[width=\textwidth, height=5cm,trim= 0 40 0 40, clip,keepaspectratio]{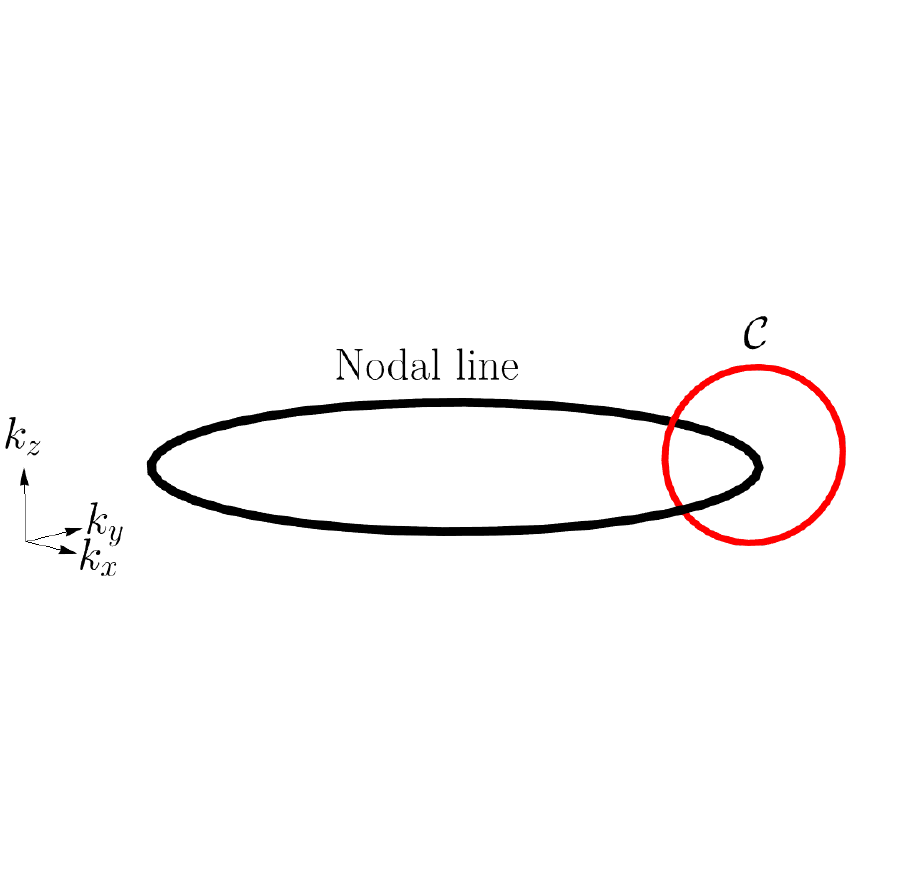}
        \caption{}
        \label{fig:nodal_loop_cartoon}
    \end{subfigure}
    \caption{\textbf{(a):} The eigenvalues of the nodal line Hamiltonian~\eqref{eq:toy_model_hamiltonian}, here plotted as functions of \(k_x\) for \(k_y = k_z = 0\) and \(M = b/2\). Two of the four eigenvalues meet at \(k_x/b = \pm \sqrt{3}/2\), which is the projection to \(k_y=0\) of the nodal line. \textbf{(b):} Cartoon of the nodal line in momentum space. The red circle \(\cC\) interlinks the nodal line. The integral of the Berry connection around \(\cC\) is \(\oint_{\cC} \cA = \pm \pi\).}
\end{figure}

The single-particle Hamiltonian following from equation~\eqref{eq:toy_model_lagrangian} is 
\begin{equation} \label{eq:toy_model_hamiltonian}
    H(\vec{k}) = \g^0 \le( \g^i k_i - i M - b_{\m\n} \g^{\m\n}\ri),
\end{equation}
where \(\vec{k} = (k_x, k_y, k_z)\) is the momentum of the fermion. To obtain a nodal line we take the only non-zero component of the tensor field to be \(b_{xy} = - b_{yx} = b\) for some constant \(b\), explicitly breaking the symmetries of rotations about the \(x\) and \(y\) axes, Lorentz boosts in the \(x\) and \(y\) directions, and time reversal, since \(\bar{\y} \g^{ij} \y \to - \bar{\y} \g^{ij} \y\) under time reversal. Without loss of generality, we will take \(b>0\). With this choice of \(b_{\m\n}\), the eigenvalues of the Hamiltonian~\eqref{eq:toy_model_hamiltonian} are
\begin{equation} \label{eq:toy_model_eigenvalues}
    \ve(\vec{k}) = \pm \sqrt{k_z^2 + \le(b \pm \sqrt{k_x^2 + k_y^2 + M^2} \ri)^2},
\end{equation}
with the two \(\pm\) signs independent. For \(|M|  < b\), two of these eigenvalues meet at \(\ve=0\) along the circle at \(k_x^2 + k_y^2 = b^2 - M^2\) and \(k_z = 0\), as illustrated in figure~\ref{fig:example_loop}. This circle is the nodal line.

The degeneracy of eigenvalues at the nodal line is protected by the non-trivial topology of the Berry connection \(\cA\), along with the discrete symmetries of the Hamiltonian. Recall that for a Hamiltonian depending on some set of parameters \(\vec{k}\), with corresponding normalised eigenstates \(\ket{u(\vec{k})}\), the Berry connection is defined as \(\cA_i = i \bra{u(\vec{k})} \p_{k_i} \ket{u(\vec{k})}\). If we take \(\ket{u(\vec{k})}\) to be either of the two eigenstates of the Hamiltonian~\eqref{eq:toy_model_hamiltonian} with eigenvalues that meet at the nodal line, then the integral of \(\cA\) about any closed contour \(\cC\) in momentum space  that winds once around the nodal line is~\cite{PhysRevB.84.235126}
\begin{equation} \label{eq:berry_connection_integral}
    \oint_{\cC} \cA = \pm \pi,
\end{equation}
depending on the relative orientation of \(\cC\) and the nodal line. An example contour \(\cC\) is depicted in figure~\ref{fig:nodal_loop_cartoon}. If the integral of \(\cA\) around any curve \(\cC\) that does not encircle a band-touching vanishes, then a small perturbation to the Hamiltonian that does not break the symmetries cannot destroy the nodal line without also changing the right-hand side of equation~\eqref{eq:berry_connection_integral} to zero. Such a large change would violate the assumption that the perturbation is small.

The stability of the nodal line therefore rests on the assumption that the left-hand side of equation~\eqref{eq:berry_connection_integral} vanishes for any \(\cC\) that does not encircle the nodal line. This is not always the case for a generic Hamiltonian, and typically requires the presence of multiple discrete symmetries. For the Hamiltonian in equation~\eqref{eq:toy_model_hamiltonian} with \(b_{xy} = -b_{yx} =b\), the relevant symmetries are parity, rotation about the \(z\) axis by angle \(\pi\), and the composition of time reversal with an \(x \to -x\) reflection.\footnote{The reflection inverts the sign of \(b\), undoing the effect of the time-reversal transformation on the Lagrangian~\eqref{eq:toy_model_lagrangian}.} For further details, see ref.~\cite{PhysRevB.84.235126}.

The non-trivial topology of the band structure becomes clearer if we convert the Hamiltonian~\eqref{eq:toy_model_hamiltonian} into a two-band Hamiltonian of the type commonly used in the condensed matter literature. This can be achieved by employing a basis in which the \(\g\) matrices are Kronecker products of Pauli matrices, for example \(\g^0 = i \s_3 \otimes \s_2\), \(\g^1 = \s_1 \otimes \id_2\), \(\g^2 = \s_2 \otimes \id_2 \), and \(\g^3 = \s_3 \otimes \s_3\)~\cite{PhysRevB.84.235126}. In this basis, the Hamiltonian with \(b_{xy} = - b_{yx} = b\) is
\begin{equation}
    H(\vec{k}) = \le(k_y \s_1  - k_x \s_2 + M \s_3 + b \id_2 \ri) \otimes \s_2 - k_z \id_2 \otimes \s_1,
\end{equation}
and the energy eigenstates take the factorised form \(\ket{\ve} = \ket{\ve_1} \otimes \ket{\ve_2}\), where \(\ket{\ve_1}\) is an eigenstate of \(\le(k_y \s_1  - k_x \s_2 + M \s_3 + b \id_2 \ri)\) with eigenvalue \(\ve_1 = b \pm \sqrt{k_x^2 + k_y^2 + M^2}\), while \(\ket{\ve_2}\) is an eigenstate of \(\le(\ve_1 \s_2 - k_z \s_1\ri)\). Hence, if we choose the minus sign in \(\ve_1\), then \(\ket{\ve_2}\) is an eigenstate of
\begin{equation}
    H_2(\vec{k}) = \le( b - \sqrt{k_x^2 + k_y^2 + M^2} \ri) \s_2 - k_z \s_1,
\end{equation}
with eigenvalues equal to the two eigenvalues in equation~\eqref{eq:toy_model_eigenvalues} that have a minus sign inside the square root. These are the two eigenvalues that meet to form the nodal line, i.e. the inner two eigenvalues in figure~\ref{fig:example_loop}, so we may regard \(H_2(\vec{k})\) as an effective Hamiltonian for these two bands. This Hamiltonian has a `chiral' symmetry as it only depends on two Pauli matrices, arising from the combination of discrete symmetries described above~\cite{PhysRevB.84.235126}. If we write the two-band Hamiltonian in the form of a Zeeman interaction, $H_2(\vec{k}) \equiv - \vec{B}(\vec{k}) \cdot \vec{\sigma}$, the field $\vec{B}(\vec{k})$ therefore has non-zero components only in the \(x\) and \(y\) directions. If we vary \(\vec{k}\) around a closed loop in momentum space encircling the nodal line, $\vec{B}(\vec{k})$ rotates about the origin of the \((x,y)\) plane an integer number of times. It is this nontrivial winding number that is measured with the integral $(1/\pi) \oint_{\cC} \cA$. Note also that adding a perturbation proportional to $\sigma_3$ to the Hamiltonian will gap out the nodal line, so it is precisely the chiral symmetry that protects it.

An important feature of the nodal line band structure is that the density of electron states \(g(\ve)\) vanishes at the Fermi surface. This leads to weak screening of the Coulomb interaction between electrons, since the screening length diverges as \(g(\ve)^{-1/2}\), see for example ref.~\cite{Stoof:2009kfa}. The electrons in an NLSM may therefore become strongly interacting, with significantly different physics from the free model in equation~\eqref{eq:toy_model_lagrangian}. Indeed, evidence has recently been found for strong correlations between electrons in the nodal line semimetal ZrSiSe~\cite{Shao:2020juf}. One tool to model strongly interacting systems is the anti-de Sitter/conformal field theory (AdS/CFT) correspondence, also known as holography, which relates strongly interacting quantum field theories (QFTs) to weakly coupled gravitational theories~\cite{Maldacena:1997re,Gubser:1998bc,Witten:1998qj}. A holographic model of an NLSM was constructed in refs.~\cite{Liu:2018bye,Liu:2018djq,Liu:2020ymx}, by writing down a gravitational theory holographically dual to a strongly interacting QFT with the same operator content and symmetries as the Lagrangian in equation~\eqref{eq:toy_model_lagrangian}.\footnote{Weyl semimetals also have vanishing density of states at the Fermi surface, so may also be strongly coupled. In the same spirit, various different holographic models of Weyl semimetals have been proposed in refs.~\cite{Gursoy:2012ie,Jacobs:2015fiv,Landsteiner:2015lsa,Landsteiner:2015pdh,Copetti:2016ewq,Liu:2018spp,Landsteiner:2019kxb,Juricic:2020sgg,Hashimoto:2016ize,Kinoshita:2017uch,BitaghsirFadafan:2020lkh}.}

The holographic NLSM model includes two couplings, \(M\) and \(b\), analogous to the couplings with the same names in the free model~\eqref{eq:toy_model_lagrangian}. At zero temperature, the model exhibits a quantum phase transition at a critical value of the dimensionless ratio \(M/b = (M/b)_\mathrm{crit.}\). Below \((M/b)_\mathrm{crit.}\), spectral functions of composite fermionic operators exhibit many circular lines of sharp peaks in momentum space at zero frequency, which are identified as the nodal lines of the system~\cite{Liu:2018djq}. Above \((M/b)_\mathrm{crit.}\) no such nodal lines appear. The \(T=0\) phase structure of the holographic nodal line model therefore appears qualitatively similar to that of the free model, except for the presence of many nodal lines in the low \(M/b\) phase, rather than just one.

The many nodal lines in the holographic model pose a challenge to the application of holography to real NLSMs, which have only a finite number of nodal lines. However, we will show in section~\ref{sec:fermions} that at non-zero temperature thermal broadening causes many of the peaks in the fermion spectral function to merge, such that there is a range of temperatures for which only a small number of nodal lines are visible. One might expect holography to provide a more realistic model of nodal line physics in this temperature range.

Another possible method to model a strongly interacting system with only a small number of nodal lines would be to use semi-holography~\cite{Contino:2004vy,Hartnoll:2009ns,Faulkner:2010tq,Gursoy:2011gz}, in which an elementary fermion undergoes strong interactions mediated by a holographic conformal field theory. A second advantage of semi-holography is that it yields fermion spectral functions \(\r\) that when integrated over frequency \(\w\) obey the sum rule
\(
    \int_{-\infty}^\infty \diff \w \, \r = 1.
\)
In contrast, in the fully holographic approach one only has access to spectral functions of composite fermions, which do not obey this sum rule, and therefore cannot be directly compared to electron spectral functions in real materials, as measured in angle-resolved photoemission spectroscopy (ARPES) experiments.

In this paper we will take the fully holographic approach. We will analyse the thermodynamics and various transport properties of a version of the holographic NLSM model of ref.~\cite{Liu:2018bye}, modified by an additional coupling in the gravitational theory that will allow us to tune some of the infrared (IR) properties of the system. In section~\ref{sec:model} we will write down the holographic NLSM model, describe its solutions, and present numerical results for its thermodynamics. As for the model of ref.~\cite{Liu:2018bye}, at zero temperature we find a quantum phase transition between topologically trivial and non-trivial phases. A key result, described in section~\ref{sec:stress_tensor_comments}, is that the explicitly broken Lorentz invariance of the NLSM implies that the stress tensor is not a symmetric tensor; it has an antisymmetric contribution proportional to the coupling \(b\) responsible for the violation of Lorentz invariance.

In section~\ref{sec:fermions} we present numerical results for the spectral function of a composite Dirac fermion operator holographically dual to a pair of bulk probe fermions, in the topologically non-trivial phase. We find that at zero temperature the spectral function exhibits multiple nodal lines, appearing as sharp peaks at zero frequency and non-zero momentum. The nodal lines gradually disappear into the continuum as the temperature is increased.

In section~\ref{sec:transport} we give results for various transport coefficients of the holographic NLSM, namely the electrical and thermal conductivities, and the shear viscosity. At zero temperature in the topologically non-trivial phase, the DC electrical conductivity in the direction orthogonal to the plane of the nodal line is non-zero, while in other directions it vanishes. All components of the DC conductivity vanish in the topologically trivial phase. The asymmetry of the stress tensor induced by the Lorentz-violating coupling \(b\) is crucial to our analysis of energy and momentum transport. It implies that the thermal conductivity at vanishing charge density is not completely fixed by a Ward identity, as it is when \(b\) vanishes, and there are multiple shear viscosities, only some of which take the value \(\h = s/4\pi\) universal to Lorentz-invariant holographic systems~\cite{Policastro:2001yc,Kovtun:2003wp,Buchel:2003tz,Kovtun:2004de,Starinets:2008fb}.

%% file: holographic_model.tex
\section{Holographic model}
\label{sec:model}

\subsection{Action and equations of motion}

In this section we describe the holographic model that we will use, based on that of refs.~\cite{Liu:2018bye,Liu:2018djq,Liu:2020ymx}. The model is a bottom-up construction, built by writing down a gravitational action for a set of fields holographically dual to a set of operators with the same spins and Ward identities as the operators of the free model~\eqref{eq:toy_model_lagrangian}. Concretely, the field content on the gravitational side is:
\begin{itemize}
    \item The metric \(G_{mn}\), holographically dual to the operator \(\tilde{T}^{\m\n}\). As we will discuss in detail in section~\ref{sec:stress_tensor_comments}, \(\tilde{T}^{\m\n}\) is closely related to the stress tensor.
    \item A complex scalar field \(\f\), dual to a complex scalar operator \(\cO\). Roughly speaking, we think of the real part of \(\cO\) as modelling the fermion mass operator \(\bar{\y}\y\), while the imaginary part of \(\cO\) models \(\bar{\y} \g^5 \y\).
    \item A real two-form field \(B_{mn}\), dual to an antisymmetric tensor operator \(\cO^{\m\n}\), to model the operator \(\bar{\y} \g^{\m\n} \y \).\footnote{It has been argued that a more realistic model may use a complex, self-dual two-form to model both \(\bar{\y} \g^{\m\n} \y\) and \(\bar{\y} \g^{\m\n} \g^5 \y\)~\cite{Liu:2020ymx}, with the self-duality constraint representing the duality relation \(\bar{\y}\g^{\m\n}\g^5\y \propto \e^{\m\n}{}_{\r\s} \bar{\y} \g^{\r\s} \y\), where \(\e\) is the Levi-Civita symbol~\cite{Alvares:2011wb}. However, it is extremely unlikely that \(B_{mn}\) is genuinely dual to an operator that may be written as \(\bar{\y} \g^{\m\n}\y\), where \(\y\) is an elementary fermion in the dual QFT, and in any case the physics of the model does not appear to depend sensitively on the choice to impose self-duality~\cite{Liu:2020ymx}, so for simplicity we will work with real \(B_{mn}\) with no self-duality constraint. When massless, the two-form field was suggested in ref.~\cite{Gursoy:2010kw} to model superfluidity, with strings winding the Euclidean time circle in the black hole phase spontaneously breaking the U(1) symmetry associated to the gauge symmetry of the two-form field. The role of this two-form symmetry in hydrodynamics, holography and superfluids was later studied in detail in \cite{Grozdanov:2016tdf,Hofman:2017vwr,Delacretaz:2019brr}.}
    \item A \(U(1)\) gauge field \(A_m\), dual to a conserved \(U(1)\) current, modelling the conserved current \(i \bar{\y} \g^\m \y\) in the dual field theory.
\end{itemize}
One might also choose to include an axial vector field \(A_m^5\) to holographically model the axial current \(i \bar{\y} \g^\m\g^5 \y\). However, the axial current will play no role in any of the physics analysed in this paper, so will be neglected.

We take the gravitational action to be
\begin{align}
    S &= \frac{1}{16 \pi \gn} \int_\mathcal{M} \diff^5 x \, \sqrt{-G} \, \le(R + \frac{12}{L^2} - \cL \ri) + \frac{1}{8\pi \gn} \int_{\p \cM} \diff^4 x \, \sqrt{-\g} \, K+ S_\mathrm{ct},
    \nonumber \\ 
    \cL &= |\p \f|^2 + \frac{m_\f^2}{L^2} |\f|^2 + \frac{\h}{2L^2} |\f|^4
        + \frac{1}{6} H^2
        + \frac{m_B^2}{2L^2} B^2
        + \frac{\s}{L^2} |\f|^2 B^2
        + \frac{\l}{4 L^2} \le(B^2\ri)^2
        + \frac{1}{4} F^2,
        \label{eq:holographic_bulk_action}
\end{align}
where \(\gn\) is Newton's constant, \(L\) is the curvature radius of the asymptotically-\ads\ spacetime \(\cM\), \(F = \diff A\), \(H = \diff B\), \(B^2 \equiv B_{mn} B^{mn}\) and similar for \(H\) and \(F\), \(\g\) is the induced metric on the conformal boundary \(\p\cM\), \(K\) is the mean curvature of \(\p\cM\), and \(\h\), \(\s\), and \(\l\) are dimensionless coupling constants. The counterterms \(S_\mathrm{ct}\) are boundary terms necessary to render the on-shell action finite and ensure a well-defined variational principle. The form of the counterterms depends on the masses \(m_\f\) and \(m_B\) of the fields \(\f\) and \(B\) respectively, so we will only give them later, once these have been fixed.

The action in equation~\eqref{eq:holographic_bulk_action} is the same as the action used in ref.~\cite{Liu:2018bye} except for our addition of the \(\l (B^2)^2\) self-coupling of the two-form field. As we will show, by tuning the value of \(\l\) we can choose some of the IR properties of the model to match expected properties of NLSMs, hopefully making the model more applicable to real-world systems. There are of course many other couplings possible, for example one could add a term proportional to \(B^{mn} B_{nk} B^{kl} B_{lm}\), or terms with higher powers of \(\f\) and/or \(B\). A fuller analysis should investigate the effects of such terms.

The equations of motion following from equation~\eqref{eq:holographic_bulk_action} are
\begin{align}
    R_{mn} - \frac{1}{2} G_{mn} R - \frac{6}{L^2} G_{mn} &= \Theta_{mn},
    \nonumber \\
    L^2 \nabla^2 \f - m_\f^2 \f - \h |\f|^2 \f - \s B^2  \f &=0,
    \nonumber \\
    L^2 \nabla^a H_{amn} - m_B^2 B_{mn} -2 \s |\f^2| B_{mn} - \l B^2 B_{mn} &= 0,
    \label{eq:eom_general}
    \\
    \nabla^m F_{mn} &= 0.
    \nonumber
\end{align}
where \(\Theta_{mn}\) is proportional to the bulk stress tensor, explicitly
\begin{equation} \label{eq:bulk_stress_tensor}
    \Theta_{mn} = \p_m \f^* \p_n \f + \frac{1}{2} H_{m ab} H_n{}^{ab} + \frac{1}{L^2} ( m_B^2 + 2 \s |\f|^2 + \l B^2) B_{ma} B_{n}{}^a + \frac{1}{2} F_{ma} F_n{}^a - \frac{1}{2} G_{mn} \cL.
\end{equation}
To solve these equations of motion we make the black-brane ansatz
\begin{equation} \label{eq:background_ansatz}
    \diff s^2 = \frac{L^2}{r^2} \diff r^2 - f(r) g(r) \diff t^2 + h(r) \le(\diff x^2 + \diff y^2 \ri) + g(r) \diff z^2, \quad
    \f = \f(r),
    \quad
    B_{xy} = B(r),
\end{equation}
with all other components of \(B_{mn}\) vanishing. Moreover, we take \(\f\) to be real. We also take \(A_m =0 \) in our ansatz, meaning that the dual field theory will have vanishing net charge density, and will not be subjected to any external electric or magnetic field. Physically, this will later on imply that the Fermi energy lies exactly at the same energy as the nodal line, i.e., our NLSM model is particle-hole symmetric. 

In the coordinates used in the ansatz~\eqref{eq:background_ansatz}, the asymptotically-AdS boundary is at \(r \to 0\). When the dual field theory is at a non-zero temperature \(T\), there is a horizon at some \(r = r_0\) where \(f(r)\) has a double zero while \(g(r)\) and \(h(r)\) are finite. The temperature \(T\) and entropy density \(s\) in the field theory are given by the Hawking temperature and the Bekenstein-Hawking entropy density of the horizon, respectively,
\begin{equation} \label{eq:hawking_temperature}
    T = \frac{r_0}{4\pi L} \sqrt{2 f''(r_0) g(r_0)},
    \qquad
    s = \frac{h(r_0)}{4 \gn} \sqrt{g(r_0)},
\end{equation}
in units where Boltzmann's constant \(k_B=1\). At \(T = 0\), the spacetime extends to \(r=\infty\).

Substituting our ansatz into the equations of motion~\eqref{eq:eom_general}, we find four second-order equations of motion
\begin{subequations} \label{eq:eom_second_order}
\begin{align}
    \frac{f''}{f'} - \frac{f'}{2f} + \frac{g'}{g} + \frac{h'}{h} + \frac{1}{r} &= 0,
    \label{eq:eom_second_order_f}
    \\[1em]
    \frac{h''}{h} + \frac{g'h'}{gh} + \frac{(r^2 f)' h'}{2 r^2 f h} + \frac{2 B'^2}{3 h^2}
    + \frac{4 B^2}{3 r^2 h^2}  \le( m_B^2 + 2 \s \f^2 \ri) \hspace{2cm} &
    \nonumber \\ 
     + \frac{2 \l B^4}{r^2 h^4} + \frac{1}{3 r^2} \le(2 m_\f^2 \f^2 + \h \f^4 - 24 \ri) & = 0,
    \\[1em]
    \f'' + \le(\frac{f'}{2f} + \frac{g'}{g} + \frac{h'}{h} + \frac{1}{r} \ri) \f' - \frac{\f}{r^2} \le(m_\f^2 + \h \f^2 + 2 \s \frac{B^2}{h^2}\ri) &= 0,
    \\[1em]
    B'' + \le( \frac{f'}{2f} + \frac{g'}{g} - \frac{h'}{h} + \frac{1}{r} \ri) B'
    - \frac{L^2 B}{r^2} \le(m_B^2 + 2 \s \f^2 + 2 \l \frac{B^2}{h^2} \ri)
    &= 0,
\end{align}
\end{subequations}
and one first-order equation
\begin{multline} \label{eq:eom_first_order}
    \frac{f'}{2f} + \frac{g'}{g} + \frac{2h'}{h} =
    \\- \sqrt{
        \frac{f'^2}{4 f^2} + \frac{3h'^2}{h^2} + \frac{2B'^2}{h^2} + \frac{24}{r^2}+ 2 \f'^2 -\frac{ 2 m_\f^2 \f^2}{r^2} -\frac{ \h \f^4}{r^2} -  \frac{2 m_B^2 B^2}{r^2h^2} - \frac{4 \s \f^2 B^2}{r^2h^2} -\frac{ 2 \l B^4}{r^2 h^4}
    } .
\end{multline}
These equations of motion are invariant under the separate, constant rescalings \(f \to \Omega_f f\), \(g \to \Omega_g g\), and \((h,B) \to \Omega_h (h,B)\), reflecting the freedom to rescale the coordinates \((t,x,y,z)\) in the metric ansatz~\eqref{eq:background_ansatz}.

As a consequence of these symmetries, the equations of motion imply two radial conservation laws~\cite{Liu:2018bye}.\footnote{One might expect three conservation laws from the three different rescalings. However, the conservation laws arising from \(f \to \Omega_f f\) and \(g \to \Omega_g g\) are identical.} The first follows directly from a rewriting of equation~\eqref{eq:eom_second_order_f},
\begin{equation} \label{eq:Ts_conserved_quantity}
    \p_r \le( \frac{r g h f'}{\sqrt{f}} \ri) = 0
    \qquad
    \Rightarrow
    \qquad
    \frac{r g h f'}{\sqrt{f}} = - 16 \pi \gn L T s,
\end{equation}
where on the right-hand side we have used equation~\eqref{eq:hawking_temperature} to evaluate the conserved quantity at the horizon. Note that at \(T=0\) equation~\eqref{eq:Ts_conserved_quantity} implies that \(f(r)\) is constant. The second conservation law is
\begin{equation} \label{eq:vanishing_conserved_quantity}
    \p_r \le[- r h^2 \sqrt{f} \,  \p_r \le( \frac{g}{h} \ri) + 2 r \sqrt{f} \frac{g}{h} B B' \ri] = 0.
\end{equation}
Evaluating the factor in square brackets at the horizon, we find that it vanishes at \(r=r_0\) since it is proportional to \(\sqrt{f}\), and therefore by equation~\eqref{eq:vanishing_conserved_quantity} it must vanish at all \(r\),
\begin{equation} \label{eq:vanishing_conserved_quantity_2}
    - r h^2 \sqrt{f} \,  \p_r \le( \frac{g}{h} \ri) + 2 r \sqrt{f} \frac{g}{h} B B' = 0.
\end{equation}

Relativistic, \((3+1)\)-dimensional renormalisation group flows obey the \(a\)-theorem, guaranteeing the existence of a quantity \(a\) that monotonically decreases along the flow~\cite{Cardy:1988cwa,OSBORN198997,Jack:1990eb,Komargodski:2011vj}. Since non-zero \(b\) explicitly breaks Lorentz invariance, the \(a\)-theorem does not apply to our system. Nevertheless, following refs.~\cite{Freedman:1999gp,Hoyos:2010at,Liu:2012wf,Giataganas:2017koz} we can use the null energy condition (NEC) to derive two combinations of metric coefficients that must be monotonically decreasing functions of \(r\) in our solutions.

The geometric form of the NEC is \(k^m k^n R_{mn} \geq 0\), where \(k^m\) is a null vector field. The first quantity is obtained by choosing the only non-zero components of the null vector to be \(k^t\) and \(k^r\), for which the NEC implies
\begin{equation}
    -\frac{\sqrt{f}}{r} \le(\frac{g}{h} \ri)^{1/3} \p_r \le[\frac{r}{\sqrt{f}} \le(\frac{h}{g} \ri)^{1/3} \le( \frac{h'}{h} + \frac{g'}{2g} \ri) \ri] - \frac{1}{6} \le( \frac{h'}{h} - \frac{g'}{g}\ri)^2 \geq 0,
\end{equation}
implying that \(\p_r \mathcal{C}_1(r) \leq 0\) for any solution of the equations of motion, where
\begin{equation}
    \mathcal{C}_1(r) = \frac{r}{\sqrt{f}} \le(\frac{h}{g} \ri)^{1/3} \le( \frac{h'}{h} + \frac{g'}{2g}\ri).
\end{equation}
Near the boundary at small \(r\), where we can use the asymptotically-AdS boundary conditions \(f \to 1\) and \(g, h \to L^2/r^2\), we find \(\mathcal{C}_1(r \to 0) = - 3 L^2\).

The second quantity is obtained by choosing the only non-zero components of the null vector to be \(k^t\) and \(k^x\), in which case the NEC yields
\begin{equation}
    - \frac{1}{r g h} \p_r \le[ r g h \le(\frac{h'}{h} - \frac{g'}{g}-\frac{f'}{f} \ri) \ri] + \frac{f'^2}{2f^2} -  \frac{f'}{2f} \le(\frac{h'}{h} - \frac{g'}{g}\ri) \geq 0.
\end{equation}
This implies that when \(T=0\), corresponding to constant \(f\), we must have \(\p_r \mathcal{C}_2(r) \leq 0\), where
\begin{equation} \label{eq:C2}
    \mathcal{C}_2(r) = r g h \le(\frac{h'}{h} - \frac{g'}{g}\ri) = - \frac{r g}{h} B B',
\end{equation}
and the second equality is obtained using equation~\eqref{eq:vanishing_conserved_quantity}. At leading order near the boundary we have \(B \approx b r^{2-\Delta}\), where \(\Delta\) is the conformal dimension of the operator dual to \(B\), and \(b\) is the source for this operator, so that \(\mathcal{C}_2(r) \approx (\Delta-2)b^2 r^{4 - 2 \Delta}\) at small \(r\). The unitarity bound implies \(\Delta > 2\)~\cite{cmp/1103900926,Grinstein:2008qk}, so we find that \(\mathcal{C}_2(r)\) diverges as \(r \to 0\), perhaps making the monotonicity of \(\mathcal{C}_1(r)\) a better candidate for a non-relativistic generalisation of the \(a\)-theorem.

\subsection{Types of solution --- zero temperature}
\label{sec:T0_solutions}

At \(T=0\), equation~\eqref{eq:Ts_conserved_quantity} implies that \(f(r)\) is constant. The equations of motion~\eqref{eq:eom_second_order} and~\eqref{eq:eom_first_order} with constant \(f(r)\) admit multiple types of solution, classified by their behaviour in the deep IR \(r \to \infty\)~\cite{Liu:2018bye}. The different types of solution correspond to different phases of the dual field theory.

\subsubsection{Topological solutions} The first class of solutions have the large-\(r\) behaviour
\begin{equation} \label{eq:ir_asymptotics_topological}
    g \approx g_0 r^{-\a},
    \quad
    h \approx h_0 r^{-\b},
    \quad
    \f \approx \f_0 r^{-\g},
    \quad
    B \approx h_0 B_0 r^{-\b},
\end{equation}
where the equations of motion imply that the constants \(\a\), \(\b\), \(\g\), and \(B_0\) satisfy
\begin{align}
    m_B^2 &= \frac{\b}{\a -\b} \le(48 - 3 \a^2 - 5 \a\b - 4 \b^2 \ri),
    \nonumber
    \\
    \l &= \frac{4 \b^2}{(\a - \b)^2} (\a^2 + \a\b + \b^2 - 12),
    \nonumber
    \\
    \g &= \sqrt{\frac{(\a+\b)^2}{4} + \s \le(\frac{\a}{\b}-1 \ri) + m_\f^2} - \frac{\a + \b}{2},
    \label{eq:topological_coefficients_general}
    \\
    B_0 &= \sqrt{ \frac{\a - \b}{2\b}}.
    \nonumber
\end{align}
This solution is distinguished from the other solutions described below by the fact that the scalar field vanishes as \(r \to \infty\), since \(\g > 0\) in equation~\eqref{eq:topological_coefficients_general}, while \(B/h\) tends to the non-zero constant value \(B_0\).

The coefficients \(g_0\) and \(h_0\) appearing in equation~\eqref{eq:ir_asymptotics_topological} are arbitrary, since the equations of motion are invariant under the rescalings \(g \to \Omega_g g\) and \((h,B) \to \Omega_h (h,B)\). This leaves a single parameter family of solutions, parameterised by \(\f_0\). In this phase, for the values of the couplings used in refs.~\cite{Liu:2018bye,Liu:2018djq} the spectral functions of composite fermionic operators exhibit multiple circular nodal lines. We show in section~\ref{sec:fermions} that the same is true for the values of couplings that we will choose, so we will refer to this phase as \textit{topological}.

The large-\(r\) behaviour of the metric functions written in equation~\eqref{eq:ir_asymptotics_topological} imply that the asymptotic metric as \(r \to \infty\) has a scaling isometry. Concretely, if we make the coordinate transformation
\begin{equation}
    r' = r^{\a/2},
    \quad
    t' = \frac{\a \sqrt{g_0}}{2L} t,
    \quad
    x' = \frac{\a \sqrt{h_0}}{2L} x,
    \quad
    y' = \frac{\a \sqrt{h_0}}{2L} y,
    \quad
    z' = \frac{\a \sqrt{g_0}}{2L} z,
\end{equation}
then at large \(r'\) the metric~\eqref{eq:background_ansatz} becomes
\begin{equation} \label{eq:IR_metric}
    \diff s^2 \approx \frac{L'^2}{r'^2} \le( \diff r'^2 - \diff t'^2 + \diff z'^2 \ri) + \frac{L'^2}{r'^{2\b/a}} \le( \diff x'^2 + \diff y'^2 \ri),
\end{equation}
where we have defined \(L' = 2 L/\a\). This metric is manifestly invariant under the combined rescaling \((r',t',z') \to \Omega (r',t',z')\) and \((x',y') \to \Omega^{\b/\a} (x,y)\). Solutions of this type therefore describe an RG flow from a conformally invariant UV fixed point to an IR fixed point with non-relativistic, anisotropic scale invariance with dynamical exponent \(\a/\b\) in the \(x\) and \(y\) directions and unit dynamical exponent in the \(z\) direction. The requirements that \(B_0\) in equation~\eqref{eq:topological_coefficients_general} is real and that the IR metric in equation~\eqref{eq:IR_metric} satisfies the NEC both imply that the dynamical exponent is bounded from below, \(\a/\b \geq 1\).

The dynamical exponent \(\a/\b\) determines many of the IR properties of this phase. In particular, we will show in section~\ref{sec:transport} that at small frequency \(\w\), the AC conductivity in the direction orthogonal to the plane of the nodal line scales as \(\s_{zz} \propto \w^{2\b/\a -1}\). We will choose to engineer a constant \(\s_{zz}\) by setting the parameters of our model such that the dynamical exponent is \(\a/\b = 2\). For example, from equation~\eqref{eq:topological_coefficients_general} we find that for \(m_B^2 =1\) and \(\l = 34/13\), one has \(\a = 2 \b = \sqrt{94/13}\). 

The physical motivation for a constant $\sigma_{zz}$ is that, due to the rotational symmetry around the $z$ axis, we may think of the nodal line semimetal as an infinite collection of graphene-like sheets, with a single sheet for each azimuthal angle, and the conductivity of graphene is known to be finite and nonzero at the charge neutrality point due to particle-hole symmetry. While crude, this intuitive picture appears to give correct results for \(\s_{zz}\) in real NLSMs~\cite{PhysRevLett.119.147402}. However, the same physical picture would also suggest that $\sigma_{xx}$ and $\sigma_{yy}$ would go to the same non-zero constant, which cannot be achieved within the current setup; we always find \(\s_{xx} = \s_{yy} \propto \w\). A more realistic holographic model should therefore include one or more additional fields to engineer \(\s_{xx} = \s_{yy} \propto \w^0\). We leave the exploration of this to future work.

\subsubsection{Topologically trivial solutions} This class of solutions has the large-\(r\) behaviour
\begin{equation}
    g \approx g_0 r^{-\a},
    \quad
    h \approx h_0 r^{-\b},
    \quad
    \f \approx \f_0,
    \quad
    B \approx h_0 B_0 r^{-\g},
\end{equation}
where
\begin{equation}  \label{eq:alpha_beta_trivial}
    \a = \b = \sqrt{4 + \frac{m_\f^4}{6\h}},
    \qquad
    \g = \sqrt{m_B^2 - \frac{2 \s m_\f^2}{\h}},
    \qquad
    \f_0 = \sqrt{- \frac{m_\f^2}{\h}}.
\end{equation}
In these solutions, the scalar field \(\f\) tends to the constant \(\f_0\) as \(r \to \infty\), while \(B/h \to 0\) in the same limit.
Once more there is a single parameter family of solutions, this time parameterised by \(B_0\).
Setting \(\a=\b\) in equation~\eqref{eq:IR_metric}, we see that the IR geometry is \ads[5] with curvature radius \(L' = 2L/\a\). These solutions therefore describe an RG flow between two conformally invariant fixed points. Note that for \(\f\) to be dual to a relevant operator we must take \(m_\f^2 < 0\), so that from equation~\eqref{eq:alpha_beta_trivial} we see that these solutions only exist for \(\h > 0\). This then implies \(\a > 2\) and therefore \(L' < L\).

\subsubsection{Critical solution} Finally there is a single solution with the large-\(r\) behaviour
\begin{equation}
    g \approx g_0 r^{-\a},
    \quad
    h \approx h_0 r^{-\b},
    \quad
    \f \approx \f_0,
    \quad
    B \approx h_0 B_0 r^{-\b},
\end{equation}
where in this case \(\a\), \(\b\), \(B_0\), and \(\f_0\) satisfy
\begin{align}
    m_B^2 &= \frac{\b}{\a - \b} \le(48 - 3 \a^2 - 5 \a \b - 4 \b^2 + \frac{2 m_\f^4}{\h} \ri) + \frac{2 \s m_\f^2}{\h},
    \nonumber \\
    \l &= \frac{4 \b^2}{(\a - \b)^2} \le(\a^2 + \a \b + \b^2- 12 - \frac{m_\f^4}{2 \h}\ri) + \frac{2 \s^2}{\h},
    \nonumber \\
    B_0 &= \sqrt{\frac{\a - \b}{2 \b}},
    \\
    \f_0 &= \sqrt{- \frac{m_\f^2}{\h} - \frac{\s(\a - \b)}{\h\b}}.
\end{align}
Since \(g_0\) and \(h_0\) are arbitrary, due to the scaling symmetries of the equations of motion, there is only one solution in this class. We will refer to this solution as the critical solution, since it turns out to describe the critical point of a second-order phase transition between the topological and topologically trivial solutions. In the critical solution, both \(\f\) and \(B/h\) tend to non-zero constants as \(r \to \infty\). Like the topological solutions, the critical solution has non-relativistic, anisotropic scale invariance in the IR, but with a different value for the dynamical exponent~\(\a/\b\).

\subsection{Thermodynamics and one-point functions}
\label{sec:thermodynamics}

We now give expressions for various thermodynamic quantities and one-point functions in the dual field theory. To do so we will need to fix the masses of the fields \(\f\) and \(B_{mn}\), as the masses will determine the near-boundary expansions of these fields, and consequently the form of the counterterms. Following ref.~\cite{Liu:2018bye}, we will take \(m_\f^2 = -3\) and \(m_B^2 = 1\). With these choices, both \(\f\) and \(B_{mn}\) are dual to operators of conformal dimension \(\Delta=3\) in the dual field theory, the same as the dimensions as the operators \(\bar{\y} \y\) and \(\bar{\y} \g^{\m\n} \y\) in the free field model~\eqref{eq:toy_model_lagrangian}, respectively. Of course, interactions will in general renormalise the dimensions of these operators, so our choice of masses has been made purely for concreteness.

For the choice of masses described above, the fields of our ansatz have the near-boundary expansions
\begin{align}
    f &= 1 - m r^4 + \dots \; ,
    \nonumber
    \\
    g &= \frac{L^2}{r^2} \le[1 - \frac{M^2 + 3 b^2}{6} r^2 - \le(\frac{2 + 3 \h}{24} M^4 + \frac{1-\l}{12} b^4 + \frac{\s}{6} M^2 b^2\ri) r^4 \log (r/L)   + g_4 r^4 \ri] + \dots \,,
    \nonumber
    \\
    h &= \frac{L^2}{r^2} \le[1 - \frac{M^2 - 3 b^2}{6}r^2 - \le(\frac{2 + 3 \h}{24} M^4 + \frac{1-\l}{12} b^4 + \frac{\s}{6} M^2 b^2\ri)  r^4 \log (r/L) + h_4 r^4 \ri]  + \dots \;,
    \nonumber
    \\
    \f &= r \le[M + \le(\frac{2 + 3 \h}{6}M^3 + \s M  b^2\ri) r^2 \log(r/L) + \f_2 r^2 \ri] + \dots \; ,
    \label{eq:bg_near_boundary}
    \\
    B &= \frac{L^2}{r} \le[b - \le((1-\l) b^3 - \s M^2 b \ri) r^2 \log(r/L) + b_2 r^2  \ri] + \dots \; ,
    \nonumber
\end{align}
where the dots denote terms of higher order in the small-\(r\) expansion, \(m\), \(M\), \(b\), \(\f_2\), and \(b_2\) are integration constants, and \(g_4\) and \(h_4\) are given by
\begin{align}
    g_4 &= \frac{m}{4} - \frac{M \f_2}{4} + \frac{b b_2}{12} + \frac{4 + 3 \h}{144}  M^4 + \frac{2 + 7 \l}{24} b^4 + \frac{\s}{3} M^2 b^2,
    \nonumber \\
    h_4 &= \frac{m}{4} - \frac{M \f_2}{4} + \frac{b b_2}{12} + \frac{4 + 3 \h}{144} M^4 + \frac{2 -5 \l}{24} b^4 - \frac{\s}{6} M^2 b^2.
\end{align}
The coefficients \(M\) and \(b\) are proportional to the sources for the operators dual to \(\f\) and \(B\), while \(\f_2\) and \(b_2\) will determine the corresponding vacuum expectation values. Substituting the expansions~\eqref{eq:bg_near_boundary} into equation~\eqref{eq:Ts_conserved_quantity}, we find that \(m\) is related to the temperature and entropy density as
\begin{equation} \label{eq:m_T_s_relation}
    m = \frac{4 \pi \gn}{L^3} T s.
\end{equation}

With \(m_\f^2 = -3\) and \(m_B^2 =1\), the counterterm action appearing in equation~\eqref{eq:holographic_bulk_action} may be taken to be~\cite{deHaro:2000vlm,Bianchi:2001kw,Liu:2018bye}
\begin{align} \label{eq:counterterms}
    S_\mathrm{ct} &= \frac{1}{16 \pi \gn L} \int_{r=\e} \diff^4 x \, \sqrt{-\g} \biggl[
        -6 - \frac{L^2}{2} R[\g] - |\f|^2 + \frac{1}{2} B_{\m\n} B^{\m\n}
    \nonumber \\ &\phantom{= \frac{1}{16 \pi \gn L} \int_{r=\e} \diff^4 x \, \sqrt{-\g} \biggl[}\hspace{1cm}
        + L^2  \hat{\nabla}_\l B_{\m\n} \hat{\nabla}^\l B^{\m\n}
     - a \log\le(\frac{r}{L}\ri)
    \biggr],
    \nonumber \\[0.5em]
    a &= 
        \frac{L^4}{6} R[\g]^2 
        - \frac{L^4}{4} R_{\m\n}[\g]  R^{\m\n}[\g]
        + \frac{L^2}{6} R[\g] \f^2
        - \frac{L^2}{4} R[\g] B_{\m\n} B^{\m\n}
        + \frac{L^2}{4} F_{\m\n} F^{\m\n}
        \\ &\phantom{=}
        + L^2 \p_\m \f^* \p^\m \f
        + \frac{L^2}{6} H_{\m\n\r} H^{\m\n\r}
        +   \frac{2+3\h}{6}|\f|^4 
        + \s |\f|^2 B_{\m\n}B^{\m\n} 
        + \frac{\l-1}{4} (B_{\m\n}B^{\m\n})^2.
        \nonumber
\end{align}
where \(\e\) is a small-\(r\) cutoff, the field theory \((\m\n)\) indices are raised and lowered with the induced metric \(\g_{\m\n}\), \(\hat{\nabla}\) is the covariant derivative calculated with respect to \(\g\), \(R_{\m\n}[\g]\) is the Ricci tensor computed with \(\g\), and \(R[\g]= \g^{\m\n} R_{\m\n}[\g]\) is the corresponding Ricci scalar. Note that there is a choice of renormalisation scheme implicit in the counterterms~\eqref{eq:counterterms}; one could choose to add additional, finite counterterms that would modify the one-point functions of operators by local functions of the sources.\footnote{The finite counterterm involving \(\hat{\nabla}_\l B_{\m\n} \hat{\nabla}^\l B^{\m\n}\) appearing in equation~\eqref{eq:counterterms} has been included to remove a contact term from the two-point functions calculated in section~\ref{sec:thermal_conductivity}.}

The Helmholtz free energy at temperature \(T\) is given by \(\cF = T I^\star\), where \(I^\star\) is the Euclidean signature on-shell action. Using the equations of motion, the on-shell action may be reduced to a boundary term. Upon substitution of the near-boundary expansion~\eqref{eq:bg_near_boundary}, one finds
\begin{equation}
    \cF = - \frac{V L^3}{16 \pi \gn} \le(m + M \f_2 + b b_2 + \frac{2 + 3 \h}{24} M^4 + \frac{\s}{2} M^2 b^2 + \frac{\l-1}{4} b^4 \ri),
\end{equation}
where \(V\) is the spatial volume of our system. The one-point function of the scalar operator dual to \(\f\) is given by \(\vev{\cO} = \d S^\star / \d M\), where \(S^\star\) is the Lorentzian-signature on-shell action. The equations of motion imply that this variation is a boundary term, and using the near-boundary expansions~\eqref{eq:bg_near_boundary} we find
\begin{equation} \label{eq:scalar_vev}
    \vev{\cO} = \frac{L^3}{16 \pi \gn} \le(4 \f_2 + \frac{2 + 3 \h}{3} M^3 + 2 \s M b^2 \ri).
\end{equation}
With our ansatz, the only non-zero components of the one-point function of the antisymmetric tensor operator dual to \(B_{mn}\) are \(\vev{\cO^{xy}} = -\vev{\cO^{yx}} = \frac{1}{2} \d S^\star/\d b\). We find\footnote{One could eliminate some of the terms that depend only on sources in equations~\eqref{eq:scalar_vev} and~\eqref{eq:antisymmetric_vev} by the addition of finite counterterms to the action. However it is not possible to remove all such terms, i.e. we cannot choose a renormalization scheme in which both \(\vev{\cO} \propto \f_2\) and \(\vev{\cO^{xy}} \propto b_2\).}
\begin{equation} \label{eq:antisymmetric_vev}
    \vev{\cO^{xy}} = \frac{L^3}{16 \pi \gn} \le[2 b_2 + (\l-1) b^3 + \s M^2 b \ri].
\end{equation}
Further details of the computation of \(\cF\), \(\vev{\cO}\), and \(\vev{\cO^{xy}}\) are given in appendix~\ref{app:holo_rg}.

Differentiation of the on-shell action with respect to the boundary metric, yields the one-point function of a symmetric tensor operator~\cite{deHaro:2000vlm}
\begin{equation} \label{eq:T_tilde_holographic}
    \vev{\tilde{T}_{\m\n}} \equiv - \lim_{\e \to 0} \frac{L^2}{\e^2} \frac{2}{\sqrt{-\g}} \frac{\d S}{\d \g^{\m\n}}
    = \lim_{\e \to 0} \frac{L^2}{\e^2}\le[ - \frac{1}{8\pi\gn} (K_{\m\n} - K \g_{\m\n}) - \frac{2}{\sqrt{-\g}} \frac{\d S_\mathrm{ct}}{\d \g^{\m\n}} \ri],
\end{equation}
where \(K_{\m\n}\) and \(K\) are the extrinsic and mean curvature of the cutoff surface at \(r=\e\), respectively. The tensor \(\tilde{T}_{\m\n}\) is not quite the stress tensor, since it is not conserved in the presence of a non-zero source \(b_{\m\n}\).\footnote{A similar phenomenon with a one-form source is described in ref.~\cite{Taylor:2015glc}.} Instead, as we will show in the next subsection, the conserved stress tensor is \(T_{\m\n} \equiv \tilde{T}_{\m\n} + 2 \cO_{\m\r} b_\n{}^\r\). For our black brane ansatz we find \(\vev{T_{\m\n}} = \mathrm{diag}(\ve,p,p,p)\), where the energy density \(\ve\) and pressure \(p\) are given by
\begin{align}
    \ve &= \frac{L^3}{16 \pi \gn} \le(3m - M \f_2 - b b_2 - \frac{2 + 3 \h}{24} M^4 - \frac{\s}{2} M^2 b^2  - \frac{\l-1}{4} b^4\ri),
    \nonumber
    \\
    p &=  \frac{L^3}{16 \pi \gn} \le(m + M \f_2 + b b_2 + \frac{2 + 3 \h}{24} M^4 + \frac{\s}{2} M^2 b^2  + \frac{\l-1}{4} b^4 \ri).
    \label{eq:energy_pressure}
\end{align}
The first law of thermodynamics for our system reads \(\diff \ve = T \diff s - \vev{\cO}\diff M - 2\vev{\cO^{xy}} \diff b \).

As consistency checks of the expressions in this section, note that \(p = - \cF/V\), and that \(\ve\) and \(p\) satisfy the expected thermodynamic relation
\begin{equation}
    \ve + p = \frac{L^3 m}{4\pi\gn} = T s,
\end{equation}
where we obtain the second equality using equation~\eqref{eq:m_T_s_relation}, and we remind the reader that we are working at vanishing chemical potential. The one-point function of the trace of the stress tensor also satisfies the appropriate Ward identity
\begin{equation}
    \vev{T^\m{}_\m} = - \b(M) \vev{\cO} - \b(b) \le(\vev{\cO^{xy}} + \vev{\cO^{yx}} \ri) + \cA,
\end{equation}
where \(\b(M) = -M\) and \(\b(b) = -b\) are the beta functions for the sources, and we find the Weyl anomaly to be
\begin{equation}
    \cA  = - \frac{L^3}{16 \pi \gn} \le[\frac{2 + 3 \h}{6} M^4 + 2 \s M^2 b^2 + (\l-1) b^4 \ri].
\end{equation}

\subsection{Conservation of the stress tensor}
\label{sec:stress_tensor_comments}

In this subsection we show that the operator \(\tilde{T}^{\m\n}\) defined in equation~\eqref{eq:T_tilde_holographic} is not the stress tensor, since it does not satisfy the appropriate Ward identity. We will consider a general quantum field theory on a spacetime with metric \(g_{\m\n}\), which we will later set to the Minkowski metric, in the presence of an external two-form source \(b_{\m\n}\). Writing the generating functional of connected correlation functions as \(W[g,b]\), the one-point function of \(\tilde{T}^{\m\n}\) may be calculated by functional differentiation of \(W\) with respect to \(g_{\m\n}\),
\begin{equation} \label{eq:almost_stress_tensor_formula}
    \vev{\tilde{T}^{\m\n}} =  - i \frac{2}{\sqrt{-g}}  \frac{\d W}{\d g_{\m\n}} ,
\end{equation}
Similarly, the one-point function of the operator \(\cO^{\m\n}\) sourced by \(b_{\m\n}\) is
\begin{equation} \label{eq:antisymmetric_tensor_vev_formula}
    \vev{\cO^{\m\n}} = - i \frac{1}{\sqrt{-g}} \frac{\d W}{\d b_{\m\n}}.
\end{equation}

The method we use to obtain the Ward identity is well known, see for example ref.~\cite{Osborn:1999az}. We begin by considering the effect of an infinitesimal coordinate transformation generated by a vector \(\xi^\m\). The corresponding infinitesimal changes in the metric and source are
\begin{equation} \label{eq:ward_identity_coupling_change}
    \d g_{\m\n} = - \nabla_\m \xi_\n - \nabla_\n \xi_\m,
    \quad
    \d b_{\m\n} = - \xi^\r \nabla_\r b_{\m\n}  - b_{\m\r} \nabla_\n \xi^\r - b_{\r\n} \nabla_\m \xi^\r.
\end{equation}
Demanding that the generating functional is invariant under this coordinate transformation, to linear order in \(\d g\) and \(\d b\) we have
\begin{align}
    W[g,b] &= W[g+\d g, b+\d b]
    \nonumber \\
    &= W[g,b] + \int \diff^4 x \le[\frac{\d W}{\d g_{\m\n}} \d g_{\m\n} + \frac{\d W}{\d b_{\m\n}} \d b_{\m\n} \ri] + \dots \; ,
    \label{eq:generating_functional_change_coordinates}
\end{align}
which implies that the second term on the right-hand side must vanish. Substituting equations~\eqref{eq:almost_stress_tensor_formula} and~\eqref{eq:antisymmetric_tensor_vev_formula} for the functional derivatives, setting the metric to be flat, and performing an integration by parts, this implies
\begin{equation}
   \int \diff^4 x \, \xi_\n \le[
        \p_\m \vev{\tilde{T}^{\m\n}}
        - \vev{\cO^{\r\s}} \p^\n b_{\r\s}
        + 2 \p_\mu \le(\vev{\cO^{\m\r}} b^{\n}{}_\r \ri)
    \ri] = 0.
    \label{eq:generating_functional_change_coordinates_expanded}
\end{equation}

In order for equation~\eqref{eq:generating_functional_change_coordinates_expanded} to hold for any choice of \(\xi\), we find that the factor in square brackets must vanish, yielding the Ward identity
\begin{equation} \label{eq:Ttilde_ward_identity}
    \p_\m \vev{\tilde{T}^{\m\n}} = \vev{\cO^{\r\s}} \p^\n b_{\r\s} - 2 \p_\mu \le(\vev{\cO^{\m\r}} b^{\n}{}_\r \ri).
\end{equation}
The first term on the right-hand side reflects the explicit breaking of translational invariance that arises if \(b\) is position-dependent. On the other hand, the second term on the right-hand side of equation~\eqref{eq:Ttilde_ward_identity} can be non-zero even if translational symmetry is not explicitly broken, in other words even if \(\p_\m b_{\r\s} = 0\), so \(\tilde{T}^{\m\n}\) is manifestly not the conserved current associated to translational invariance. Instead, the conserved stress tensor is
\begin{equation} \label{eq:conserved_stress_tensor}
    T^{\m\n} = \tilde{T}^{\m\n} + 2 \cO^{\m\r} b^\n{}_\r,
\end{equation}
which satisfies the correct Ward identity
\begin{equation} \label{eq:translation_ward_identity}
    \p_\m \vev{T^{\m\n}} = \vev{\cO^{\r\s}} \p^\n b_{\r\s}.
\end{equation}
We show in appendix~\ref{app:holo_ward_identity} that the \(T^{\m\n}\) that we obtain in holography satisfies this Ward identity. Note that \(\tilde{T}^{\m\n}\) is manifestly symmetric in its indices, since it is calculated from a functional derivative with respect to the symmetric \(g_{\m\n}\). However, the second term on the right-hand side of equation~\eqref{eq:conserved_stress_tensor} is not symmetric in general, so the stress tensor \(T^{\m\n}\) is \textit{not} a symmetric tensor. This is an example of the general principle that the stress tensor is asymmetric in any theory with explicitly broken Lorentz invariance~\cite{Guica:2010sw}, see also refs.~\cite{Taylor:2015glc,PhysRevB.100.045114}.

\subsection{Numerical results}
\label{sec:numerical_solutions}

We now present our numerical results for the thermodynamic quantities and one-point functions of the holographic NLSM. In addition to fixing the masses of the scalar and two-form fields to be \(m_\f^2 = - 3\) and \(m_B^2 = 1\), as discussed in section~\ref{sec:thermodynamics}, we need to fix the various couplings appearing in the action~\eqref{eq:holographic_bulk_action}. We choose the two-form self-coupling to be \(\l = 34/13\). As discussed in section~\ref{sec:T0_solutions} this implies that the dynamical exponent in the topological phase is \(\a/\b = 2\). With these choices of \(m_\f^2\), \(m_B^2\) and \(\l\), we must take the coupling between the scalar and two-form to satisfy \(\s \geq3\), so that \(\g \geq 0\) in equation~\eqref{eq:topological_coefficients_general}, while the scalar self-coupling must obey \(\h > 0\) so that \(\f_0\) in equation~\eqref{eq:alpha_beta_trivial} is real. It will be convenient for numerical purposes to choose these couplings so that the various coefficients and exponents in section~\ref{sec:T0_solutions} are relatively simple. We will take \(\h = 2\) and \(\s = 8\).

At \(T=0\) we solve the equations of motion numerically by integrating out from the horizon, using the large-\(r\) asymptotics in section~\ref{sec:T0_solutions} to set boundary conditions. We can then fit the numerical solutions to the near-boundary behaviour~\eqref{eq:bg_near_boundary} at small \(r\) to determine the coefficients \((M,b,\f_2,b_2)\). At \(T=0\) the dual field theory contains two independent dimensionful quantities, \(M\) and \(b\), with all other quantities depending on the dimensionless ratio \(M/b\). The topological and topologically trivial solutions each depend on a single parameter, \(B_0\) and \(\f_0\) respectively, which map to \(M/b\) through the bulk equations of motion. We find that the critical solution has \(M/b = (M/b)_\mathrm{crit.} \simeq 0.9493\). The topological solutions exist only for \(M/b < (M/b)_\mathrm{crit.}\), while the topologically trivial solutions exist only for \(M/b > (M/b)_\mathrm{crit.}\). 

\begin{figure}
    \begin{subfigure}{0.5\textwidth}
        \includegraphics[width=\textwidth]{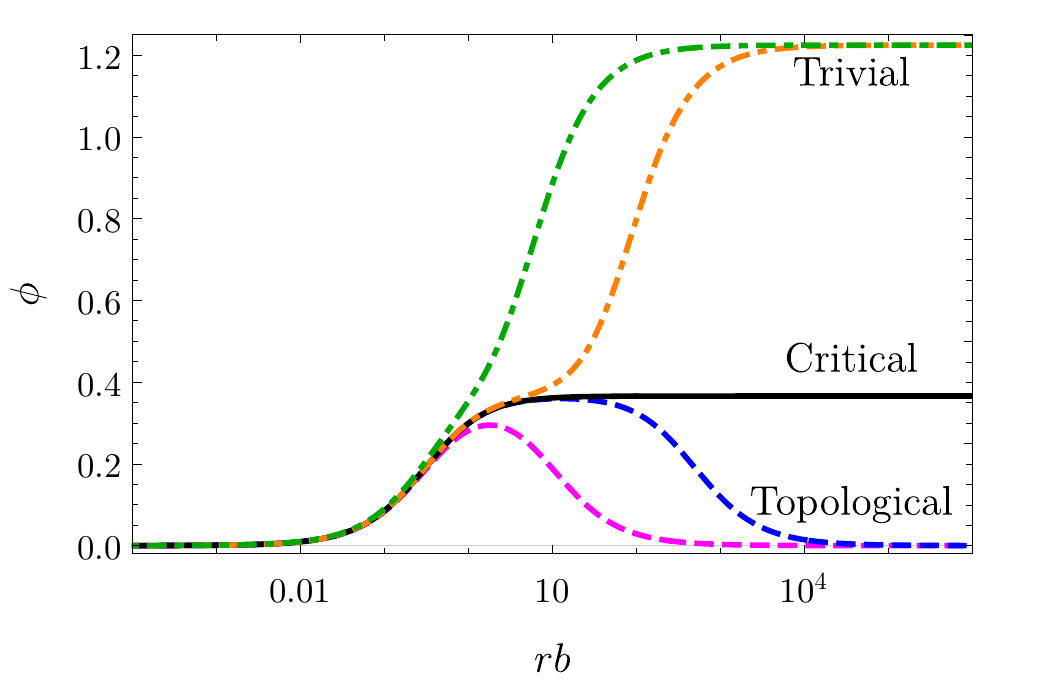}
        \caption{Scalar profile}
    \end{subfigure}
    \begin{subfigure}{0.5\textwidth}
        \includegraphics[width=\textwidth]{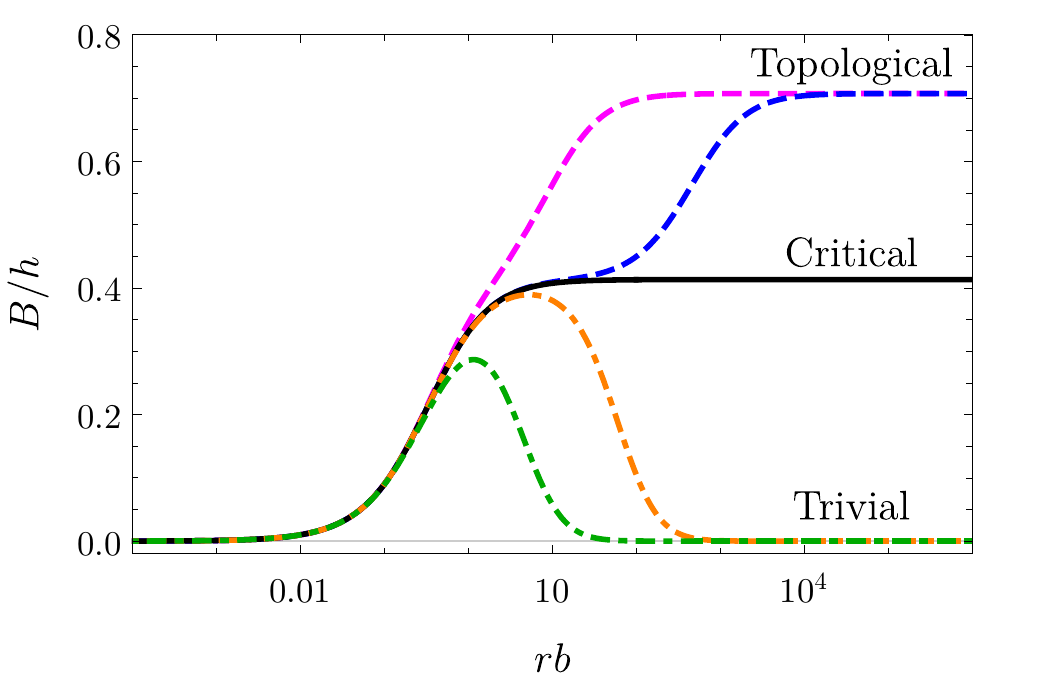}
        \caption{Two-form profile}
    \end{subfigure}
    \begin{subfigure}{0.5\textwidth}
        \includegraphics[width=\textwidth]{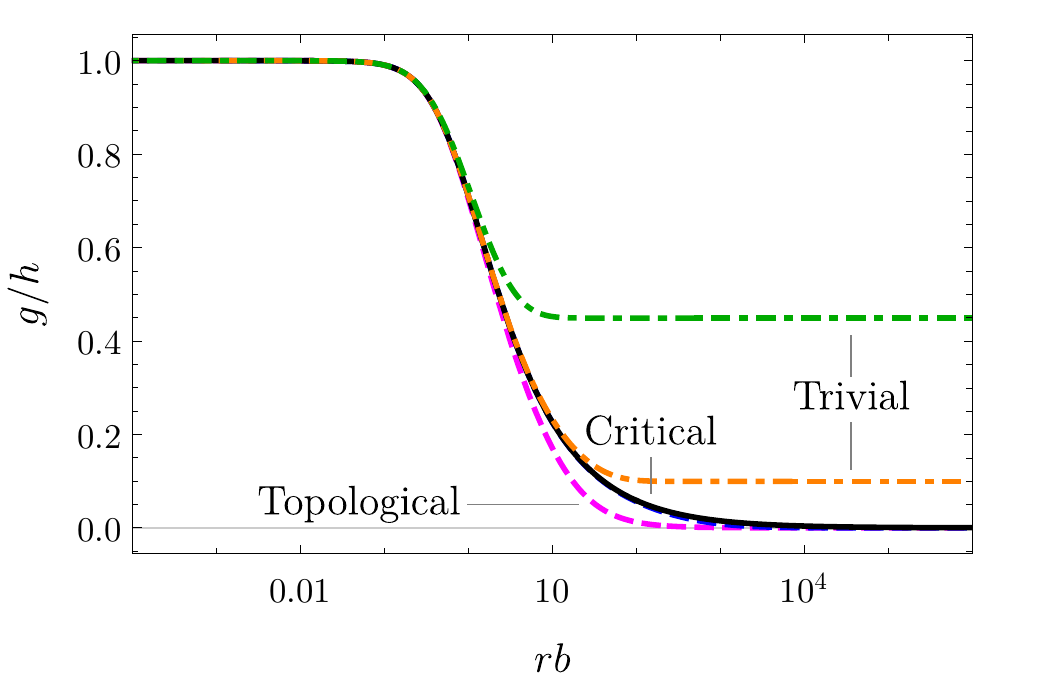}
        \caption{Ratio of metric functions}
    \end{subfigure}
    \begin{subfigure}{0.5\textwidth}
        \includegraphics[width=\textwidth]{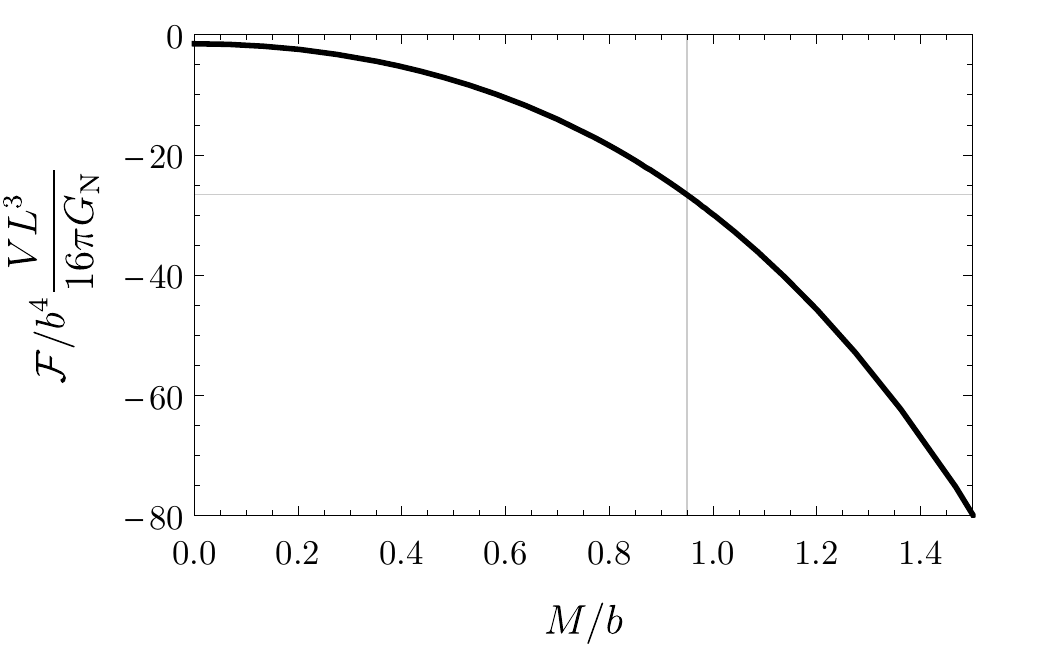}
        \caption{Free energy}
    \end{subfigure}
    \begin{subfigure}{0.5\textwidth}
        \includegraphics[width=\textwidth]{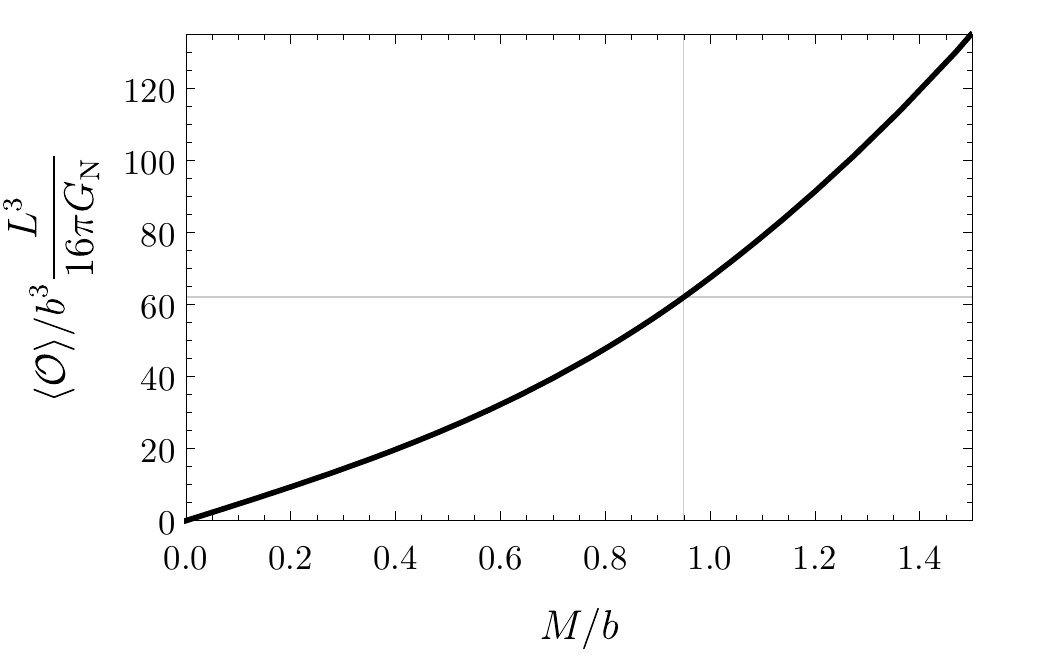}
        \caption{Scalar one-point function}
    \end{subfigure}
    \begin{subfigure}{0.5\textwidth}
        \includegraphics[width=\textwidth]{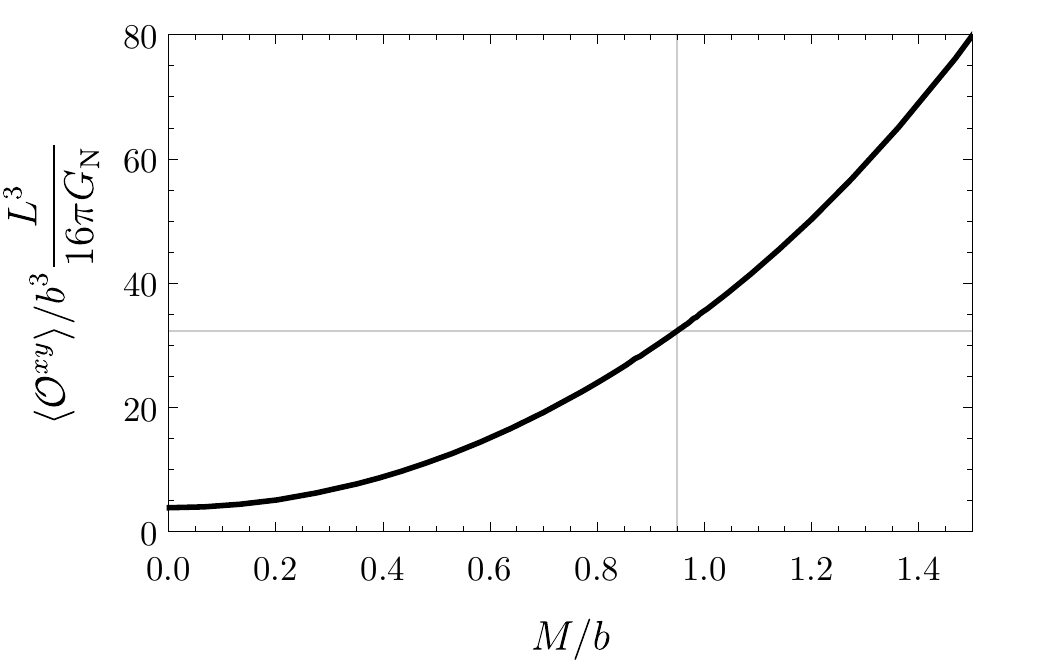}
        \caption{Antisymmetric tensor one-point function}
    \end{subfigure}
    \caption{
        \textbf{(a,\,b\,c):} Sample radial profiles of the fields in the different types of solutions at \(T=0\). The dashed magenta and blue curves are sample solutions in the topological phase, the dot-dashed orange and green curves are sample solutions in the topologically trivial phase, while the solid black curve is the critical solution. \textbf{(d,\,e,\,f):} The free energy, scalar one-point function and antisymmetric tensor one-point function, as functions of \(M/b\) at \(T=0\). All three quantities are continuous across the phase transition at \((M/b)_{\mathrm{crit.}} \simeq 0.9493\), the location of which is indicated by the thin grey lines in the plots.
    }
    \label{fig:T0_numerics}
\end{figure}

We plot some sample numerical solutions for different values of \(M/b\) in figure~\ref{fig:T0_numerics}. The dashed magenta and blue curves correspond to topological solutions, characterised by the vanishing of \(\f\) as \(r \to \infty\). The dot-dashed orange and green curves correspond to topologically trivial solutions, for which \(B/h \to 0\) as \(r \to \infty\). The solid black curve corresponds to the critical solution, for which both \(\f\) and \(B/h\) are non-zero as \(r \to \infty\). In figure~\ref{fig:T0_numerics} we also plot the free energy and the one-point functions of the scalar and antisymmetric tensor operators as functions of \(M/b\). All three of these quantities are continuous across \(\le(M/b\ri)_\mathrm{crit.}\). However, not every physical quantity is continuous across \(\le(M/b\ri)_\mathrm{crit.}\). In particular, we find in section~\ref{sec:conductivity} that the DC electrical conductivity in the \(z\) direction has a discontinuous first derivative with respect to \(M/b\) at \(\le(M/b\ri)_\mathrm{crit.}\), so we identify \(\le(M/b\ri)_\mathrm{crit.}\) as the location of a second-order quantum phase transition between the two phases dual to the topological and topologically trivial solutions.

At non-zero \(T\) we solve the equations of motion by integrating out from the horizon at \(r=r_0\), imposing the boundary conditions that \(f\) has a double zero at \(r_0\) while \(g\), \(h\), \(\f\), and \(B\) are finite and non-zero. Since the values of \(g\), \(h\), and the second derivative of \(f\) at the horizon may be set to any desired numbers using the scaling symmetries mentioned under equation~\eqref{eq:eom_first_order}, there are only two free parameters in the boundary conditions: the values of \(\f\) and \(B\) at the horizon. These parameters map to the two dimensionless ratios in the dual field theory, \(T/b\) and \(M/b\), through the bulk equations of motion.

\begin{figure}
    \begin{subfigure}{\textwidth}
    \begin{center}
        \includegraphics{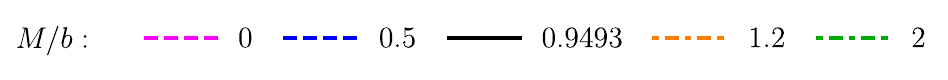}
    \end{center}
    \end{subfigure}
    \begin{subfigure}{0.5\textwidth}
        \includegraphics[width=\textwidth]{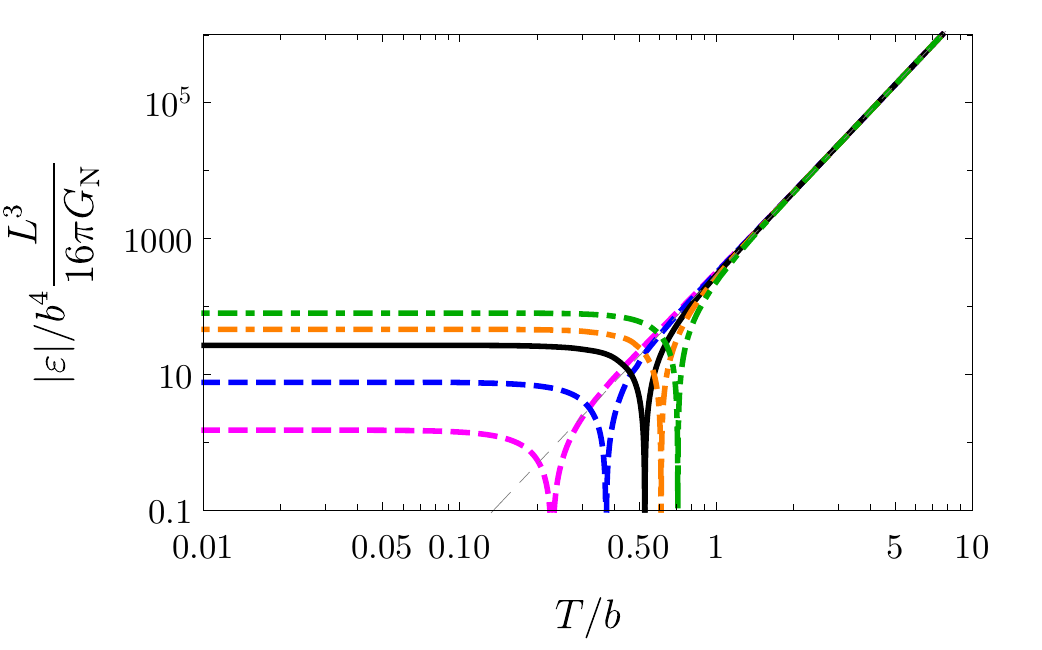}
        \caption{Energy density}
        \label{fig:energy_density}
    \end{subfigure}
    \begin{subfigure}{0.5\textwidth}
        \includegraphics[width=\textwidth]{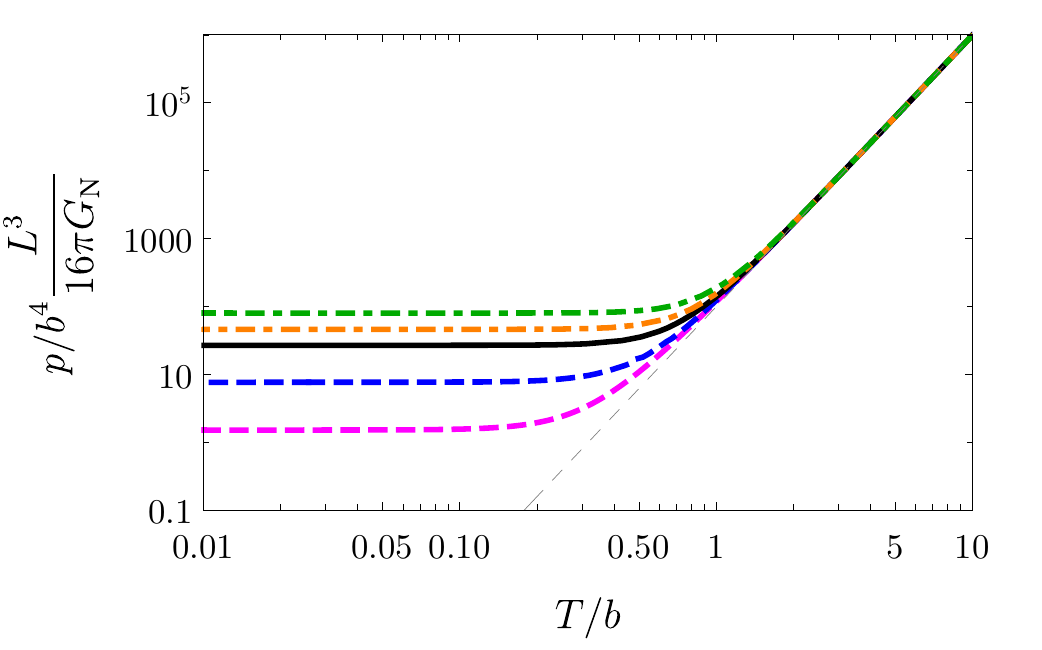}
        \caption{Pressure}
        \label{fig:pressure}
    \end{subfigure}
    \begin{subfigure}{0.5\textwidth}
        \includegraphics[width=\textwidth]{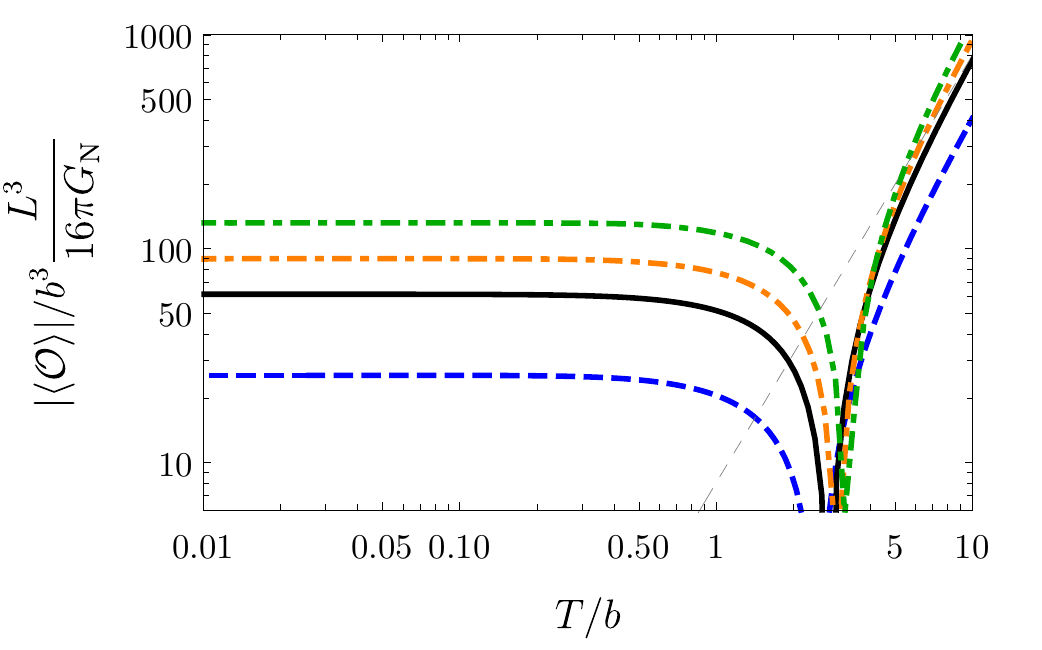}
        \caption{Scalar one-point function}
        \label{fig:O_phi}
    \end{subfigure}
    \begin{subfigure}{0.5\textwidth}
        \includegraphics[width=\textwidth]{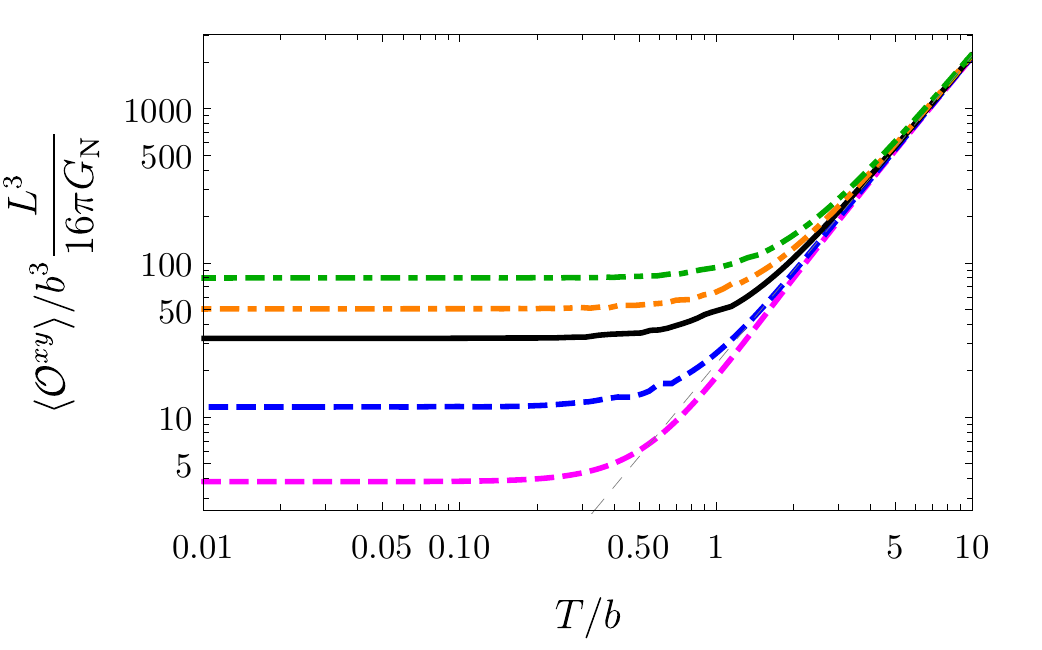}
        \caption{Antisymmetric tensor one-point function}
        \label{fig:O_B}
    \end{subfigure}
    \begin{subfigure}{0.5\textwidth}
        \includegraphics[width=\textwidth]{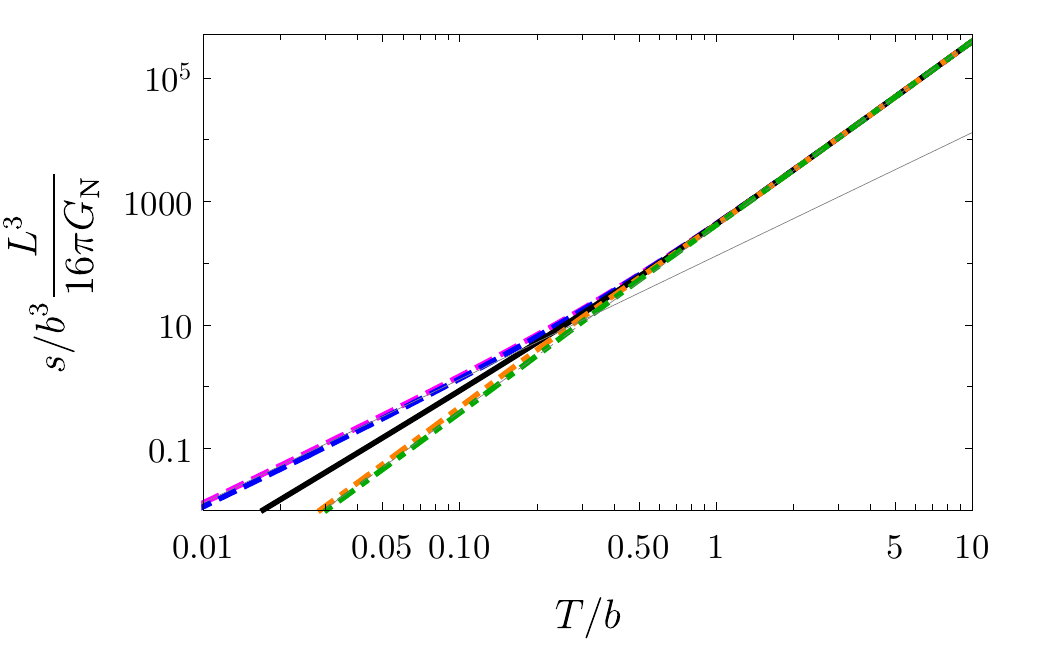}
        \caption{Entropy density}
        \label{fig:entropy}
    \end{subfigure}
    \begin{subfigure}{0.5\textwidth}
        \includegraphics[width=\textwidth]{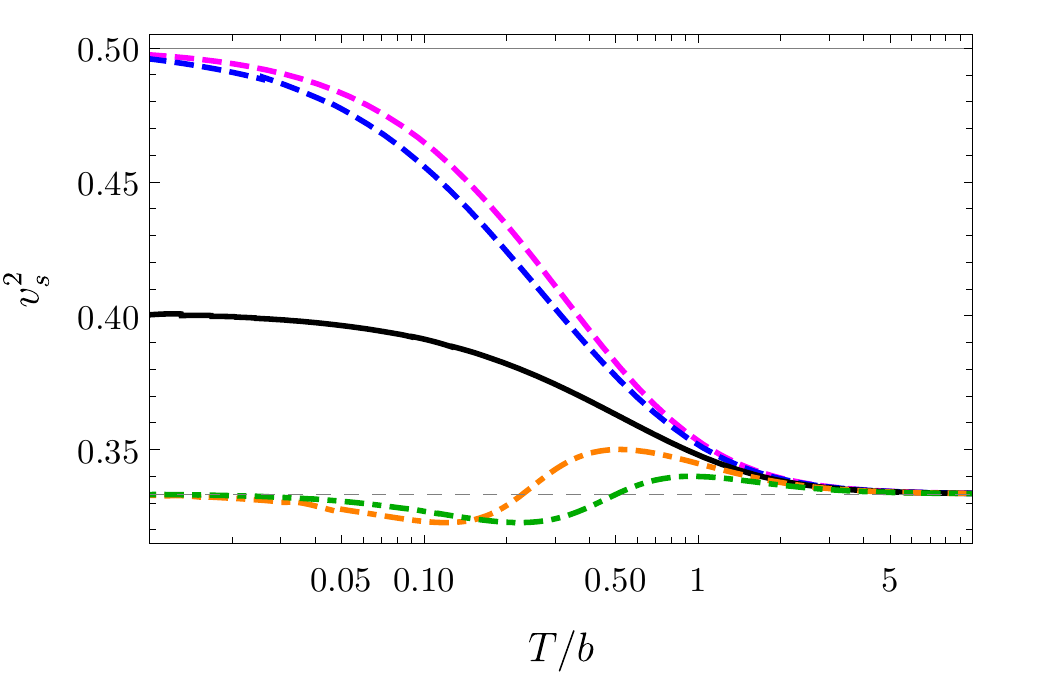}
        \caption{Speed of sound squared}
        \label{fig:sound_speed}
    \end{subfigure}
    \caption{
        Numerical results for physical quantities as functions of temperature, for selected values of \(M/b\). The dashed magenta and blue curves are values of \(M/b\) for which the system is in the topological phase at \(T=0\), while the dot-dashed orange and green curves correspond to the topologically trivial phase. The solid black line corresponds to the critical point at \(T=0\). Notes: \textbf{(a,\,c):} Absolute values have been taken in the plots of \(\ve\) and \(\vev{\cO}\), since they are not of fixed sign. The energy density is negative for small \(T/b\) and positive for large \(T/b\), and vice versa for \(\vev{\cO}\). \textbf{(e,\,f):} The entropy density scales as \(s \propto T^3\) at large \(T/b\), and \(s \propto T^{1 + 2 \b/\a}\) at small \(T/b\), as discussed in the text. The speed of sound squared therefore interpolates between the three-dimensional conformal value \(v_s^2 = 1/3\) as \(T \to \infty\) and \(v_s^2 = 1/(1+2 \b/\a)\) as~\(T \to 0\).
    }
    \label{fig:finite_T_results}
\end{figure}

Figure~\ref{fig:finite_T_results} shows the energy density, pressure, speed of sound squared \(v_s^2\), scalar one-point function, and antisymmetric tensor one-point function in units of \(b\) for sample values of \(M/b\), all as functions of \(T/b\). The speed of sound can be calculated from
\begin{equation} \label{eq:sound_speed}
    v_s^2 = \frac{\p p}{\p \ve} = \frac{s}{T} \frac{\p T}{\p s}.
\end{equation}
where the partial derivatives are to be taken at fixed \(M\) and \(b\). In the figure, the dashed magenta and blue curves correspond to \(M/b =0\) and \(0.5\) respectively, meaning that at \(T = 0\) the system is in the topological phase. The solid black curve has \(M/b=0.9493\), corresponding to the critical phase at \(T=0\). Finally, the dot-dashed orange and green curves are for \(M/b = 1.2\) and \(2\), respectively, corresponding to the topologically trivial phase at \(T=0\).

The energy density and pressure, plotted in figures~\ref{fig:energy_density} and~\ref{fig:pressure}, as well as \(\vev{\cO}\) and \(\vev{\cO^{xy}}\) in figures~\ref{fig:O_phi} and~\ref{fig:O_B}, behave qualitatively similarly for all values of \(M/b\). They interpolate between constant values independent of temperature at small \(T/b\) and power law growth at large \(T/b\), with an exponent determined by dimensional analysis: \(\ve , \, p  \propto T^4\) and \(\vev{\cO},\,\vev{\cO^{xy}} \propto T^3\). The high temperature scalings are indicated by the thin dashed grey lines in figure~\ref{fig:finite_T_results}.

On the other hand, the entropy density in figure~\ref{fig:entropy} and the squared sound speed in figure~\ref{fig:sound_speed} behave qualitatively differently at low \(T\) depending on the value of \(M/b\). When the system is in the topological phase at \(T=0\), we find \(s \propto T^2\) at small \(T\), as indicated by the thin solid grey line in figure~\ref{fig:entropy}, and consequently \(v_s^2 = 1/2\) from equation~\eqref{eq:sound_speed}. When the system is in the topologically trivial phase at \(T=0\) we instead find \(s \propto T^3\) at small \(T\), leading to \(v_s^2 =1/3\). At the critical value of \(M/b\), \(s\) scales with an intermediate power of \(T\), approximately \(s \propto T^{2.49}\) at small \(T\), and therefore \(v_s^2 \simeq 0.401\). At high temperatures we find \(s \propto T^3\) and \(v_s^2 \simeq 1/3\) for all values of \(M/b\), as required by dimensional analysis.

To derive these different exponents and values for \(v_s^2\), we construct an approximate near-horizon solution for \(f(r)\) valid at low temperatures by solving equation~\eqref{eq:eom_second_order_f} using the zero temperature, large-\(r\) results \(g(r) \approx g_0 r^{-\a}\) and \(h(r) \approx h_0 r^{-\b}\), where the use of the large-\(r\) forms is justified by the fact that sending \(T \to 0\) is equivalent to sending \(r_0 \to \infty\). The resulting near-horizon solution for \(f\) is \(f(r) \approx f_0 \le[ 1 - (r/r_0)^{\a + \b} \ri]^2\) for some integration constant \(f_0\). Substituting these solutions into equation~\eqref{eq:hawking_temperature} we can determine how the temperature and entropy density scale with \(r_0\) at leading order at low temperatures,
\begin{equation} \label{eq:temperature_entropy_low_T}
    T \approx \frac{(\a + \b)\sqrt{f_0 g_0}}{2 \pi L} r_0^{-\a/2},
    \qquad
    s \approx \frac{\sqrt{g_0} h_0}{4 \gn} r_0^{-\frac{\a}{2} - \b}.
\end{equation}
From the \(r_0\) dependence of these two results, we see that \(s \propto T^{1 + 2 \b/\a}\) at small \(T\), and consequently
\begin{equation} \label{eq:sound_speed_low_temperature}
    v_s^2 = \frac{1}{1 + 2 \b/\a},
    \qquad \text{for } T \ll b, M.
\end{equation}
For our choice of couplings and masses, the topological phase has \(\b/\a=1/2\), the critical phase has \(\b/\a \simeq 0.745\), and the topologically trivial phase has \(\b/\a = 1\). Substituting these ratios into equation~\eqref{eq:sound_speed_low_temperature}, we reproduce the low temperature results for \(v_s^2\) quoted above.

The result for \(v_s^2\) in equation~\eqref{eq:sound_speed_low_temperature} may be understood as arising from the non-relativistic scaling symmetry of the infrared fixed points. Recall from the discussion in section~\ref{sec:T0_solutions} that the asymptotic metric at large \(r\) is invariant under the rescaling \((t,z) \to \Omega (t,z)\) and \((x,y) \to \Omega^{\b/\a} (x,y)\), with a different value of \(\b/\a\) in each phase. The corresponding Ward identity is then \(\vev{T^t{}_t} + \vev{T^z{}_z}+ \frac{\b}{\a} (\vev{T^x{}_x} + \vev{T^y{}_y}) = 0\), which we can solve to find the pressure as a function of the energy density, \(p = \ve / (1 + 2\b/\a)\). Substituting this into the formula~\eqref{eq:sound_speed} for \(v_s^2\), we recover equation~\eqref{eq:sound_speed_low_temperature}.

%% file: fermions.tex
\section{Fermion spectral functions}
\label{sec:fermions}

\subsection{Fermion equations of motion and boundary conditions}

In this section we calculate the spectral function of a composite fermionic operator dual to a pair of probe fermions in the gravitational background described in the previous section. The spectral function is defined as
\begin{equation} \label{eq:fermion_spectral_function_definition}
    \r(k) = \frac{1}{\pi} \Im \tr G_\mathrm{R}(k),
\end{equation}
where \(G_\mathrm{R}(k)\) is the retarded two-point function of the fermionic operator, at four-momentum \(k\). Nodal lines should appear in the spectral function as rings of sharp peaks in the \((k_x,k_y)\) plane near zero frequency. Since our objective is simply to demonstrate the presence of these peaks in the low \(M/b\) phase, for simplicity we will work at \(M=0\), meaning that we set the scalar field \(\f\) of the background to zero. 

We compute the fermion Green's function holographically following refs.~\cite{Iqbal:2009fd,Liu:2009dm,Cubrovic:2009ye,Gursoy:2011gz,Plantz:2018tqf}. In order to introduce a Dirac fermion on the boundary we need two bulk Dirac fermions \(\Psi_{1,2}\). We denote the five-dimensional Dirac matrices as \(\G^{m}\), they satisfy \(\{\G^m,\G^n\} = 2 G^{mn}\). Denoting the vielbeins as \(e_{\underline{a}}^m\), where we use underlines to denote the tangent space indices, we define \(\G^{\underline{a}} = e^{\underline{a}}_m \G^m\), satisfying \(\{\G^{\underline{a}}, \G^{\underline{b}}\} = 2 \h^{\underline{ab}}\).\footnote{Our sign conventions for the vielbeins are
\[
 \qquad e^r_{\underline{r}} = - \frac{r}{L},\quad e^t_{\underline{t}} = \frac{1}{\sqrt{fg}},\quad e^x_{\underline{x}} = e^y_{\underline{y}} = \frac{1}{\sqrt{h}},\quad e^z_{\underline{z}} = \frac{1}{\sqrt{g}}.
\]
The minus sign in \(e^r_{\underline{r}}\) is convenient for comparison to other works that place the boundary at \(r=\infty\) rather than \(r=0\).
} We also define \(\G^{\underline{ab}} = \frac{1}{2} \le[ \G^{\underline{a}}, \G^{\underline{b}} \ri]\). It will be useful to define projectors \(P_\pm = \frac{1}{2} (1 \pm \Gamma^{\underline{r}})\). On the boundary \(\Gamma^{\underline{r}}\) is the chirality operator, and these projectors isolate the right- and left-handed components of Dirac fermions. In the bulk, we define \(\Psi_{I,\pm} = P_\pm \Psi_I\). We choose \(\G^{\underline{r}}\) to be hermitian.

We take the action for the fermions to be\footnote{The choice of how to couple the bulk fermions to the two-form field \(B\) is not unique. However, it appears to be the only choice that leads to nodal lines in the spectral function~\cite{Liu:2018djq}.}
\begin{align}
    S &= i \int \diff^5 x \, \sqrt{-G} \, \le[\bar{\Psi}_1 (\slashed{D} - m_f) \Psi_1 + \bar{\Psi}_2 (\slashed{D} + m_f) \Psi_2 \ri]
    \nonumber \\ &\phantom{=}
    + \Lambda \int \diff^5 x \, \sqrt{-G} \, \le( \bar{\Psi}_1 \slashed{B} \G^{\underline{r}} \Psi_2  - \bar{\Psi}_2 \slashed{B} \G^{\underline{r}}  \Psi_1 \ri)
    \label{eq:fermion_action}
    \\ &\phantom{=}
    - i \int_{r=\e} \diff^4 x\,  \sqrt{-\g} \, \le( \bar{\Psi}_{1,+} \Psi_{1,-} - \bar{\Psi}_{2,-} \Psi_{2,+}\ri),
    \nonumber
\end{align}
where \(\Lambda\) is a coupling constant, \(\slashed{D} \equiv \G^{\underline{a}} e^m_{\underline{a}} D_m\), \(\slashed{B}= B_{\underline{ab}} \G^{\underline{ab}}\), and \(D_m = \p_m + \frac{1}{4} \w_{m\underline{ab}} \G^{\underline{ab}}\), with \(\w_{m\underline{ab}}\) the spin connection. The corresponding equations of motion are
\begin{equation}     \label{eq:fermion_eom}
    \le(\slashed{D} - m_f\ri) \Psi_1 = i \Lambda \slashed{B} \G^{\underline{r}} \Psi_2,
    \qquad
    \le(\slashed{D} + m_f \ri) \Psi_2 = - i \Lambda \slashed{B} \G^{\underline{r}} \Psi_1.
\end{equation}
The boundary term in equation~\eqref{eq:fermion_action} is evaluated at some small-\(r\) cutoff \(\e\). We leave implicit in our expressions that \(\e\) is to be taken to zero at the end of calculations. The form of the boundary term is chosen such that the variational principle requires the boundary values of \(\Psi_{1,+}\) and \(\Psi_{2,-}\) to be fixed.  Holographically, we therefore interpret the boundary values of \(\Psi_{1,+}\) and \(\Psi_{2,-}\) as the right- and left-handed components of a source for a composite Dirac fermion operator in the boundary field theory, with the conjugate momenta \(\Psi_{1,-}\) and \(\Psi_{2,+}\) determining the corresponding one-point functions. It will be convenient to repackage the spinors into
\begin{equation}
    \Psi = \Psi_{1,+} + \Psi_{2,-}, \qquad \Pi = \Psi_{1,-} - \Psi_{2,+},
\end{equation}
i.e. \(\Psi\) contains the sources and \(\Pi\) contains the one-point functions.

The two-point functions of the boundary fermion operator are determined from the on-shell action \(S^\star\). Since the bulk part of the action vanishes when the equations of motion are satisfied, only the boundary term in equation~\eqref{eq:fermion_action} contributes, yielding
\begin{equation}
    S^\star = - i \int_{r=\e} \diff^4 x \, \sqrt{-\g} \, \le( \bar{\Psi}_{1,+} \Psi_{1,-} - \bar{\Psi}_{2,-} \Psi_{2,+} \ri)=  - i \int_{r=\e} \diff^4 x \, \sqrt{-\g} \, \bar{\Psi} \Pi.
\end{equation}
To obtain the momentum space Green's function, we Fourier transform in the field theory directions, writing \(\Psi(x,r) = \int \frac{\diff^4 k}{(2\pi)^4} e^{i k \cdot x} \Psi(k;r)\) and similar for \(\Pi\). The momentum space versions of the equations of motion~\eqref{eq:fermion_action} may then be solved for \(\Pi\), yielding \(\Pi(k;r) = - i \xi(k;r) \Psi(k;r)\) for some matrix \(\xi(k;r)\). The on-shell action becomes
\begin{equation}
    S^\star = - \int_{r=\e} \diff^4 x \, \sqrt{-\g} \, \bar{\Psi}(-k;r) \xi(k;r) \Psi(k;r).
\end{equation}
Applying the Minkowski space correlator prescription of refs.~\cite{Son:2002sd,Herzog:2002pc}, the matrix of fermion Green's functions is then
\begin{equation} \label{eq:fermion_greens_function_formula}
    G_\mathrm{R}(k) =- \lim_{r \to 0} r^{-2 m_f L} \Gamma^{\underline{0}} \xi(k;r),
\end{equation}
where the factor of \(r^{-2 m_f L}\) arises because the equations of motion~\eqref{eq:fermion_eom} imply that \(\Psi \approx r^{2 - m_f L }\) near the boundary, while \(\sqrt{-\g} \approx r^{-4}\).

In order to actually determine \(\xi(k;r)\), and therefore the fermion Green's functions, it will  be convenient to rescale the spinors. We define
\begin{equation}
    \Psi = \exp\le[- \int_c^r \diff r' F(r')\ri] \y,
    \qquad
    \Pi = \exp\le[- \int_c^r \diff r' F(r')\ri] \chi,
\end{equation}
where \(c\) is some arbitrary reference point, and \(F(r) =  \frac{f'}{4f} + \frac{g'}{2g} + \frac{h'}{2h}\). Since \(\Psi\) and \(\Pi\) are rescaled by the same amount, we have \(\y(k;r) = - i \xi(k;r) \chi(k;r)\). The rescaling effectively eliminates the spin connection from the equations of motion for \(\y\) and \(\chi\). Indeed, applying the projectors \(P_\pm\) to the fermion equations of motion~\eqref{eq:fermion_eom}, it is straightforward to show that \(\y\) and \(\chi\) satisfy
\begin{align} 
    \le(e^r_{\underline{r}} \p_r - m_f \ri)\y(k;r) + i \le(\slashed{k} + \Lambda \slashed{B} \ri) \chi(k;r) &= 0,
    \nonumber \\
    \le(e^r_{\underline{r}} \p_r + m_f \ri) \chi(k;r) -i \le( \slashed{k} + \Lambda \slashed{B} \ri) \y(k;r) &= 0,
    \label{eq:fermion_eom_FT}
\end{align}
where \(\slashed{k} = e^\m_{\underline{a}} \G^{\underline{a}} k_\m\). Finally, substituting \(\chi = - i \xi \y\) into the second line of equation~\eqref{eq:fermion_eom_FT}, and using the first line to eliminate first derivatives of \(\y\), we obtain a first-order matrix equation for \(\xi(k;r)\),
\begin{equation} \label{eq:xi_eom}
    e^r_{\underline{r}} \p_r \xi + 2 m_f \xi + \slashed{k} - \xi \slashed{k} \xi + \Lambda \slashed{B} -  \Lambda \xi \slashed{B} \xi = 0.
\end{equation}

Since we expect to see the nodal lines at \(k_z=0\), using rotational symmetry in the \((x,y)\) plane we can take \(k^\m = (\w,k_x,0,0)\). For this choice of \(k^\m\), we can solve the first line of equation~\eqref{eq:fermion_eom_FT} for \(\chi\) to determine the matrix structure of \(\xi\), finding
\begin{equation}
    \xi = \xi_0 \G^{\underline{0}} + \xi_1 \G^{\underline{1}} + \xi_{2} \G^{\underline{02}} + \xi_{3} \G^{\underline{12}},
\end{equation}
where the coefficients \(\xi_{0,1,2,3}\) satisfy four coupled first-order differential equations. These equations may be decoupled by defining the four independent linear combinations
\begin{align} \label{eq:xi_pm_pm}
    \xi_{s_1,s_2} = \xi_0 + s_1 \xi_1 + s_2 \xi_2 + s_1 s_2 \xi_3,
\end{align}
where \(s_1 = \pm 1\) and \(s_2 = \pm 1\) are uncorrelated signs. In terms of these variables, it is straightforward to show that equation~\eqref{eq:xi_eom} becomes,
\begin{equation} \label{eq:fermion_eom_decoupled}
    \le( -\frac{r}{L} \p_r + 2 m_f \ri) \xi_{s_1,s_2} = \le(\frac{\w}{\sqrt{f g}} - 2 s_1 s_2 \Lambda \frac{B}{h}\ri) \le(\xi_{s_1,s_2}^2 + 1\ri) + s_1 \frac{k_x}{\sqrt{h}} \le(\xi_{s_1,s_2}^2 - 1\ri).
\end{equation}
where we have substituted the explicit expressions for the vielbeins.

The variables \(\xi_{s_1,s_2}\) are holographically dual to the eigenvalues of the fermion Green's function; substituting the expansion~\eqref{eq:xi_pm_pm} into equation~\eqref{eq:fermion_greens_function_formula}, one finds that the four eigenvalues of \(G_\mathrm{R}(\w,k_x)\) are
\begin{equation} \label{eq:greens_function_eigenvalues}
    \l_{s_1,s_2} = \lim_{r\to0} r^{-2m_fL} \xi_{s_1,s_2}.
\end{equation}
The fermion spectral function~\eqref{eq:fermion_spectral_function_definition} is proportional to the sum of the imaginary parts of these eigenvalues, allowing us to determine various properties of the spectral function by inspection of the equations of motion~\eqref{eq:fermion_eom_decoupled}. For example, sending \(\Lambda \to - \Lambda\) maps the equations of motion of \(\xi_{\pm,\pm}\) to those of \(\xi_{\pm,\mp}\), so the spectral function is invariant under this change in the sign of \(\Lambda\). Similarly, the spectral function is invariant under \(k_x \to - k_x\), since this interchanges the equations of motion for \(\xi_{\pm,\pm}\) and \(\xi_{\mp,\mp}\).

\paragraph{Boundary conditions}

We now determine the appropriate boundary conditions for \(\xi_{s_1,s_1}\), beginning at \(T=0\). In the deep IR we have \(g \approx g_0 r^{-\a}\), \(h \approx h_0 r^{-\b}\), and \(B \approx h_0 B_0 r^{-\b}\). For the low \(M/b\) phase \(\a > \b\), so as \(r \to \infty\) the terms proportional to \(\w/\sqrt{g(r)}\) dominate the equations of motion~\eqref{eq:fermion_eom_FT}, such that they become approximately
\begin{equation} \label{eq:fermion_eom_IR}
    \y' + \frac{i \w L}{\sqrt{g_0}} r^{1-\a/2} \Gamma^{\underline{0}} \chi \approx 0,
    \qquad
    \chi' - \frac{i \w L}{\sqrt{g_0}} r^{1-\a/2} \Gamma^{\underline{0}} \y \approx 0.
\end{equation}
We can use the first equation to eliminate \(\chi\) from the second, yielding a second-order equation of motion just for \(\y\),
\begin{equation}
    r^{1-\a/2} \p_r \le( r^{1-\a/2} \y' \ri) + \frac{\w^2 L^2}{g_0} \y \approx 0,
\end{equation}
with the solutions
\begin{equation} \label{eq:fermion_T0_IR}
    \y \approx \exp\le(\pm \frac{2 i \w L}{\a \sqrt{g_0}}  r^{\a/2}\ri) \y_0,
\end{equation}
for some constant spinor \(\y_0\). In order to compute the retarded Green's function we impose ingoing boundary conditions on \(\y\) at \(r \to \infty\), corresponding to the plus sign in the exponent. If we then substitute this solution back into the first equation in~\eqref{eq:fermion_eom_IR}, we find \(\chi \approx \Gamma^0 \y\).  Comparing to \(\chi = - i \xi \y\) we then find \(\xi \approx i \Gamma^0\). This means we must impose \(\xi_{s_1,s_2} = i\) as \(r\to\infty\) on each of the decoupled variables.

The same procedure can be used to find the boundary conditions at non-zero temperature. Near the horizon, where \(f(r)\) has a double zero, the equations of motion~\eqref{eq:fermion_eom_FT} are dominated by the terms proportional to \(\w/\sqrt{f(r) g(r)}\). Taylor expanding near the horizon and using the expression for the Hawking temperature in equation~\eqref{eq:hawking_temperature}, we find that near the horizon the equations of motion become
\begin{equation} \label{eq:fermion_near_horizon_eom}
    \y' + \frac{i \w}{2 \pi T (r_0 -r)} \G^{\underline{0}} \chi \approx 0,
    \qquad
    \chi' - \frac{i \w}{2 \pi T (r_0 -r)} \G^{\underline{0}} \y \approx 0.
\end{equation}
Again using the first equation to eliminate \(\chi\) from the second, we find
\begin{equation}
    (r_0 - r) \, \p_r \le[ (r_0-r) \y'\ri]  + \frac{\w^2}{4 \pi^2 T^2} \y \approx 0,
\end{equation}
which we can solve to obtain the near-horizon behaviour \(
    \y \approx (r_0 - r)^{\pm i \w/2\pi T} \y_0
\)
for some constant spinor \(\y_0\). Ingoing boundary conditions correspond to choosing the minus sign in the exponent. Substituting the ingoing solution into the first equation in~\eqref{eq:fermion_near_horizon_eom} we then find the near horizon behaviour of \(\chi\) to be \(\chi \approx \Gamma^{\underline{0}} \y\), as for at \(T=0\). So the appropriate boundary conditions are \(\xi_{s_1,s_2} = i\) at the horizon.

Suppose we instead chose outgoing boundary conditions for the bulk fermion, meaning we would compute the advanced, rather than retarded, Green's function. Following the same steps as presented above, we find that the appropriate boundary condition on \(\xi\) becomes \(\xi_{s_1,s_2} = - i\), either at \(r \to \infty\) for \(T=0\) or at \(r = r_0\) for \(T \neq 0\). In other words, to compute the advanced Green's function we flip the sign of the boundary condition. Notice that the left- and right-hand sides of equation~\eqref{eq:fermion_eom_decoupled} are odd and even functions of \(\xi_{s_1,s_2}\), respectively. Moreover, the sign of the right-hand side may be inverted by simultaneously sending \(\w \to - \w\) and \(s_1 \to - s_1\). Taken together, these two facts imply that if a given set of functions \(\xi_{\pm,\pm}\) solve equation~\eqref{eq:fermion_eom_decoupled} at frequency \(\w\), then \(-\xi_{\mp,\pm}\) is also a solution, but at frequency \(-\w\). If \(\xi_{\pm,\pm}\) satisfies ingoing boundary conditions, so that its \(r \to 0\) limit yields an eigenvalue of the retarded Green's function, then \(-\xi_{\mp,\mp}\) satisfies outgoing boundary conditions, and its \(r \to 0\) limit yields an eigenvalue of the advanced Green's function. Since the retarded and advanced Green's functions are related to each other by complex conjugation, we therefore find that the eigenvalues of the retarded Green's function satisfy
\begin{equation} \label{eq:eigenvalue_frequency_flip}
    \l_{\pm,\pm}(\w,k_x) = - \l_{\mp,\pm}^*(-\w,k_x).
\end{equation}
This means that the spectral function, being proportional to the sum of the imaginary parts of the eigenvalues, satisfies
\begin{equation} \label{eq:spectral_function_frequency_flip}
    \r(\w,k_x) = \r(-\w,k_x).
\end{equation}

\subsection{Numerical results}

We compute the fermion spectral function by numerically solving the equation of motion~\eqref{eq:fermion_eom_decoupled}, imposing ingoing boundary conditions \(\xi_{s_1,s_2} = i\) at the horizon (or at \(r\to\infty\), in the case of \(T=0\)). The spectral function is determined by subsituting the resulting numerical solution for \(\xi\) into equation~\eqref{eq:fermion_greens_function_formula}. We expect any nodal lines to appear as sharp peaks in the spectral function at \(\w = 0\) and \(k_x = K\), for some \(K \neq 0\). By rotational symmetry in the \((x,y)\) plane, such a peak would exist everywhere along the circle \(k_x^2 + k_y^2 = K^2\), hence forming a nodal line. To obtain numerical results we will need to choose definite values of the fermion mass \(m_f\) and the coupling \(\Lambda\) to the two-form field. We will choose \(m_f = - 1/4L\) and \(\Lambda = 1\).

\begin{figure}
    \begin{subfigure}{0.5\textwidth}
        \includegraphics[width=\textwidth]{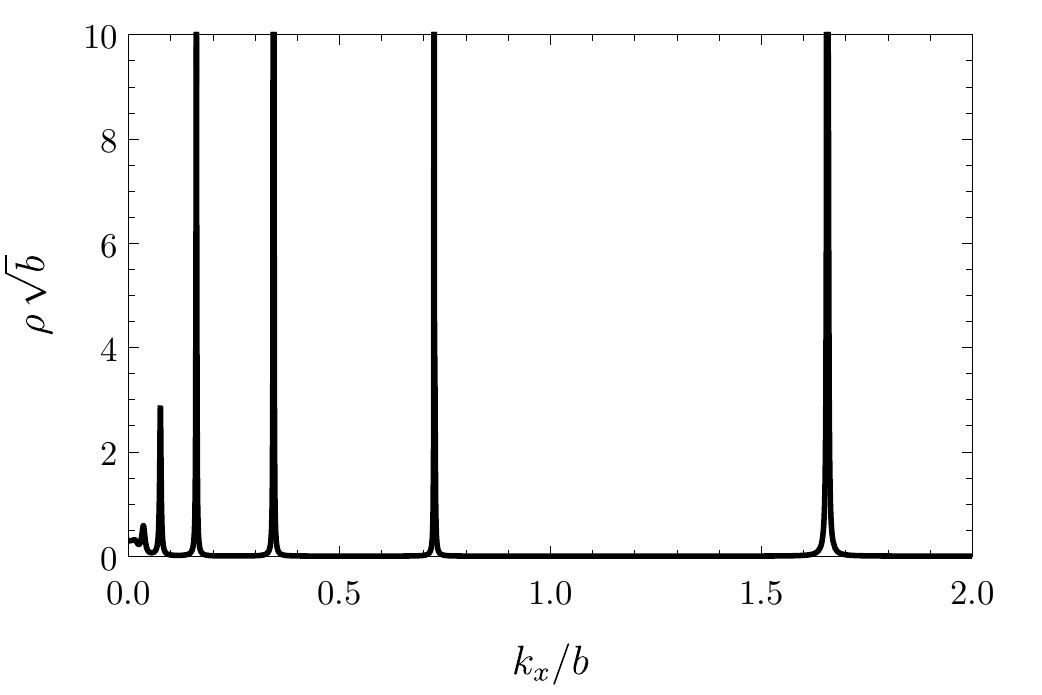}
        \caption{\(\Re \w =0 \)}
        \label{fig:fermion_spectral_function_T0_w0}
    \end{subfigure}
    \begin{subfigure}{0.5\textwidth}
        \includegraphics[width=\textwidth]{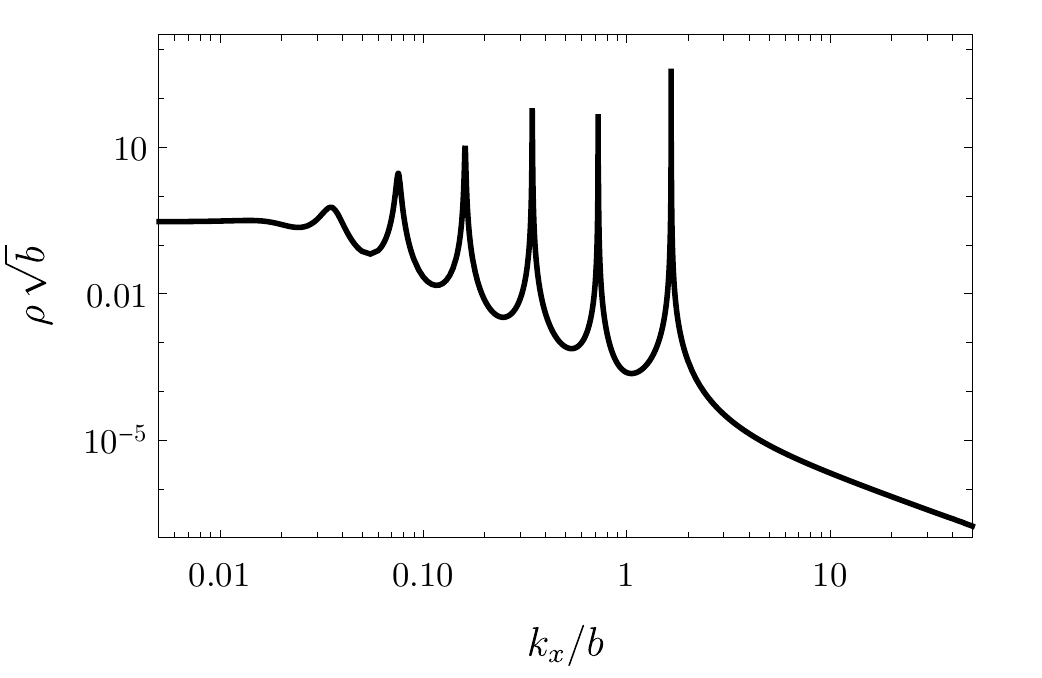}
        \caption{\(\Re \w = 0\)}
        \label{fig:fermion_spectral_function_T0_w0_log}
    \end{subfigure}
    \begin{subfigure}{\textwidth}
        \includegraphics[width=\textwidth]{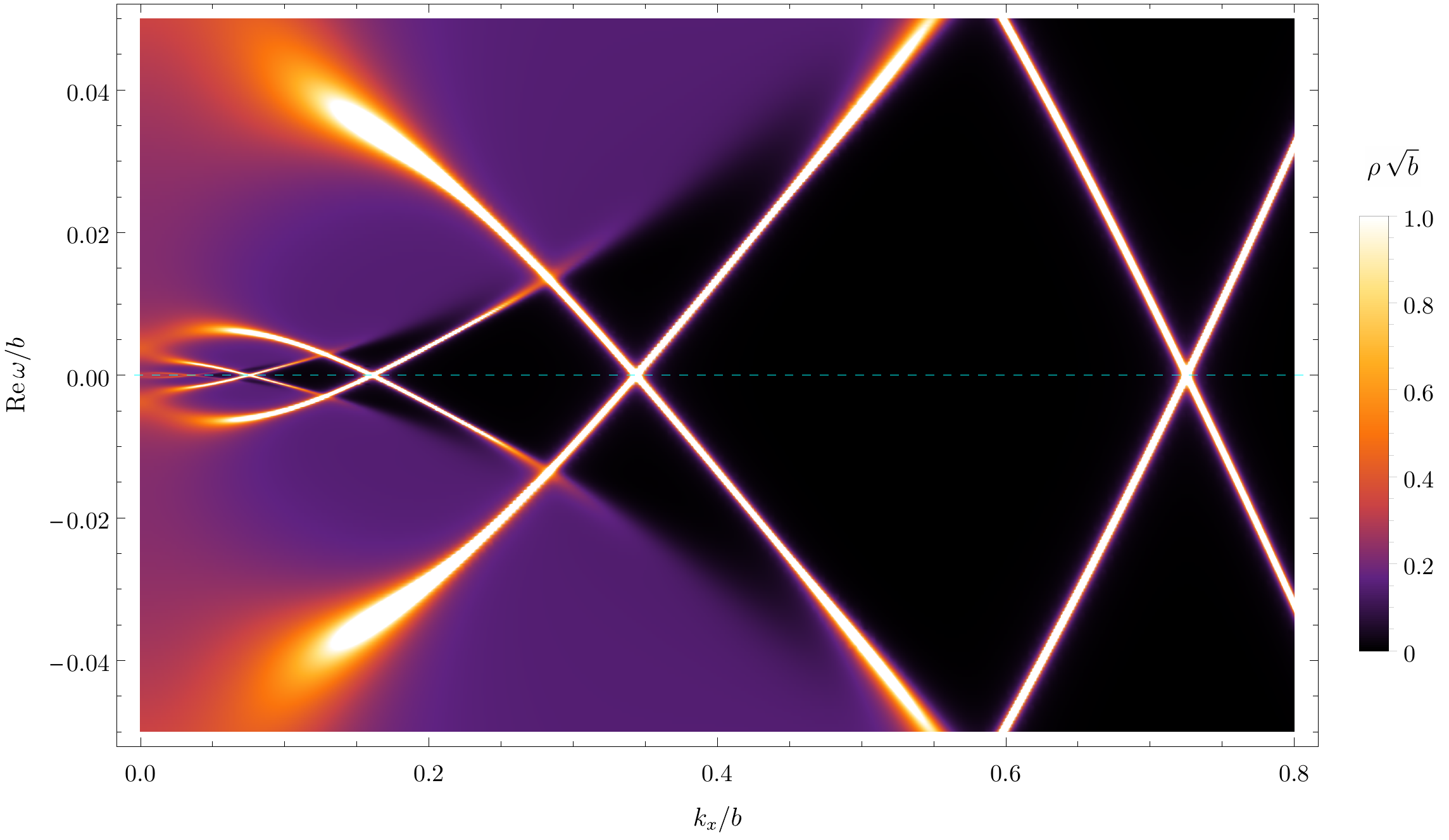}
        \caption{}
        \label{fig:fermion_spectral_function_T0_plane}
    \end{subfigure}
    \caption{The fermion spectral function in units of \(1/\sqrt{b}\) at \(k_y = k_z = 0\), for \(m_f = -1/4 L\) and \(\Lambda=1\), at \(T=0\). The frequency has been given a small imaginary part \(\Im \w /b = 10^{-4}\) in order to broaden the peaks of the spectral function, making them easier to resolve numerically. \textbf{(a):} The spectral function at \(\Re \w=0\), as a function of \(k_x/b\). The spectral function exhibits several sharp peaks at non-zero values of \(k_x\). Rotational symmetry implies that these peaks are the projections to \(k_y=0\) of circular nodal lines in the \((k_x,k_y)\) plane. \textbf{(b):} A log-log plot of the data in figure (a). With a logarithmic \(k_x\) axis the radii of peaks in the spectral function appear evenly spaced. Indeed, we find that the radius of the \(n\)-th line, counting inwards from large \(k_x\) and beginning at \(n=0\), is well approximated by \(K_n \simeq 1.65 e^{-0.782 \, n} b\). \textbf{(c):} Density plot of the fermion spectral function in the \((k_x,\Re \omega)\) plane, with the horizontal, dashed line indicating \(\Re\w =0 \).  Four nodal lines are clearly visible on the scale of this plot as bright peaks at \(\Re \w = 0\), with a fifth nodal line visible upon zooming in near \(\Re \w = k_x = 0\).}
    \label{fig:fermion_spectral_function_T0}
\end{figure}

Figure~\ref{fig:fermion_spectral_function_T0} shows our results for the fermion spectral function at \(T=0\). Since the spectral function is invariant under \(k_x \to - k_x\), we only show results for \(k_x \geq 0\). In all of the plots in the figure, the frequency has been given a small imaginary part \(\Im \w/b = 10^{-4}\). This broadens the peaks in the spectral function, making them easier to resolve numerically. Outside the light cone, we expect these peaks to become delta functions at \(\Im \w = 0\). Figure~\ref{fig:fermion_spectral_function_T0_w0} shows the spectral function in units of \(1/\sqrt{b}\) as a function of \(k_x/b\), at \(\Re \w = 0\). We observe multiple nodal lines, visible as the extremely sharp peaks in the figure.

Figure~\ref{fig:fermion_spectral_function_T0_w0_log} shows the same data as figure~\ref{fig:fermion_spectral_function_T0_w0} but with logarithmic axes, making it easier to see most of the peaks. With the logarithmic \(k_x/b\) axis the peaks appear equally spaced. Indeed, from a fit to our results we find that the locations of the peaks (i.e. the radii of the nodal lines) are very well approximated by the Efimov-like spectrum \(K_n \simeq 1.65 e^{-0.782 \, n} b\), where the integer \(n\) labels the different peaks, starting from the outermost at \(n=0\). We find a total of seven peaks in the spectral function,\footnote{Six of the seven peaks are very clearly visible in figure~\ref{fig:fermion_spectral_function_T0_w0_log}. The innermost peak at \(k_x/b \simeq 0.015\) is difficult to see on the scale of the plot.} however it is plausible that this finite number of peaks arises due to the artificial broadening introduced by the non-zero imaginary part of the frequency, which washes out the very closely-spaced peaks at large \(n\), corresponding to small \(K_n/b\), and that the spectrum of peaks may continue to \(n \to \infty\).

Figure~\ref{fig:fermion_spectral_function_T0_plane} is a density plot of the fermion spectral function in part of the \((k_x/b,\Re \w/b)\) plane, showing how some of the peaks evolve as we move away from \(\Re\w=0\). Our results for the spectral function are invariant under \(\Re \w \to - \Re \w\), as expected from equation~\eqref{eq:spectral_function_frequency_flip}. We find that each peak splits in two at non-zero \(\Re\w\), with one of the daughter peaks moving to small \(k_x\) as we increase \(\Re\w\), while the other moves to larger \(k_x\). This splitting qualitatively resembles the momentum dependence of the inner two eigenvalues of the non-interacting toy model, plotted in figure~\ref{fig:example_loop}. Also visible in figure~\ref{fig:fermion_spectral_function_T0_plane} is an apparent continuum of states, where the spectral function is non-zero but varies smoothly, rather than exhibiting a peak (this is the region coloured purple in the plot). We expect that this is an artifact of the broadening induced by the non-zero imaginary part of the frequency.

\begin{figure}
    \includegraphics[width=\textwidth]{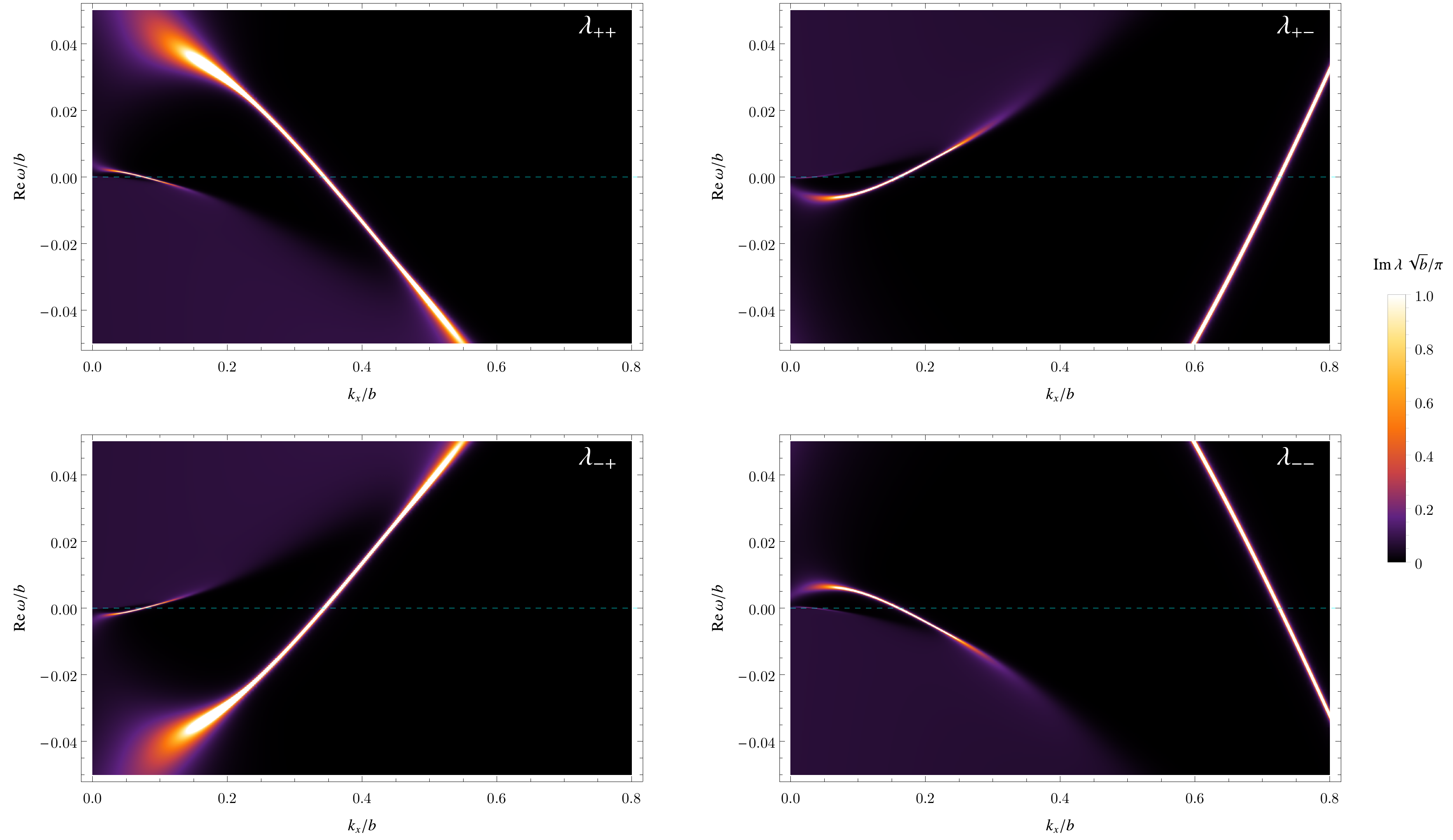}
    \caption{Density plots of the imaginary parts of the eigenvalues of the fermion Green's function, obtained using equation~\eqref{eq:greens_function_eigenvalues}, in the \((k_x,\Re\w)\) plane. Half of the nodal lines are formed by the intersection of peaks in \(\l_{++}\) and \(\l_{-+}\). The other half are formed by the intersection of peaks in \(\l_{+-}\) and \(\l_{--}\).}
    \label{fig:fermion_eigenvalues}
\end{figure}
The behavior of the individual four components of the spectral function clearly shows the chiral nature of the excitations around the nodal line. We can straightforwardly use equation~\eqref{eq:greens_function_eigenvalues} to compute the individual eigenvalues \(\l_{s_1,s_2}\) of the fermion Green's function. In figure~\ref{fig:fermion_eigenvalues} we show density plots of the imaginary parts of these eigenvalues over the same range of \(\Re \w\) and \(k_x\) as figure~\ref{fig:fermion_spectral_function_T0_plane}. Notice that the imaginary parts of \(\l_{\pm,\pm}\) and \(\l_{\mp,\pm}\) are related by \(\Re \w \to -\Re \w\), as expected from equation~\eqref{eq:eigenvalue_frequency_flip}. From figure~\ref{fig:fermion_eigenvalues} we see that each nodal line is formed by one of the two pairs of eigenvalues: half of the nodal lines are formed by the intersection of peaks in \(\l_{++}\) and \(\l_{-+}\) at \(\Re\w=0\), while the other half are formed by the intersection of peaks in \(\l_{+-}\) and \(\l_{--}\), with each pair alternating as we increase momentum from \(k_x=0\). Similar behaviour was seen in earlier holographic work~\cite{Liu:2018djq}.

\begin{figure}
    \begin{subfigure}{0.5\textwidth}
        \includegraphics[width=\textwidth]{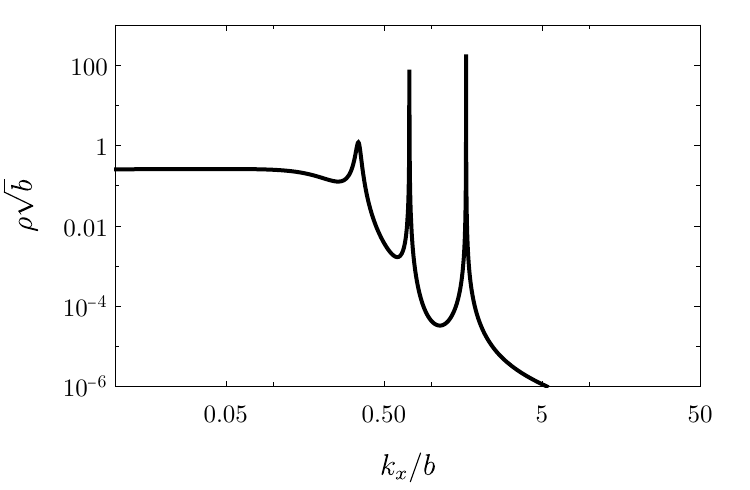}
        \caption{\(T/b = 0.01\)}
    \end{subfigure}
    \begin{subfigure}{0.5\textwidth}
        \includegraphics[width=\textwidth]{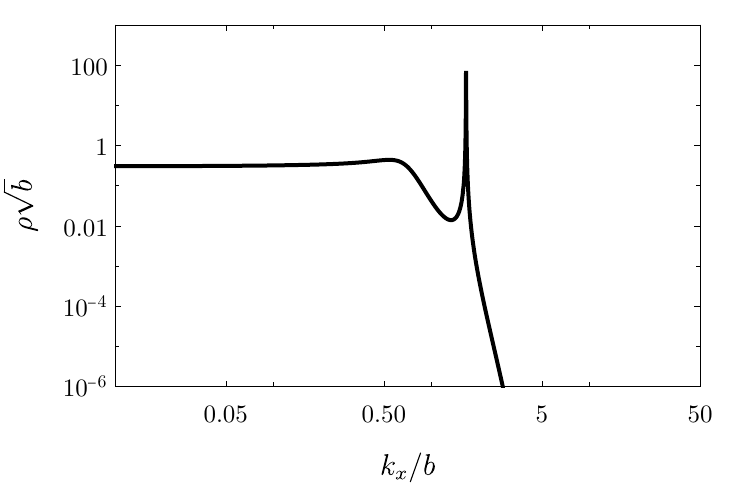}
        \caption{\(T/b = 0.1\)}
    \end{subfigure}
    \begin{subfigure}{0.5\textwidth}
        \includegraphics[width=\textwidth]{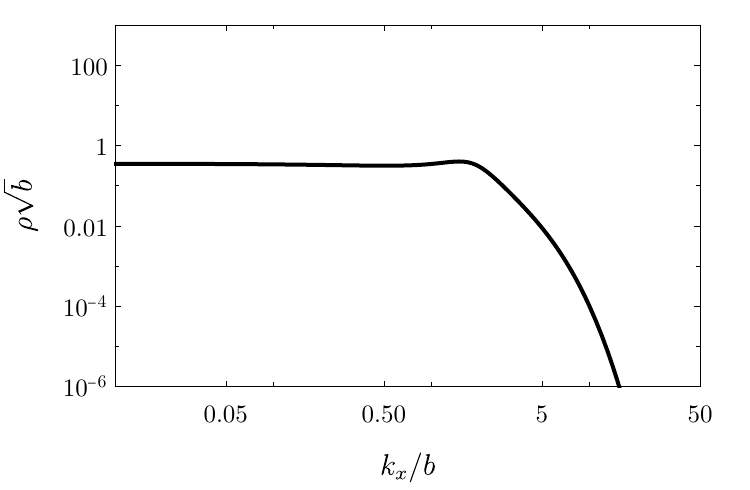}
        \caption{\(T/b = 1\)}
    \end{subfigure}
    \begin{subfigure}{0.5\textwidth}
        \includegraphics[width=\textwidth]{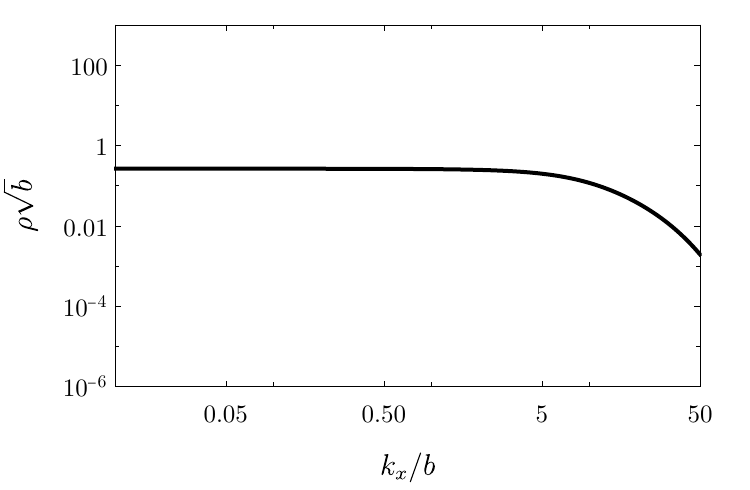}
        \caption{\(T/b = 10\)}
    \end{subfigure}
    \caption{Logarithmic plots of the fermion spectral function  as a function of \(k_x/b\) for sample values of temperature at \(\w/b = 10^{-4} i\) and \(k_y = k_z = 0\). As the temperature is increased, more and more of the peaks in the spectral function are washed out, beginning at small \(k_x/b\). Eventually, for \(T/b \gtrsim 1.5\) we find no peaks at all.}
    \label{fig:fermion_spectral_function}
\end{figure}

Figure~\ref{fig:fermion_spectral_function} shows our numerical results for the fermion spectral function at non-zero temperature. The plots in the figure show the fermion spectral function in units of \(1/\sqrt{b}\) as a function of \(k_x/b\) at \(\w/b = 10^{-4}i\), each at a different value of \(T/b\). We find that as the temperature is increased, the peaks broaden and merge into a continuum, starting at small \(k_x/b\) and moving outwards. Eventually, for sufficiently large temperatures there are no peaks in the spectral function at all. We find that this occurs for \(T/b\gtrsim 1.5\). Notice that there is a range of temperatures for which there is only a single sharp peak in the fermion spectral function.

%% file: transport.tex
\pagebreak
\section{Transport}
\label{sec:transport}

\subsection{Conductivity}
\label{sec:conductivity}

In this section we compute the electrical conductivities of our system, using the Kubo formula
\begin{equation} \label{eq:conductivity_kubo}
    \sigma_{ij}(\w) = \frac{\vev{J^i J^j}_\mathrm{R}(\w,\vec{k}=0)}{i \w},
\end{equation}
where \(\vev{A B}_\mathrm{R}(\w,\vec{k})\) denotes the retarded Green's function of operators \(A\) and \(B\) at frequency \(\w\) and momentum \(\vec{k}\). To compute this two-point function holographically we consider linearised fluctuations of the bulk gauge field \(A_m\). We take the fluctuations to be of the form \(A_m(t,r) = \int \frac{\diff \w}{2\pi} e^{-i \w t} A_m(\w;r)\). In the radial gauge \(A_r=0\), the equations of motion fix \(A_t\) to be a constant, which we set to be zero by a further gauge transformation. The equations of motion for \(A_{x,y,z}\) are then
\begin{align}
    \frac{r}{\sqrt{f}} \p_r \le[ r \sqrt{f} g \, \p_r A_{x,y}(\w;r) \ri] &= -\frac{\w^2 L^2}{f} A_{x,y}(\w; r) ,
    \nonumber
    \\
    \frac{r}{ \sqrt{f}} \p_r \le[ r \sqrt{f} h  \, \p_r A_{z}(\w;r) \ri] &= -\frac{h \w^2  L^2}{f g} A_{z} (\w;r),
    \label{eq:gauge_fluctuation_eom_no_bg}
\end{align}

Using the near boundary expansions of \(f\), \(g\), and \(h\) written in equation~\eqref{eq:bg_near_boundary}, one finds that for small \(r\), solutions to the gauge field equations of motion~\eqref{eq:gauge_fluctuation_eom_no_bg} take the form
\begin{equation} \label{eq:gauge_fluctuation_near_boundary}
    A_i(\w;r) = A_i^{(0)}(\w) \le[1 - \frac{\w^2 r^2 }{2} \log (r/L) \ri] + r^2 A_i^{(2)}(\w)  + \cO(r^4 \log r) \;.
\end{equation}
with coefficients \(A_i^{(0)}(\w)\) and \(A_i^{(2)}(\w)\) determined by the boundary conditions. The on-shell action for the gauge field fluctuations is obtained from equation~\eqref{eq:holographic_bulk_action}. Using the equations of motion for the \(A_i\), it reduces to a boundary term
\begin{align}
    S^*_A = \frac{L}{32 \pi \gn} \int \diff^3 x \int \frac{\diff \w}{2\pi} \biggl[&
        2 A_x^{(0)}(-\w) A_x^{(2)}(\w) + 2 A_y^{(0)}(-\w) A_y^{(2)}(\w) 
        \nonumber \\ &
        + 2 A_z^{(0)}(-\w) A_z^{(2)}(\w)
        -  \frac{\w^2}{2} A_i^{(0)}(-\w) A_i^{(0)}(\w) 
    \biggr],
    \label{eq:gauge_fluctuation_action}
\end{align}
Applying the Lorentzian correlator prescription of refs.~\cite{Son:2002sd,Herzog:2002pc}, we can then read off the expressions for the current-current Green's functions at zero momentum,
\begin{equation}
    \vev{J^i J^j}_{\mathrm{R}}(\w, 0) = \begin{cases}
         \dfrac{L}{8 \pi \gn} \le(\dfrac{A_i^{(2)}(\w)}{A_j^{(0)}(\w)} - \dfrac{\w^2}{4} \ri), & i=j,
        \\
    0, & \text{otherwise}.
    \end{cases}
    \label{eq:no_bg_greens_function_formula}
\end{equation}
Rotational symmetry in the \((x,y)\) plane implies that \(\vev{J^x J^x}_{\mathrm{R}}(\w,0) = \vev{J^y J^y}_{\mathrm{R}}(\w,0)\). We will therefore focus only on the calculation of \(\vev{J^x J^x}_{\mathrm{R}}\) and \(\vev{J^z J^z}_{\mathrm{R}}\).

In order to compute the retarded Green's functions we must impose ingoing boundary conditions on the fluctuations of the gauge field at the horizon. Near the horizon, we find solutions to the equations of motion~\eqref{eq:gauge_fluctuation_eom_no_bg} take the form
\begin{equation} \label{eq:finite_T_gauge_fluctuation_IR_bcs}
    A_{x,z}(\w;r) \propto (r_0 - r)^{\pm i \w/2\pi T}.
\end{equation}
Ingoing boundary conditions correspond to choosing the minus sign in the exponent. At \(T=0\) there is no horizon. In this case, at large \(r\) where \(g \approx g_0 r^{-\a}\) and \(h \approx h_0 r^{-\b}\) we find that the gauge field equations of motion have the approximate solutions
\begin{equation} \label{eq:T0_gauge_fluctuation_IR_bcs}
    A_x(\w;r) \propto r^{\a/4} \exp\le(\pm \frac{2 i \w L r^{\a/2}}{\a \sqrt{g_0}} \ri),
    \qquad
    A_z(\w;r) \propto r^{(2\b-\a)/4} \exp\le(\pm \frac{2 i \w L r^{\a/2}}{\a \sqrt{g_0}} \ri),
\end{equation}
where ingoing boundary conditions correspond to the plus sign in the exponents.

\subsubsection{DC conductivity}

Substituting the expressions for the Green's functions in equation~\eqref{eq:no_bg_greens_function_formula} into the Kubo formula~\eqref{eq:conductivity_kubo} and making use of the near-boundary expansions written in equation~\eqref{eq:gauge_fluctuation_near_boundary}, it is straightforward to show that the DC (zero frequency) limits of the conductivities are
\begin{equation} \label{eq:dc_conductivity_formulas}
    \s_{xx}^\mathrm{DC} = \frac{L}{16 \pi \gn} \lim_{\w \to 0} \, \lim_{r \to 0}  \frac{r \sqrt{f} g \, \p_r A_x(\w;r)}{i \w  A_x(\w;r)},
    \qquad
    \s_{zz}^\mathrm{DC} =   \frac{L}{16 \pi \gn}  \lim_{\w \to 0} \, \lim_{r \to 0} \frac{r \sqrt{f} h\, \p_r A_z(\w;r)}{i \w A_z(\w;r)}.
\end{equation}
The equations of motion~\eqref{eq:gauge_fluctuation_eom_no_bg} imply that the combinations \(r \sqrt{f} g   \, \p_r A_x(\w;r)/i \w A_x(\w;r)\) and \(r \sqrt{f} h  \,  \p_r A_z(\w;r)/i \w A_z(\w;r)\) are independent of \(r\) up to \(\cO(\w)\) corrections~\cite{Iqbal:2008by}. We can therefore relax the \(r \to 0\) limits in equation~\eqref{eq:dc_conductivity_formulas}, instead evaluating the right-hand sides at any convenient value of \(r\).

At \(T=0\) we evaluate the conductivities at \(r \to \infty\). Using the large-\(r\) behaviour of the gauge fluctuations written in equation~\eqref{eq:T0_gauge_fluctuation_IR_bcs} we find \(\p_r A_x/i \w A_x = \p_r A_z/i \w A_z = L/ (r \sqrt{g})\) at large \(r\). Substituting this solution into equation~\eqref{eq:dc_conductivity_formulas} we find that the DC conductivities in the plane of the nodal line vanish for all phases,
\begin{equation}
    \s_{xx}^\mathrm{DC} \propto \lim_{r \to \infty} \sqrt{g(r)} = 0,
\end{equation}
since \(g(r) \approx g_0 r^{-\a}\) at large \(r\), with \(\a \geq 0\). On the other hand, the DC conductivity normal to the plane of the nodal line depends on the phase of the system through the exponents \(\a\) and \(\b\),
\begin{align}
    \s_{zz}^\mathrm{DC} = \frac{L^2}{16 \pi \gn} \lim_{r \to \infty} \frac{h(r)}{\sqrt{g(r)}} = \frac{L^2}{16 \pi \gn} \frac{h_0}{\sqrt{g_0}} \lim_{r \to \infty} r^{\frac{\a}{2}-\b}.
\end{align}
In the trivial and critical phases, we have \(\a < 2 \b\), so that \(\s_{zz}^\mathrm{DC}=0\). However, recall that by tuning \(\l=34/13\) we were able to fix \(\a = 2\b\) in the topological phase, which then gives a finite, non-zero value for \(\s_{zz}^\mathrm{DC}\),
\begin{equation} \label{eq:DC_conductivity_T0}
    \s_{zz}^\mathrm{DC} = \begin{cases}
        \dfrac{L^2}{16 \pi \gn} \dfrac{h_0}{\sqrt{g_0}}, & \text{topological phase},
        \\
        0, & \text{other phases}.
    \end{cases}
\end{equation}
This behaviour is very special to our particular choice of \(\l\). For \(\l < 34/13\) we find \(\a > 2 \b\), and consequently \(\s_{zz}^\mathrm{DC}\) diverges. Conversely, for \(\l > 34/13\) we find \(\a < 2 \b\), leading to \(\s_{zz}^\mathrm{DC}=0\).

At non-zero temperature it is convenient to evaluate equation~\eqref{eq:dc_conductivity_formulas} at the horizon. Using the near-horizon behaviour written in equation~\eqref{eq:finite_T_gauge_fluctuation_IR_bcs} we find
\begin{equation} \label{eq:DC_conductivity_formulas}
    \s_{xx}^\mathrm{DC} = \frac{L^2}{16 \pi \gn} \sqrt{g(r_0)},
    \qquad
    \s_{zz}^\mathrm{DC} = \frac{L^2}{16 \pi \gn} \frac{h(r_0)}{\sqrt{g(r_0)}}.
\end{equation}
We can use these expressions to determine the DC conductivities from the numerical solutions presented in section~\ref{sec:numerical_solutions}.

We can determine the low-temperature behaviour of the DC conductivities using the \(T=0\) solutions \(g(r)\approx g_0 r^{-\a}\) and \(h(r) \approx h_0 r^{-\b}\) in equation~\eqref{eq:DC_conductivity_formulas} and making use of the relationship between \(r_0\) and \(T\), valid at low temperatures, given in equation~\eqref{eq:temperature_entropy_low_T}. We find \(\s_{xx}^\mathrm{DC} \propto T\) for all values of \(M/b\), while the behaviour of \(\s_{zz}^\mathrm{DC}\) for a given value of \(M/b\) depends on which phase the system is in at \(T = 0\), through the dynamical exponent \(\a/\b\), \(\s_{zz}^\mathrm{DC} \propto T^{2\frac{\b}{\a} - 1}\). In the topological phase we have \(\a = 2\b\), and consequently \(\s_{zz}^\mathrm{DC}\) is finite and non-zero as \(T \to 0\), given by the result in equation~\eqref{eq:DC_conductivity_T0}. In the topologically trivial phase we have \(\a = \b\), leading to \(\s_{zz}^\mathrm{DC} \propto T\). Finally, in the critical phase we have \(\a/\b \simeq 1.34\), implying \(\s_{zz}^\mathrm{DC}\propto T^{0.491}\).

\begin{figure}
    \begin{center}
        \includegraphics{sigma_temperature_mass_legend}
    \end{center}
    \begin{subfigure}{0.5\textwidth}
        \includegraphics[width=\textwidth]{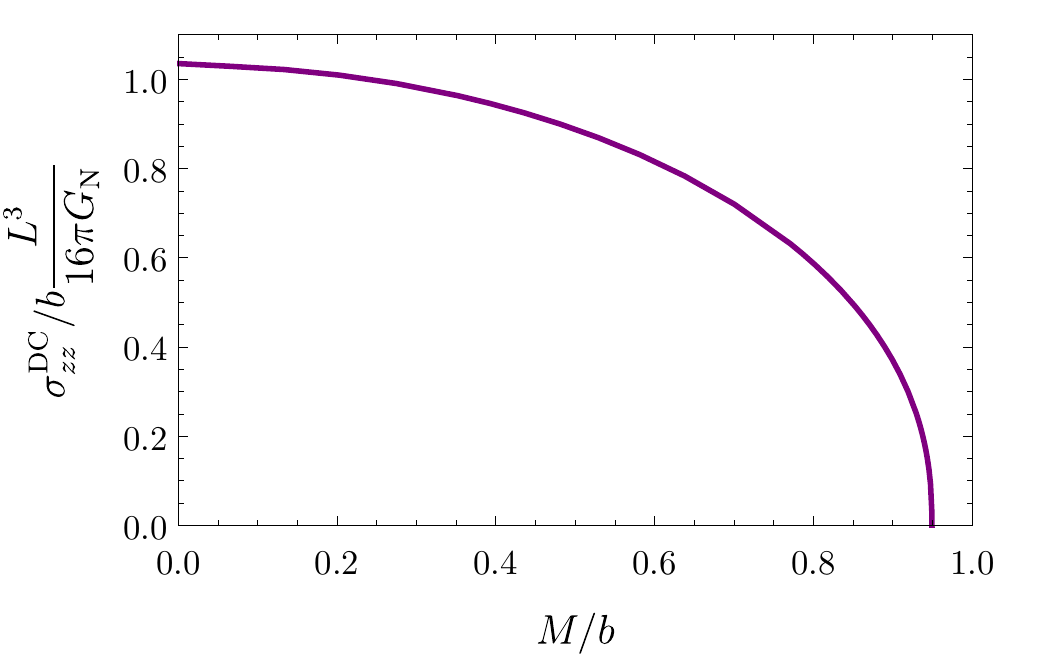}
        \caption{\(\s_{zz}^\mathrm{DC}\) at \(T=0\)}
        \label{fig:sigma_zz_DC_T0}
    \end{subfigure}
    \begin{subfigure}{0.5\textwidth}
        \includegraphics[width=\textwidth]{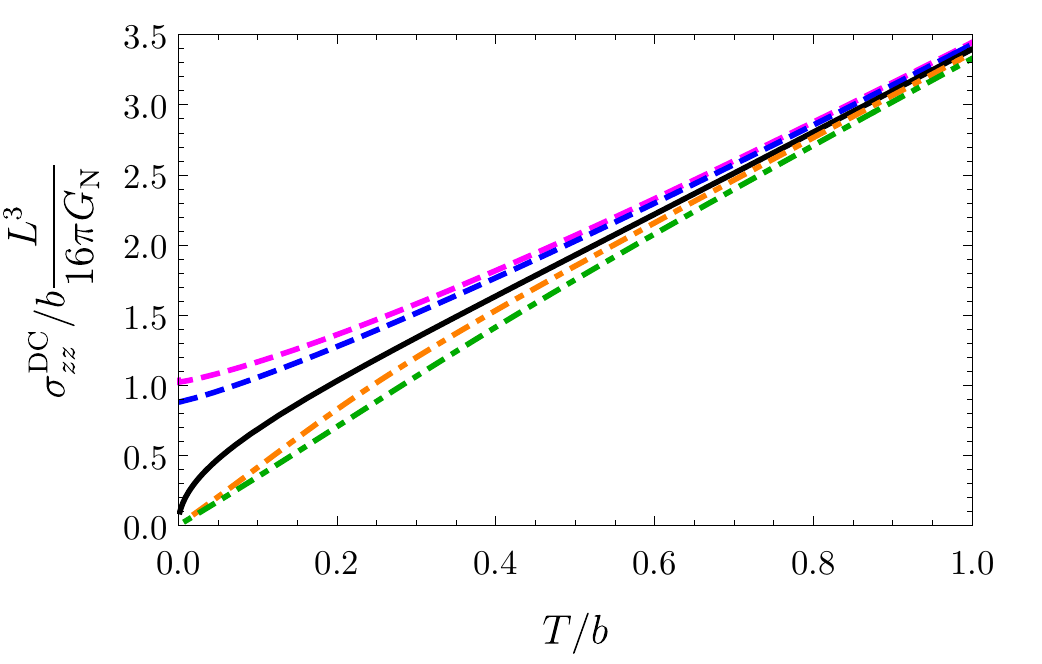}
        \caption{\(\s_{zz}^\mathrm{DC}\)}
        \label{fig:sigma_zz_DC}
    \end{subfigure}
    \begin{subfigure}{0.5\textwidth}
        \includegraphics[width=\textwidth]{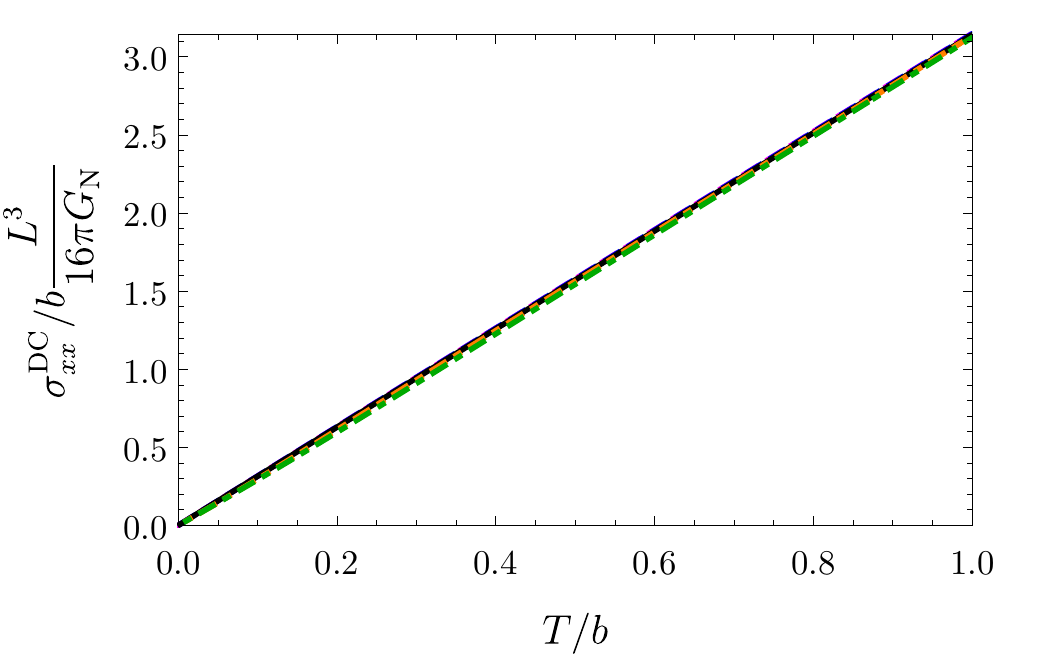}
        \caption{\(\s_{xx}^\mathrm{DC}\)}
        \label{fig:sigma_xx_DC}
    \end{subfigure}
    \begin{subfigure}{0.5\textwidth}
        \includegraphics[width=\textwidth]{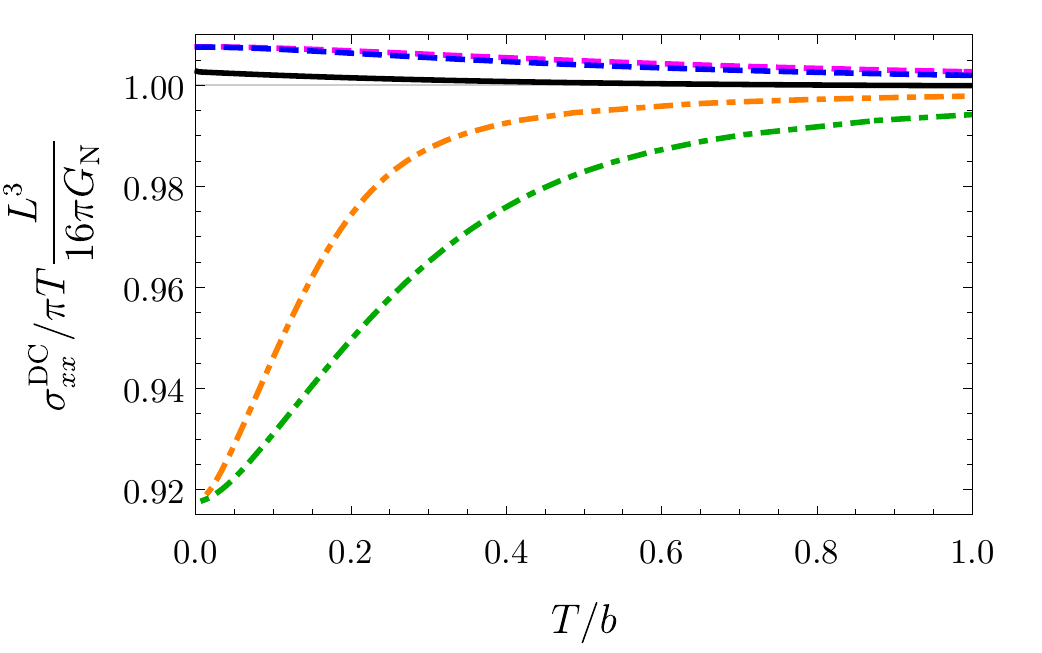}
        \caption{\(\s_{xx}^\mathrm{DC}/\pi T\)}
        \label{fig:sigma_xx_DC_over_T}
    \end{subfigure}
    \caption{
    Results for the DC conductivities.
    \textbf{(a):} The DC conductivity \(\s_{zz}^\mathrm{DC}\) in the direction perpendicular to the plane of the nodal line as a function of \(M/b\) at \(T=0\). It is non-zero only in the topological phase.
    \textbf{(b):} The behaviour of \(\s_{zz}^\mathrm{DC}\) at non-zero temperature, with colour coding given in the legend at the top of the figure. The dashed magenta and blue curves correspond to values of \(M/b\) for which the system is in the topological phase at \(T=0\). The solid black curve corresponds to the critical phase at \(T=0\). The dot-dashed orange and green curves correspond to the topologically trivial phase at \(T=0\).
    \textbf{(c):} The DC conductivity \(\s_{xx}^\mathrm{DC}\) in the plane of the nodal line does not depend sensitively on \(M/b\)
    \textbf{(d):} Dividing by the temperature makes the difference in \(\s_{xx}^\mathrm{DC}\) between the phases clearer. The colour coding and dashing is the same as in figure (b).}
    \label{fig:dc_conductivity_M0}
\end{figure}

To obtain the DC conductivities away from small \(T\) we must evaluate equation~\eqref{eq:DC_conductivity_formulas} numerically. In figure~\ref{fig:dc_conductivity_M0} we plot our numerical results for the DC conductivity. Figure~\ref{fig:sigma_zz_DC_T0} shows \(\s_{zz}^\mathrm{DC}\) as a function of \(M/b\) at \(T=0\). It is non-zero in the topological phase, and vanishes as we approach the phase transition at \((M/b)_\mathrm{crit.} \simeq 0.9493\). From a fit to the our results for \(\s_{zz}^\mathrm{DC}\) near \((M/b)_\mathrm{crit.}\), we find that it vanishes as \(\s_{zz}^\mathrm{DC} \propto \le[(M/b)_\mathrm{crit.} - (M/b)\ri]^{0.42}\).

Figure~\ref{fig:sigma_zz_DC} shows how \(\s_{zz}^\mathrm{DC}\) evolves with increasing temperature for sample values of \(M/b\). For each value of \(M/b\), we find that for \(T/b \ll 1\) the conductivity grows with the expected power of \(T\). Concretely, when the system is in the topological phase at \(T=0\), we find \(\s_{zz}^\mathrm{DC} \propto T^0\) at small \(T/b\) (the dashed magenta and blue curves in the figure), in the topologically trivial phase we find \(\s_{zz}^\mathrm{DC} \propto T\) (the dot-dashed orange and green curves), and in the critical phase \(\s_{zz}^\mathrm{DC} \propto T^{0.49}\) (the black curve). For all values of \(M/b\), at \(T/b \gg 1\) we find that the DC conductivity is well approximated by the \ads[5]-Schwarzschild result \(\s_{zz}^\mathrm{DC} \simeq T L^3/16 \gn\).

Figure~\ref{fig:sigma_xx_DC} shows \(\s_{xx}^\mathrm{DC}\) as a function of \(T/b\) for sample values of \(M/b\). We find that \(\s_{xx}^\mathrm{DC}\) is proportional to \(T\) at both small and large \(T/b\), with different proportionality coefficients. It is very difficult to distinguish the different curves in figure~\ref{fig:sigma_xx_DC} as \(\s_{xx}^\mathrm{DC}\) does not depend sensitively on \(M/b\). To make the differences clearer, in figure~\ref{fig:sigma_xx_DC_over_T} we show \(\s_{xx}^\mathrm{DC}\) divided by temperature. From figure~\ref{fig:sigma_xx_DC_over_T} we see that the DC conductivity in the plane of the nodal loop takes the form \(\s_{xx}^\mathrm{DC} = T \bar{\sigma}(T/b, M/b)\), where the function \(\bar{\sigma}\) depends only weakly on both \(T/b\) and \(M/b\).

\begin{figure}
    \begin{center}
    \includegraphics{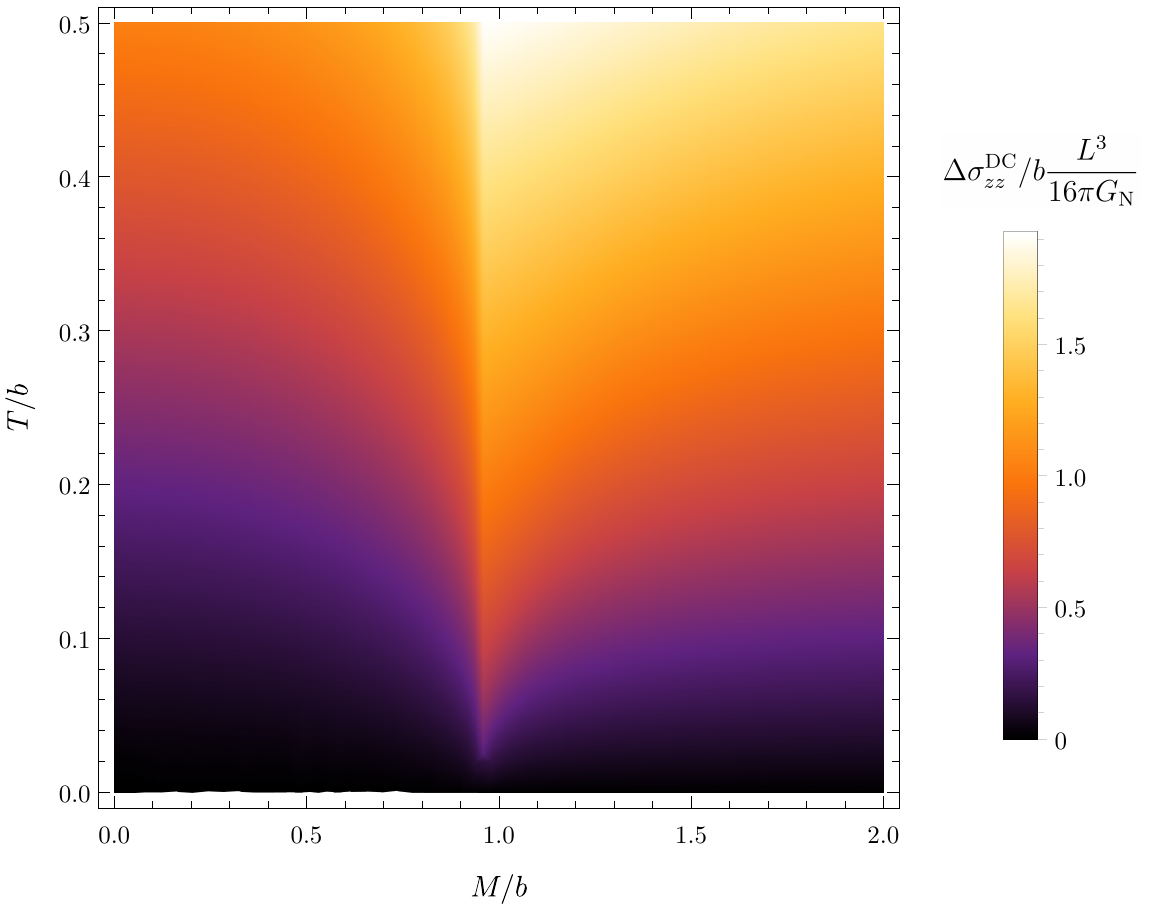}
    \end{center}
    \caption{Density plot of \(\Delta \s_{zz}^\mathrm{DC} = \s_{zz}^\mathrm{DC}- \le.\s_{zz}^\mathrm{DC}\ri|_{T=0}\) in the \((M/b,T/b)\) plane. At small \(T/b\) this quantity exhibits a structure reminiscent of a quantum critical fan, with the tip of the fan at \((M/b)_\mathrm{crit.} \simeq 0.9493\), suggesting that the DC conductivity may provide a probe of the quantum phase transition at \((M/b)_\mathrm{crit.} \).}
    \label{fig:sigma_zz_density}
\end{figure}

From figure~\ref{fig:sigma_zz_DC} we see that at small \(T/b\), \(\s_{zz}^\mathrm{DC}\) increases much more rapidly with increasing temperature for values of \(M/b\) close the quantum phase transition. This suggests that the DC conductivity in the \(z\) direction at \(T \neq 0\) may provide a probe of the quantum phase transition. Indeed, in figure~\ref{fig:sigma_zz_density} we plot \(\Delta \s_{zz}^\mathrm{DC}\), defined by
\begin{equation}
    \Delta \s_{zz}^\mathrm{DC}(M,b,T) = \s_{zz}^\mathrm{DC}(M,b,T) - \s_{zz}^\mathrm{DC}(M,b,0),
\end{equation}
in the \((M/b,T/b)\) plane. Expanding out from the quantum phase transition at \((M/b)_\mathrm{crit.} \simeq 0.9493\) we observe a typical quantum critical fan-like structure at small \(T/b\), with \(\Delta \s_{zz}^\mathrm{DC}\) larger inside the fan than outside for fixed \(T/b\). For larger \(T/b\) this structure is washed out by the universal \(\s_{zz}^\mathrm{DC} \propto T\) behaviour observed for all \(M/b\) in figure~\ref{fig:sigma_zz_DC}.

The bright vertical line in figure~\ref{fig:sigma_zz_density} is an artifact of the \(T=0\) subtraction. For any fixed \(T/b \neq 0\), \(\s_{zz}^\mathrm{DC}\) is a smooth function of \(M/b\), while at \(T=0\) it has a discontinuous first derivative as shown in figure~\ref{fig:sigma_zz_DC_T0}, leading to a cusp in \(\Delta \s_{zz}^\mathrm{DC}\). To avoid this it may be better to compute \(T \, \p \s_{zz}^\mathrm{DC}/\p T\), with the derivative taken at fixed \(M\) and \(b\).  This also has the advantage that one could compute the derivative without knowing the DC conductivity exactly at \(T=0\), so it may be more easily experimentally accessible. However, it would be computationally intensive to evaluate \(\p \s_{zz}^\mathrm{DC}/\p T\) numerically for enough data points to obtain a plot comparable to figure~\ref{fig:sigma_zz_density}, and other than the cusp the qualitative features of such a plot should not be too different, so we have only computed \(\Delta \s_{zz}^\mathrm{DC}\).

\subsubsection{AC conductivity}

We now compute the AC conductivities of our system. To keep the discussion concise we will restrict to \(T=0\). We will begin by deriving approximate formulas for the AC conductivities at small and large frequencies. We will then present numerical results for a wider range of frequencies.

\paragraph{Small frequency}

Using the near-boundary expansions in equation~\eqref{eq:gauge_fluctuation_near_boundary}, it is straightforward to rewrite the imaginary parts of the Green's functions~\eqref{eq:no_bg_greens_function_formula} as
\begin{align}
    \Im \, \vev{J^x J^x}_\mathrm{R}(\w, 0) &= \frac{1}{32 \pi \gn L |A_x^{(0)}(\w)|^2}  \cF_x(\w),
    \nonumber \\
    \Im \, \vev{J^z J^z}_\mathrm{R}(\w,0) &= \frac{1}{32 \pi \gn L |A_z^{(0)}(\w)|^2} \cF_z(\w),
    \label{eq:T0_no_bg_spectral_functions}
\end{align}
where
\begin{align}
    \cF_x(\w) &= -i r \sqrt{f} g \le[ A_x(-\w;r) \p_r A_x(\w;r) - A_x(\w,r) \p_r A_x(-\w;r) \ri],
    \nonumber \\
    \cF_z(\w) &= -i r \sqrt{f} h \le[ A_z(-\w;r) \p_r A_z(\w;r) - A_z(\w,r) \p_r A_z(-\w;r) \ri] .
    \label{eq:T0_no_bg_fluxes}
\end{align}
and we have used that \(A_i(\w;r)^* = A_i(-\w;r)\), as required by reality of \(A_i(t,r)\). Notice that we have not written \(r\to 0\) limits in equation~\eqref{eq:T0_no_bg_spectral_functions}. This is because the equations of motion~\eqref{eq:gauge_fluctuation_eom_no_bg} imply that the right-hand sides of equation~\eqref{eq:T0_no_bg_fluxes} are independent of \(r\), allowing us to evaluate \(\cF_x(\w)\) and \(\cF_z(\w)\) at any value of of \(r\). It will be convenient to evaluate them in the limit \(r \to \infty\).

We can determine the small-frequency (\(\w \ll M,b\)) behaviours of the real parts of the conductivities following ref.~\cite{Gubser:2008wz}. These are determined by the behaviour of the solutions in the deep IR, where \(g \approx g_0 r^{-\a}\) and \(h \approx h_0 r^{-\b}\). In this region, the equations of motion~\eqref{eq:gauge_fluctuation_eom_no_bg} become
\begin{equation}
    r \p_r \le[r^{1-\a} \p_r A_x(\w;r)\ri] = - \frac{\w^2 L^2}{g_0} A_x(\w;r),
    \qquad
    r^{1-\a+\b} \p_r \le[r^{1-\b} \p_r  A_z(\w;r) \ri] = - \frac{\w^2 L^2}{g_0} A_z(\w;r).
\end{equation}
For \(\w>0\), the solutions to these equations of motion obeying ingoing boundary conditions at \(r \to \infty\) are
\begin{equation} \label{eq:T0_gauge_fluctuation_IR_solutions}
    A_x(\w;r) = r^{\a/2} H_1^{(1)} \le(\frac{2 \w L}{\a \sqrt{g_0}} r^{\a/2} \ri),
    \qquad
    A_z(\w;r) =  r^{\b/2} H_{\b/\a}^{(1)} \le( \frac{2 \w L}{\a \sqrt{g_0}} r^{\a/2} \ri),
\end{equation}
where \(H_n^{(1)}\) are Hankel functions of the first kind. For \(\w<0\), \(H_n^{(1)}\) should be replaced by Hankel functions of the second kind \(H_n^{(2)}\). Substituting these solutions into equation~\eqref{eq:T0_no_bg_fluxes}, we find that the \(r\)-independent fluxes are \(\cF_x = 2 g_0 \a/\pi\) and \(\cF_z = 2 h_0 \a/\pi\). The real parts of the conductivities are then
\begin{equation}
    \Re \s_{xx}(\w) = \frac{L g_0 \a}{16 \pi^2 \gn \w |A_x^{(0)}(\w)|^2},
    \qquad
    \Re \s_{zz}(\w) = \frac{L h_0 \a}{16 \pi^2 \gn \w |A_z^{(0)}(\w)|^2}.
    \label{eq:conductivities_small_frequency_intermediate}
\end{equation}

We now need to determine the near-boundary coefficients \(A_x^{(0)}(\w)\) and \(A_z^{(0)}(\w)\). In general these should be determined by matching the IR solutions in equation~\eqref{eq:T0_gauge_fluctuation_IR_solutions} to asymptotic solutions computed at small \(r\). However, in the simple case under consideration the small-\(r\) solution is just \(A_{x,z}(\w;r) =\text{constant}\), so we can simply obtain \(A_{x,z}^{(0)}(\w)\) by taking the \(r\to0\) limit of the solutions in equation~\eqref{eq:T0_gauge_fluctuation_IR_solutions}, yielding \(A_x^{(0)}(\w) = - i \a \sqrt{g_0} / L \pi \w\) and \(A_z^{(0)}(\w) = - i \Gamma(\b/\a) (\a \sqrt{g_0}/L \w)^{\b/\a} / \pi\). Substituting these results into equation~\eqref{eq:conductivities_small_frequency_intermediate}, we find that the real part of \(\s_{xx}\) is linear in \(\w\) for all phases, while the real part of \(\s_{zz}\) depends on the phase of the system through the exponents \(\a\) and \(\b\),
\begin{equation} \label{eq:no_bg_sigma_small_frequency}
    \Re \s_{xx}(\w) = \frac{L^3}{16 \gn \a} \w,
    \qquad
    \Re \s_{zz}(\w) = \frac{L^2 h_0}{16 \gn  \G(\b/\a)^2 \sqrt{g_0}} \le(\frac{L\w}{\a \sqrt{g_0}} \ri)^{2\frac{\b}{\a} - 1},
\end{equation}
for \(\w \ll M,\, b\). In the topologically trivial phase \(\a = \b\), and so \(\Re \s_{zz}\) is also linear in \(\w\) at small frequency. For our choice of \(\l = 34/13\), the topological phase has \(\a = 2\b\), and \(\Re \s_{zz}\) reduces to the DC conductivity written in equation~\eqref{eq:DC_conductivity_T0}. Finally, in the critical phase we have \(\a \simeq 2.327\) and \(\b \simeq 1.7346\), leading to \(\Re \s_{zz} \propto \w^{0.491}\).

\paragraph{Large frequency}

At large frequencies, \(\w \gg M,b\), the conductivity is determined by the physics of the UV fixed point. Holographically, this means that we can find the large \(\w\) limit of the conductivities by taking the metric to be that of \ads[5], i.e. \(f(r)=1\) and \(g(r) = h(r) = L^2/r^2\). The equations of motion for all three gauge field components are then the same,
\begin{equation}
    r \p_r \le[r^{-1} \p_r A_i(\w;r)\ri] = - \w^2 A_i(\w;r).
\end{equation}
For \(\w > 0\), the solution obeying ingoing boundary conditions at \(r\to \infty\) is
\begin{equation}
    A_i(\w;r) = r H_1^{(1)}\le( \w r \ri).
\end{equation}
Again, for \(\w < 0\) one should replace \(H_1^{(1)}\) with \(H_1^{(2)}\). Expanding the large frequency solution at small \(r\), we find
\begin{equation}
    A_i(\w;r) = - \frac{2 i}{\pi \w} \le[1 - \frac{\w^2 r^2}{2} \log (r/L) \ri] + \frac{\w r^2}{2 \pi} \le[\pi - i + 2 i \g_\mathrm{E} + 2 i \log(\w L/2) \ri] + \dots \; ,
\end{equation}
where \(\g_\mathrm{E} = 0.577... \) is the Euler-Mascheroni constant. Reading off the coefficients \(A_i^{(0)}(\w)\) and \(A_i^{(2)}(\w)\) and substituting into equation~\eqref{eq:no_bg_greens_function_formula} we obtain the Green's functions, from which we obtain the conductivities using the Kubo formula~\eqref{eq:conductivity_kubo},\footnote{The factor of \(L\) inside the logarithm in equation~\eqref{eq:no_bg_sigma_large_frequency} arises from the holographic renormalisation, and can be shifted by the addition of a finite counterterm proportional to \(F_{\m\n} F^{\m\n}\). For comparison to condensed matter systems, one should replace this factor \(L\) with some typical ultraviolet length scale.}
\begin{equation}
    \s_{xx}(\w) = \s_{zz}(\w) = \frac{L^3}{16 \pi \gn} \w \le[ 
        \frac{\pi}{2} + i \le(\g_\mathrm{E} + \log(L\w/2) \ri)
    \ri],
    \quad \text{for } \w \gg M,\,b
    \label{eq:no_bg_sigma_large_frequency}
\end{equation}

\paragraph{Numerical results}

Away from the limits of small or large \(\w\) we obtain the conductivities by solving the equations of motion~\eqref{eq:gauge_fluctuation_near_boundary} numerically, using the ingoing boundary conditions at large \(r\) written in equation~\eqref{eq:T0_gauge_fluctuation_IR_bcs}. We then perform a fit to the solution at small \(r\) to obtain the near-boundary coefficients \(A_i^{(0)}(\w)\) and \(A_i^{(2)}(\w)\), which determine the conductivities through equations~\eqref{eq:conductivity_kubo} and~\eqref{eq:no_bg_greens_function_formula}.

\begin{figure}
    \begin{subfigure}{0.5\textwidth}
        \includegraphics[width=\textwidth]{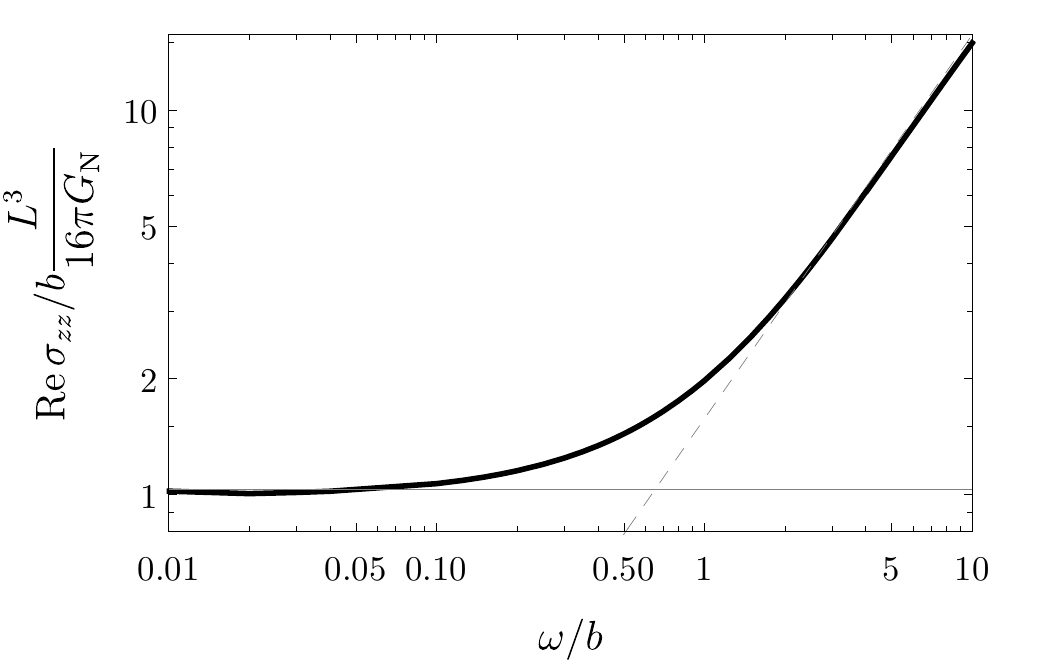}
        \caption{Real part of \(\s_{zz}\)}
    \end{subfigure}
    \begin{subfigure}{0.5\textwidth}
        \includegraphics[width=\textwidth]{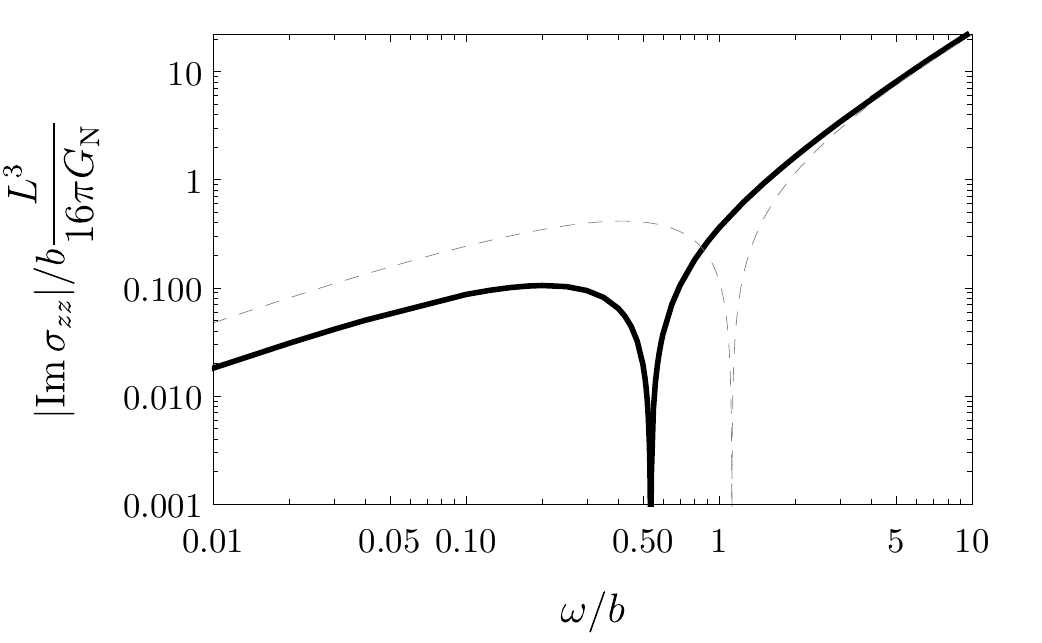}
        \caption{Imaginary part of \(\s_{zz}\)}
        \label{fig:sigma_zz_imaginary}
    \end{subfigure}
    \begin{subfigure}{0.5\textwidth}
        \includegraphics[width=\textwidth]{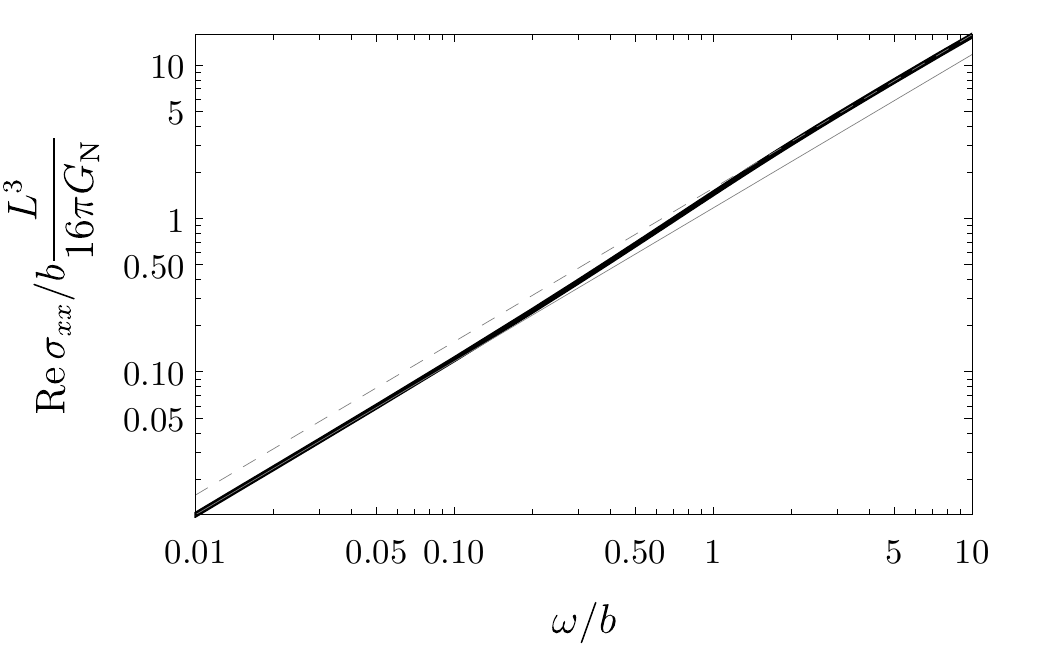}
        \caption{Real part of \(\s_{xx}\)}
    \end{subfigure}
    \begin{subfigure}{0.5\textwidth}
        \includegraphics[width=\textwidth]{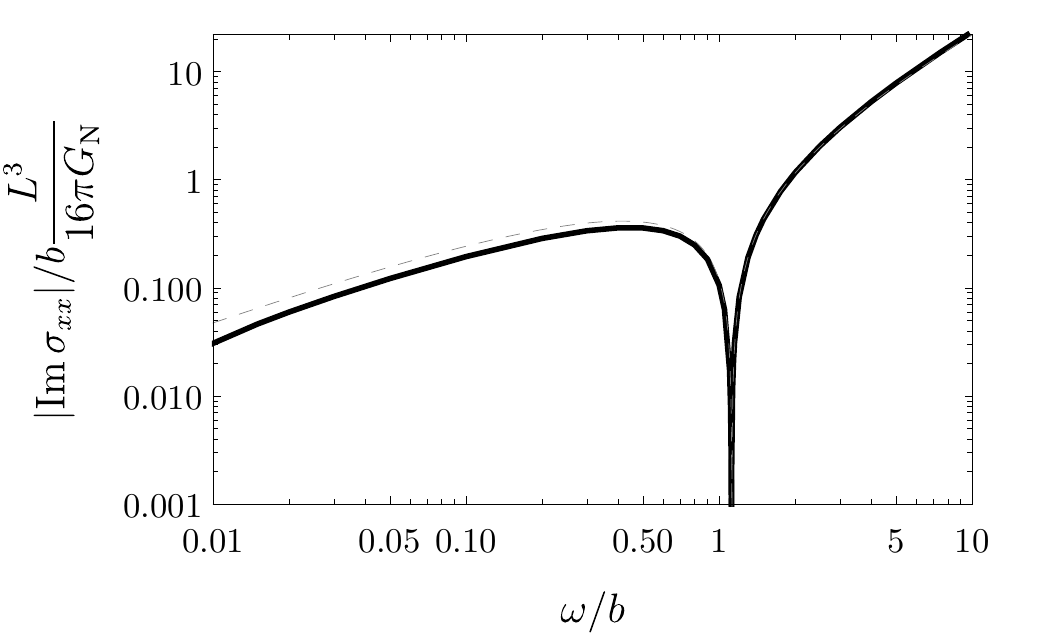}
        \caption{Imaginary part of \(\s_{xx}\)}
        \label{fig:sigma_xx_imaginary}
    \end{subfigure}
    \caption{The solid black curves show the AC conductivities at \(M=0\) and \(T=0\). The solid grey lines show the small frequency approximation~\eqref{eq:no_bg_sigma_small_frequency}, while the dashed grey lines show the large frequency approximation~\eqref{eq:no_bg_sigma_large_frequency}.
    \textbf{(a,\,b):}~The conductivity orthogonal to the plane of the nodal line.
    \textbf{(c,\,d):}~The conductivity in the plane of the nodal line.
    }
    \label{fig:ac_conductivities}
\end{figure}

Since we are most interested in the physics of the topological phase, for simplicity we restrict to \(M=0\). Logarithmic plots of our results for \(\s_{zz}\) and \(\s_{xx}\) as functions of \(\w/b\) are shown in figure~\ref{fig:ac_conductivities}. The thick black curves are our numerical results, while the solid grey lines show the small frequency approximation~\eqref{eq:no_bg_sigma_small_frequency} and the dashed grey lines show the large frequency approximation~\eqref{eq:no_bg_sigma_large_frequency}. Both approximations work well within their regimes of validity. An absolute value has been taken in the figures~\ref{fig:sigma_zz_imaginary} and~\ref{fig:sigma_xx_imaginary}, as the imaginary parts of the conductivities are not of fixed sign. Both \(\Im \s_{zz}\) and \(\Im \s_{xx}\) are negative at small frequencies and positive at large frequencies.

\subsection{Thermal conductivity}
\label{sec:thermal_conductivity}

We now calculate the thermal conductivity matrix \(\kappa\), given by~\cite{Hartnoll:2016apf}\footnote{The thermoelectric conductivity \(\a_{ij}\) vanishes for our system, since we have vanishing net charge density.}
\begin{equation} \label{eq:thermal_conductivity_kubo_formula}
    \kappa_{ij}(\w) = -\frac{1}{i \w T} \le[\vev{T^{i0} T^{j0}}_\mathrm{R}(\w,\vec{k}=0) - \vev{T^{i0} T^{j0}}_\mathrm{R}(0,\vec{k}\to 0)\ri],
\end{equation}
where in \(\vev{T^{i0} T^{j0}}_\mathrm{R}(0,\vec{k}\to 0)\) one should send the frequency to zero first, followed by momentum. Due to the asymmetry of the stress tensor, the order of indices is crucial. Replacing either \(T^{i0}\) with \(T^{0i}\) and/or \(T^{j0}\) with \(T^{0j}\) yields a Kubo formula for a different transport coefficient, special to systems with broken Lorentz invariance, that is discussed in ref.~\cite{Hoyos:2013qna}.

At \(b_{\m\n}=0\), the limits of Green's functions appearing in equation~\eqref{eq:thermal_conductivity_kubo_formula} are fixed by a Ward identity~\cite{Hartnoll:2008hs}. This is not the case at non-zero \(b_{\m\n}\). However, it will still be useful to see what the Ward identities of our system tell us about the thermal conductivity.

\subsubsection{Ward identities}

As we show in appendix~\ref{app:two_point_ward_identities}, translational symmetry implies that two-point functions of the stress tensor \(T^{\m\n}\) and antisymmetric tensor \(\cO^{\m\n}\) satisfy the Ward identities\footnote{Technically the Green's functions that satisfy equation~\eqref{eq:GTT_translation_ward_identity} are not the retarded Green's functions. However, they are related to them by contact terms that cancel in the difference appearing in the Kubo formula~\eqref{eq:thermal_conductivity_kubo_formula}~\cite{Policastro:2002tn,Herzog:2003ke}.}
\begin{subequations} \label{eq:two_point_function_ward_identities}
\begin{align}
    k_\m \le[ \vev{T^{\m\n} T^{\r\s}}_\mathrm{R}(\w,\vec{k}) - \h^{\n\r} \vev{T^{\m\s}} - \h^{\n\s} \vev{T^{\m\r}} + \h^{\m\n} \vev{T^{\r\s}} \ri] &= 0,
    \label{eq:GTT_translation_ward_identity}
    \\
    k_\m \vev{T^{\m\n} \cO^{\r\s}}_\mathrm{R}(\w,\vec{k}) + k^\n \vev{\cO^{\r\s}} &= 0,
    \label{eq:GTO_translation_ward_identity}
\end{align}
\end{subequations}
where \(k_\m = (-\w,\vec{k})\) is the four-momentum.

At \(\vec{k}=0\), the \((\n,\r,\s) = (i,j,0)\) component of the Ward identity~\eqref{eq:GTT_translation_ward_identity} reads
\begin{equation} \label{eq:ward_identity_energy}
    \w \le[ \vev{T^{0i} T^{j0}}_\mathrm{R}(\w,0) - \d^{ij} \ve \ri] = 0,
\end{equation}
so for non-zero \(\w\) we obtain \(\vev{T^{0i} T^{j0}}_\mathrm{R}(\w,0) = \d^{ij} \ve\). We wish to relate this two-point function to the one appearing in the thermal conductivity~\eqref{eq:thermal_conductivity_kubo_formula}, which involves \(T^{i0}\) rather than \(T^{0i}\). We can do so using the relation \(T^{\m\n} = \tilde{T}^{\m\n} + 2 \cO^{\m\r} b^\n{}_\r\) from equation~\eqref{eq:conserved_stress_tensor}. Since the operator \(\tilde{T}^{\m\n}\) sourced by the metric is a symmetric tensor, this implies that 
\begin{equation} \label{eq:T0i_Ti0}
    T^{i0} = T^{0i} - 2 b_{ik} \cO^{0k},
\end{equation}
where we have made use of the fact that \(b^0{}_\m =0\) for our system. We then have
\begin{equation}
    \vev{T^{i0} T^{j0}}_\mathrm{R}(\w,0) = \d^{ij} \ve - 2 b_{ik} \vev{T^{j0} \cO^{0k}}_\mathrm{R}(\w,0),
    \quad \text{for }
    \w \neq 0.
\end{equation}
Now consider the \((\n,\r,\s) = (j,0,i)\) component of equation~\eqref{eq:GTO_translation_ward_identity} at \(\vec{k}=0\), which for \(\w \neq 0\) implies \(\vev{T^{0j}\cO^{0k}}_\mathrm{R}(\w,0) = 0\). Relating \(T^{0j}\) to \(T^{j0}\) through equation~\eqref{eq:T0i_Ti0}, we find
\begin{equation} \label{eq:Ti0_Tj0}
    \vev{T^{i0} T^{j0}}_\mathrm{R}(\w,0) = \d^{ij} \ve + 4 b_{ik} b_{jl} \vev{\cO^{0k} \cO^{0l}}_\mathrm{R}(\w,0),
    \quad
    \text{for } \w \neq 0.
\end{equation}

Finally, for \(\w=0\) the \((\n,\r,\s) = (0,j,0)\) component of the Ward identity is
\begin{equation} \label{eq:Ti0_Tj0_w0}
    k_i \le[\vev{T^{i0} T^{j0}}_\mathrm{R}(0,\vec{k} \to 0) - \d^{ij} p \ri] = 0.
\end{equation}
Assuming that the \(\vec{k} \to 0 \) limit is smooth, we then have \(\vev{T^{i0} T^{j0}}_\mathrm{R}(0,\vec{k} \to 0) = - \d^{ij} p\). Substituting this into the Kubo formula~\eqref{eq:thermal_conductivity_kubo_formula} and using equation~\eqref{eq:Ti0_Tj0}, we obtain
\begin{equation} \label{eq:thermal_conductivity_two_point_function}
    \k_{ij}(\w) =  \frac{i}{\w T} \le[s T \d_{ij} +4 b_{ik} b_{jl} \vev{\cO^{0k} \cO^{0l}}_\mathrm{R}(\w, 0) \ri], \quad \text{for }\w \neq 0,
\end{equation}
where we have used the thermodynamic relation \(\ve + p = T s\).

For the holographic NLSM, the only non-zero components of the two-form source are \(b_{xy} = - b_{yx} = b\). The form of the thermal conductivity in the direction orthogonal to the plane of the nodal line is therefore unchanged from the well-known \(b=0\) result, \(\k_{zz} = i s/\w\) for \(\w \neq 0\)~\cite{Hartnoll:2009sz}. Through the Kramers-Kronig relations, the \(1/\w\) pole in the imaginary part of the thermal conductivity implies that there should be a delta-function contribution to the real part at \(\w=0\), arising from conservation of momentum. The final expression for \(\k_{zz}\), valid at all frequencies, is then
\begin{equation} \label{eq:thermal_conductivity_zz}
    \k_{zz}(\w) = s \le[\frac{i}{\w} + \pi \d(\w) \ri].
\end{equation}

On the other hand, the thermal conductivities in the plane of the nodal line depend on the two-point function of \(\cO^{0y}\) at zero momentum,
\begin{multline} \label{eq:kappa_xx_formula}
    \k_{xx}(\w) = \k_{yy}(\w) = \frac{i}{\w T} \le[s T + 4 b^2 \vev{\cO^{0y} \cO^{0y}}_\mathrm{R} (\w,0) \ri]
    \\
    + \frac{\pi}{T} \le[ s T + 4 b^2 \vev{\cO^{0y} \cO^{0y}}_\mathrm{R}(\w \to 0,0)\ri] \d(\w),
\end{multline}
where \(\k_{xx} = \k_{yy}\) due to rotational symmetry in the \((x,y)\) plane, and we have included a delta-function contribution at \(\w=0\), arising from the Kramers-Kronig relations. We will calculate the two-point function \(\vev{\cO^{0y} \cO^{0y}}_\mathrm{R} (\w,0)\) holographically. We find that this two-point function is non-zero at \(T=0\), and hence it provides the dominant contribution to equation~\eqref{eq:kappa_xx_formula} at low \(T/b\). We therefore find that the thermal conductivities \(\k_{xx} = \k_{yy}\) in the plane of the nodal line are significantly enhanced compared to the out-of-plane thermal conductivity \(\k_{zz}\) at low temperature. In particular, the Drude weight of \(\k_{xx}\) diverges as \(1/T\) as \(T \to 0\).

One should be slightly careful when applying the Kramers-Kronig relations to \(\k_{xx}(\w)\) in order to obtain the coefficient of \(\d(\w)\). The relations only hold for functions which vanish when \(|\w| \to \infty\) with \(\Im \w > 0\), while the two-point function \(\vev{\cO^{0y} \cO^{0y}}_\mathrm{R} (\w,0)\) may diverge at large \(\w\). One should first subtract these divergences before applying the Kramers-Kronig relations. However, the subtraction does not change the coefficient of the \(1/\w\) pole in the first line of equation~\eqref{eq:kappa_xx_formula}, and therefore does not change the coefficient of \(\d(\w)\).

\subsubsection{Holographic computation}

We wish to use holography to compute the two-point function \(\vev{\cO^{0y} \cO^{0y}} (\w,\vec{k}=0)\) that determines \(\k_{zz}(\w)\). We will need to consider linearised fluctuations of the metric \(\d G_{mn}\) and two-form \(\d B_{mn}\). The time dependence of the fluctuations will be written as \(\d G_{mn}(t,r) = \int \frac{\diff \w}{2\pi} e^{-i \w t} \d G_{mn}(\w;r)\), and similar for \(\d B_{mn}\). In a gauge in which \(\d G_{rm} = 0\), there are three fluctuation components relevant for computing the two-point function in question: \(\d G_{tx}\), \(\d B_{ty}\), and \(\d B_{ry}\). They couple to each other, and not to any other fluctuations. It will be convenient to define \(\cG_{tx} = \d G_{tx}/h\), \(\cB_{ry} = \d B_{tr}/L^2\), and \(\cB_{ty} =r  \d B_{ty}/L^2\). The linearised equations of motion for these fluctuations are
\begin{align}
    i \w h \, \p_r \cG_{tx} - 2 f g \frac{B}{h} \le(m_B^2 + 2 \s \f^2 + 2 \l \frac{B^2}{h^2} \ri) \cB_{ry} &= 0,
    \nonumber \\[1em]
    \frac{r^2}{\sqrt{f} g} \p_r \le(\frac{\p_r \cB_{ty}}{r \sqrt{f}} \ri) - \frac{r^2 B'}{L^2h} \p_r \le( \frac{h}{f g} \cG_{tx}\ri) + \frac{i \w r}{\sqrt{f} g} \p_r \le( \frac{r}{\sqrt{f}} \cB_{ry} \ri) \hspace{1.5cm} &
    \nonumber \\
    + \frac{1}{2 r f g} \le(\frac{r f'}{f} + 2 - 2 m_B^2 - 4 \s \f^2 - 4 \l \frac{B^2}{h^2} \ri) \cB_{ty}  &= 0,
    \label{eq:thermal_conductivity_fluctuation_equations}
    \\[1em]
    i \w L^2 \p_r \le(\frac{ \cB_{ty}}{r} \ri) - i \w \, \p_r B \cG_{tx} - \le[\w^2 L^2 - f g \le( m_B^2 + 2 \l \frac{B^2}{h^2} + 2 \s \f^2 \ri) \ri] \cB_{ry} &= 0.
    \nonumber 
\end{align}
Note that \(\cB_{ry}\) is fixed algebraically by these equations.

In order to solve the equations of motion and determine the two-point functions, it is useful to work with linear combinations of the fluctuations that are invariant under gauge transformations of the background solution~\cite{Kovtun:2005ev}. Under an infinitesimal diffeomorphism generated by a vector \(\xi^m\), the metric and two-form fluctuations transform as
\begin{equation}
    \d G_{mn} \to \d G_{mn} - \nabla_m \xi_n - \nabla_n \xi_m,
    \qquad
    \d B_{mn} \to \d B_{mn} - \xi^k \nabla_k B_{mn} - B_{mk} \nabla_n \xi^k - B_{kn} \nabla_m \xi^k.
\end{equation}
We find that \(\cB_{ry}\) is invariant under the set of such diffeomorphisms that preserve the radial gauge condition \(\d G_{rm} = 0\), while \(\cG_{tx}\) and \(\cB_{ty}\) transform non-trivially. We can form a gauge-invariant combination
\begin{align} \label{eq:thermal_conductivity_gauge_inv_fluctuation}
    \cZ(\w;r) = \cG_{tx}(\w;r) - \frac{L^2}{r B} \cB_{ty}(\w;r).
\end{align}
The equations of motion~\eqref{eq:thermal_conductivity_fluctuation_equations} imply that \(\cZ\) satisfies a second order ODE, that does not depend on the other fluctuations,
\begin{equation} \label{eq:thermal_conductivity_ZEOM}
    \p_r^2 \cZ(\w;r) + c_1 \, \p_r \cZ(\w;r) + c_2 \cZ(\w;r) = 0.
\end{equation}
The forms of the coefficients \(c_1\) and \(c_2\) appearing in this equation are rather complicated, we give them explicitly in appendix~\ref{app:coefficients}. Once a solution for \(\cZ\) has been found, the remaining fluctuations are determined by the equations
\begin{align}
    \p_r \cG_{tx}(\w;r) &= \frac{2 B \le[ h^2\le(m_B^2 + 2 \s \f^2 \ri) + 2 \l B^2\ri] \p_r[B \cZ(\w;r)]}{\le(h^2 + 2 B^2\ri)\le[ h^2\le(m_B^2 + 2 \s \f^2 \ri) + 2 \l B^2\ri] - \frac{\w^2 L^2}{f g} h^4 },
    \nonumber
    \\
    \cB_{ry}(\w;r) &= \frac{i \w h^4 \p_r[B \cZ(\w;r)]}{f g \le\{ \le(h^2 + 2 B^2\ri)\le[ h^2\le(m_B^2 + 2 \s \f^2 \ri) + 2 \l B^2\ri] - \frac{\w^2 L^2}{f g} h^4\ri\} },
    \label{eq:fluctuations_from_gauge_invariant}
\end{align}
which follow from equation~\eqref{eq:thermal_conductivity_fluctuation_equations}.

Near the horizon at \(r=r_0\), we find that solutions to the equation of motion~\eqref{eq:thermal_conductivity_ZEOM} take the form
\begin{equation}
    \cZ(\w; r) \propto (r_0 - r)^{\pm i \w/2\pi T}.
\end{equation}
To determine the retarded Green's functions we impose ingoing boundary conditions at the horizon, corresponding to choosing the minus sign in the exponent. Near the boundary, \(\cZ\) has the small-\(r\) expansion
\begin{equation} \label{eq:thermal_conductivity_gauge_invariant_near_boundary}
    \cZ(\w; r) = \cZ^{(0)}(\w) \le[1 + \frac{\w^2 r^2}{2} \log\le(\frac{r}{L}\ri)\ri] + \cZ^{(2)}(\w) r^2  + \cO\le(r^4 \log^2 r\ri),
\end{equation}
with coefficients \(\cZ^{(0)}(\w)\) and \(\cZ^{(2)}(\w)\) determined by the boundary conditions. Crucially, the leading-order coefficient in this expansion is a linear combination of the boundary values of the metric and two-point fluctuations,
\begin{equation} \label{eq:thermal_conductivity_sources}
    \cZ^{(0)}(\w) = \cG_{tx}^{(0)}(\w) - \frac{1}{b} \cB_{yz}^{(0)}(\w).
\end{equation}
where \(\cG_{tx}^{(0)}(\w) \equiv \cG_{tx}(\w;r\to0)\) and \(\cB_{ty}^{(0)} \equiv \cB_{ty}(\w;r\to0)\).

Expanding the action~\eqref{eq:holographic_bulk_action} to quadratic order in the fluctuations, and using the equations of motion, we find that the on-shell action reads
\begin{multline}
    S^\star = -\frac{L^3}{16 \pi \gn} \int \diff^3 x \int_{-\infty}^\infty \frac{\diff \w}{2\pi} \biggl[
        2 b^2 \cZ^{(0)}(-\w) \cZ^{(2)}(\w)
        + C_{\cG\cG} \cG_{tx}^{(0)}(-\w) \cG_{tx}^{(0)}(\w)
    \\ 
    -C_{\cB\cG} \cB_{ty}^{(0)}(-\w) \le( \cG_{tx}^{(0)}(\w) - \frac{1}{2b} \cB_{ty}^{(0)}(\w) \ri)
    \biggr],
\end{multline}
where the coefficients are
\begin{align}
    C_{\cG\cG} &=\frac{1}{2} \biggl(3m + 3 b b_2 - M \f_2 + \frac{7 \l + 9}{4} b^4 + \frac{3}{2} \s b^2 M^2 - \frac{2+3 \h}{24} M^4 - b^2 \w^2
    \biggr),
    \nonumber \\
    C_{\cB\cG} &= 4 b_2 + 2 (\l+1) b^3 + 2 \s b M^2 - b \w^2.
\end{align}
Applying the Minkowski space correlator prescription of refs.~\cite{Son:2002sd,Herzog:2002pc}, we can then read off expressions for the two-point functions of the operators \(\tilde{T}^{0x}\) and \(\cO^{0y}\), dual to \(\cG_{tx}\) and \(\cB_{ty}\), respectively, at zero momentum
\begin{align}
    \vev{\tilde{T}^{0x} \tilde{T}^{0x}}_\mathrm{R}(\w,0) 
    &= \frac{L^3}{8 \pi \gn} \le[
        2 b^2 \frac{\cZ^{(2)}(\w)}{\cZ^{(0)}(\w)} + C_{\cG\cG}
    \ri],
    \nonumber \\
    \vev{\tilde{T}^{0x} \cO^{0y}}_\mathrm{R}(\w,0) 
    &=- \frac{L^3}{32 \pi \gn} \le[
        4 b \frac{\cZ^{(2)}(\w)}{\cZ^{(0)}(\w)} + C_{\cB\cG}
    \ri],
    \label{eq:thermal_conductivity_greens_functions}
    \\
    \vev{\cO^{0y} \cO^{0y}}_\mathrm{R}(\w,0) &=  \frac{L^3}{32 \pi \gn} \le[2 \frac{\cZ^{(2)}(\w)}{\cZ^{(0)}(\w)} + \frac{1}{2b} C_{\cB\cG} \ri],
    \nonumber
\end{align}
where we use equation~\eqref{eq:thermal_conductivity_sources} to relate \(\cZ^{(0)}\) to the sources \(\cG_{0x}^{(0)}\) and \(\cB_{0y}^{(0)}\). Notice that we have \(\vev{T^{0x} \cO^{0y}}_\mathrm{R}(\w,0) \equiv \vev{\tilde{T}^{0x} \cO^{0y}}_\mathrm{R}(\w,0) + 2 b \vev{\cO^{0y} \cO^{0y}}(\w,0) = 0\), as demanded by the Ward identity~\eqref{eq:GTO_translation_ward_identity}. Further, using equation~\eqref{eq:energy_pressure} we can confirm
\begin{equation}
    \vev{T^{0x} T^{0x}}_\mathrm{R}(\w,0) = \vev{T^{0x} T^{x0}} _\mathrm{R}(\w,0) \equiv \vev{\tilde{T}^{0x}\tilde{T}^{0x}}_\mathrm{R}(\w,0) + 2 b \vev{\tilde{T}^{0x} \cO^{0y}}_\mathrm{R}(\w,0) = \ve,
\end{equation}
as expected from the Ward identity~\eqref{eq:ward_identity_energy}.\footnote{The reproduction of the two-point functions fixed by the Ward identities only occurs due to the presence of the counterterm involving \(\hat{\nabla}_\l B_{\m\n} \hat{\nabla}^\l B^{\m\n}\) in equation~\eqref{eq:counterterms}. Without this term present, we find \(\vev{T^{0x}\cO^{0y}}_\mathrm{R}(\w,0) \propto \w^2\) and \(\vev{T^{0x} T^{0x}}_\mathrm{R} - \ve \propto \w^2\). In other words, without this counterterm, the two point functions that we compute holographically satisfy Ward identities that differ from those in equation~\eqref{eq:two_point_function_ward_identities} by contact terms. We note that \(\vev{T^{x0}T^{x0}}_\mathrm{R}(\w,0)\), and therefore the thermal conductivity \(\k_{xx}\), is independent of the coefficient of this counterterm.}

We can now compute the thermal conductivity \(\k_{xx}\) holographically by solving the equation of motion~\eqref{eq:thermal_conductivity_ZEOM} for \(\cZ(\w;r)\) numerically, imposing ingoing boundary conditions at the horizon. Fitting the resulting solution to the near boundary expansion~\eqref{eq:thermal_conductivity_gauge_invariant_near_boundary} at small \(r\), we can determine \(\cZ^{(0)}(\w)\) and \(\cZ^{(2)}(\w)\). These coefficients may then be substituted into equation~\eqref{eq:thermal_conductivity_greens_functions} to obtain the two-point function \(\vev{\cO^{0y}\cO^{0y}}_\mathrm{R}(\w,0)\), which then determines \(\k_{xx}\) through equation~\eqref{eq:kappa_xx_formula}.

\begin{figure}
    \begin{subfigure}{\textwidth}
    \begin{center}
        \includegraphics{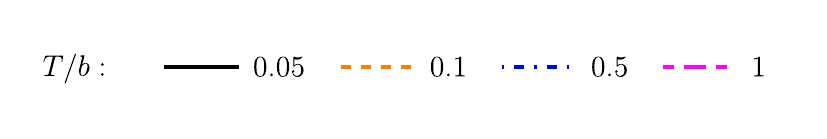}
        \vspace{-1em}
    \end{center}
    \end{subfigure}
    \begin{subfigure}{0.5\textwidth}
        \includegraphics[width=\textwidth]{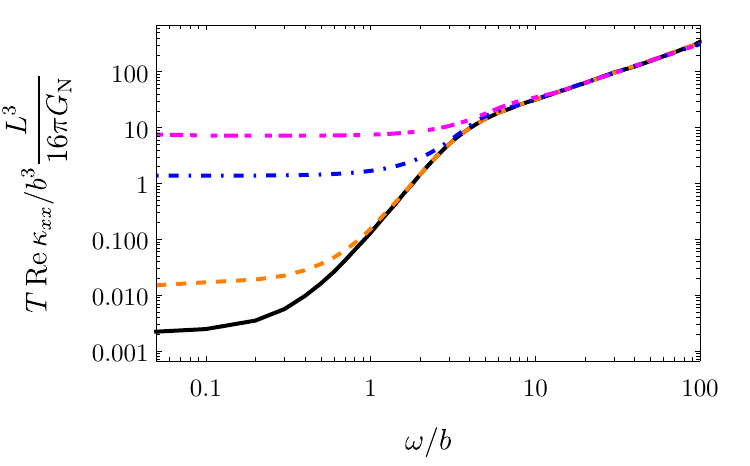}
        \caption{Real part of \(\k_{xx}\)}
        \label{fig:thermal_conductivity_real}
    \end{subfigure}
    \begin{subfigure}{0.5\textwidth}
        \includegraphics[width=\textwidth]{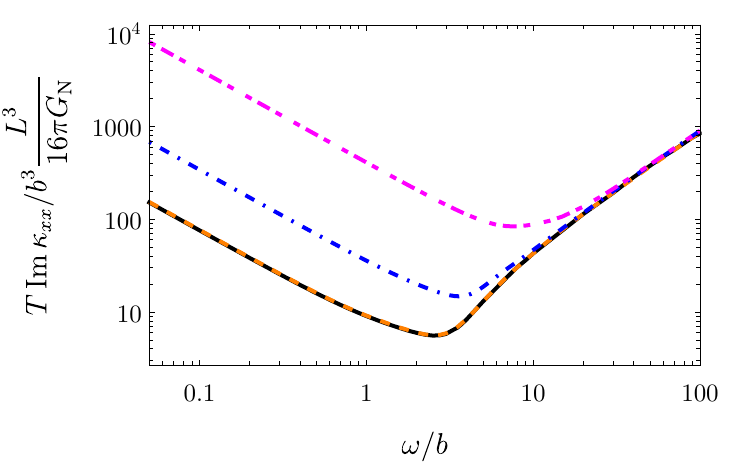}
        \caption{Imaginary part of \(\k_{xx}\)}
        \label{fig:thermal_conductivity_imaginary}
    \end{subfigure}
    \begin{subfigure}{0.5\textwidth}
        \includegraphics[width=\textwidth]{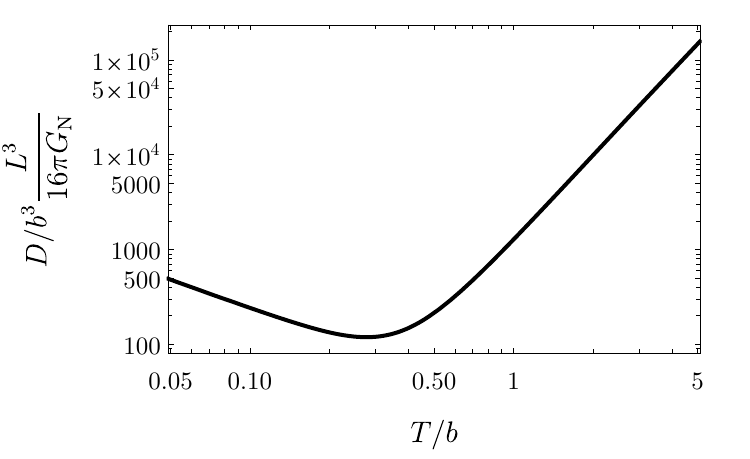}
        \caption{Drude weight}
        \label{fig:thermal_conductivity_drude}
    \end{subfigure}
    \begin{subfigure}{0.5\textwidth}
        \includegraphics[width=\textwidth]{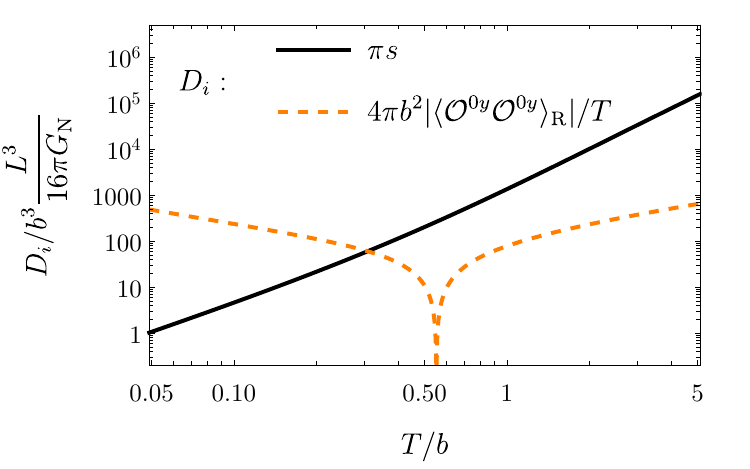}
        \caption{Contributions to Drude weight}
        \label{fig:thermal_conductivity_contributions}
    \end{subfigure}
    \caption{\textbf{(a,\,b):} The real and imaginary parts of the thermal conductivity \(\k_{xx}\) multiplied by temperature \(T\), in units of \(b^3 L^3/16 \pi \gn\), plotted as a function of frequency \(\w/b\) for sample values of \(T/b\). \textbf{(c):} The Drude weight of \(\k_{xx}\), i.e. the coefficient of \(\d(\w)\) in equation~\eqref{eq:kappa_xx_formula}, as a function of \(T/b\). It diverges at both small and large \(T/b\). \textbf{(d):} The two separate contributions to the Drude weight as functions of \(T/b\). The solid black curve shows the entropy density contribution \(\pi s\). The dashed orange curve shows the two-point function contribution \(4 \pi b^2 \vev{\cO^{0y}\cO^{0y}}_\mathrm{R}(\w \to 0,0)/T\). Although the \(\w \to 0\) limit of the two-point function is real, it is not of fixed sign, being positive for small \(T/b\) and negative for large \(T/b\). We therefore plot its absolute value. The two-point function makes the dominant contribution to the thermal conductivity at low temperatures, while the entropy density dominates at high temperatures.}
\end{figure}

For simplicity, we have only computed the thermal conductivity at \(M=0\). We show our numerical results for \(\k_{xx}\) as a function of frequency in figures~\ref{fig:thermal_conductivity_real} and~\ref{fig:thermal_conductivity_imaginary}, for different sample values of \(T/b\). In contrast to when \(b=0\), we find that the real part of \(\k_{xx}\) is non-zero for \(\w \neq 0\), taking a finite, approximately \(\w\)-independent value at \(\w/b \ll 1\), and growing as \(\Re \k_{xx} \simeq 3.2 \, b^2 \w L^3 / 16 \pi \gn T\) when \(\w/b \gg 1\). In between these two regimes we observe an intermediate region at \(\w / b\) of order one, in which \(\Re \k_{xx}\) grows more rapidly with frequency. The size of this intermediate region decreases with increasing \(T/b\).

At small \(\w/b \ll 1\) we find \(\Im \k_{xx} \propto 1/\w\), as for at \(b=0\). However, the proportionality coefficient is no longer just the entropy density, instead receiving a contribution from the real part of the two-point function appearing in equation~\eqref{eq:kappa_xx_formula}. At large \(\w/b\), the imaginary part grows as \( \Im \k_{xx} \simeq 1.9 \, b^2 \w \log (\w/b) L^3 / 16 \pi \gn T \), with a proportionality coefficient independent of \(T/b\). At low temperatures, we find that the imaginary part of \(\k_{xx} T\) is approximately independent of \(T/b\): notice that the solid black and dashed orange curves coincide in figure~\ref{fig:thermal_conductivity_imaginary}.

The approximate independence of \(T \Im \k_{xx}\) on temperature at \(T/b \ll 1\) arises because at low temperatures the thermal conductivity is dominated by \(\vev{\cO^{0y} \cO^{0y}}_\mathrm{R} (\w,0)\) in equation~\eqref{eq:kappa_xx_formula}, which is non-zero as \(T\to0\). From equation~\eqref{eq:kappa_xx_formula}, this has the significant consequence that the Drude weight diverges as \(1/T\) in the limit \(T\to0\), with a coefficient determined by the zero frequency limit of the two-point function \(\vev{\cO^{0y} \cO^{0y}}_\mathrm{R} (\w,0)\). We plot the Drude weight as a function of \(T/b\) in figure~\ref{fig:thermal_conductivity_drude}, which clearly shows the expected \(1/T\) divergence as \(T \to 0\), as well as a \(T^3\) divergence as \(T \to \infty\).

In figure~\ref{fig:thermal_conductivity_contributions} we plot the two separate contributions to the Drude weight as functions of \(T/b\). As temperature goes to zero, the entropy density vanishes as \(s \propto T^2\) for fixed \(b\), and is therefore much smaller than \(b^2\vev{\cO^{0y} \cO^{0y}}_\mathrm{R}(\w\to0,0)/T\), which diverges as \(1/T\). Conversely, at high temperatures the entropy density grows as \(s \propto T^3\), whereas the two-point function grows more slowly, \(b^2\vev{\cO^{0y} \cO^{0y}}_\mathrm{R}(\w\to0,0)/T \propto T\). The entropy density therefore dominates in this limit, so we have \(\Im \k_{xx} \simeq \Im \k_{zz}\) at high temperatures.

\subsection{Shear viscosity}

In this section we compute the shear viscosities of the nodal line system from the Kubo formula
\begin{equation} \label{eq:shear_viscosity_kubo}
    \h_{ij,kl} = - \lim_{\w\to0} \frac{\Im \, \vev{T^{ij} T^{kl}}_\mathrm{R}(\w,0)}{\w}.
\end{equation}
The indices should be such that \(i \neq j\) and \((k,l)=(i,j)\) or \((j,i)\). 

For rotationally invariant holographic systems, all shear viscosities take the universal value \(\h = s/4\pi\)~\cite{Policastro:2001yc,Kovtun:2003wp,Buchel:2003tz,Kovtun:2004de,Starinets:2008fb}. Due to the anisotropy introduced by \(b\) this is not the case for the nodal line system. As discussed in section~\ref{sec:stress_tensor_comments}, a second consequence of non-zero \(b\) is that the stress tensor is not symmetric in its indices. Concretely, from equation~\eqref{eq:conserved_stress_tensor} we find that two-point functions of the stress tensor take the form
\begin{equation} \label{eq:TT_two_point_function}
    \vev{T^{\m\n} T^{\r\s}}_\mathrm{R} = \vev{\tilde{T}^{\m\n} \tilde{T}^{\r\s}}_\mathrm{R} - 2 \vev{\cO^{\m\l} \tilde{T}^{\r\s}}_\mathrm{R} b_\l{}^\n - 2 \vev{\tilde{T}^{\m\n} \cO^{\r\l}}_\mathrm{R} b_\l{}^\s + 4 \vev{\cO^{\m\l} \cO^{\r\h}}_\mathrm{R} b_\l{}^\n b_\h{}^\s.
\end{equation}
Through the Kubo formula~\eqref{eq:shear_viscosity_kubo} this implies that \(\h_{ij,kl}\), \(\h_{ij,lk}\), and \(\h_{ji,lk}\) are not necessarily equal. In a hydrodynamic expansion of the stress tensor, the stress tensor's asymmetry implies that there are more possible tensor structures in an anisotropic system compared to a relativistic system. We expect different linear combinations of \(\h_{ij,kl}\), \(\h_{ij,lk}\), and \(\h_{ji,lk}\) will couple to different tensor structures, although we leave an analysis of the hydrodynamics of our system to future work.

To compute two-point functions of the stress tensor holographically, we consider linearised fluctuations \(\d G_{mn}\) of the metric~\cite{Son:2002sd}. The components of the metric fluctuations relevant for the computation of the shear viscosity are \(\d G_{xy}\), \(\d G_{xz}\), and \(\d G_{yz}\), which decouple from other metric fluctuations due to the subgroup of rotational symmetry preserved by our system~\cite{Policastro:2002se,Kovtun:2005ev}. These three metric fluctuations are also decoupled from one another. However, at non-zero \(b\), \(\d G_{xz}\) and \(\d G_{yz}\) do couple to certain components of linearised fluctuations \(\d B_{mn}\) of the two-form field: \(\d G_{xz}\) couples to \(\d B_{yz}\), and \(\d G_{yz}\) couples to \(\d B_{xz}\).

Since the Kubo formula~\eqref{eq:shear_viscosity_kubo} involves the two-point function at zero momentum, throughout this subsection we will assume our linearised fluctuations have the spacetime-dependence \(\d G_{mn}(t,r) = \int \frac{\diff \w}{2\pi} e^{-i\w t} \d G_{mn}(\w;r)\), and similar for \(\d B_{mn}\). It will also be convenient to define \(\cG_{ij} = \d G^i{}_j\), i.e. \(\cG_{xy} = \d G_{xy}/h\), \(\cG_{xz} = \d G_{xz}/h\), and \(\cG_{yz} = \d G_{yz}/g\). Similarly we define \(\cB_{ij} = r^{-1} \d B^i{}_j\), where the prefactor of \(r^{-1}\) is chosen so that \(\cB_{ij}\) goes to a constant at the boundary.

\subsubsection{\(\h_{xy,xy}\)}

We begin by computing \(\h_{xy,xy}\), dual to the metric fluctuation \(\cG_{xy}\). Note that since the only non-zero component of \(b_{\m\n}\) in our system is \(b_{xy}=-b_{yx}\), from equation~\eqref{eq:conserved_stress_tensor} we have \(T^{xy} = T^{yx}\), and therefore \(\h_{xy,xy} = \h_{yx,yx} = \h_{xy,yx}\). We will show that the shear viscosity in this channel takes the universal value for isotropic, holographic systems, i.e. \(\h_{xy,xy}=s/4\pi\).

The metric fluctuation \(\cG_{xy}\) satisfies the equation of motion
\begin{equation} \label{eq:Gxy_eom}
    \frac{r}{L\sqrt{f} g h} \p_r \le[\frac{r}{L} \sqrt{f} g h \, \p_r \cG_{xy}(\w;r) \ri] + \frac{\w^2}{f g} \cG_{xy}(\w;r) = 0,
\end{equation}
which is just the equation of motion for a massless scalar field in the black-brane background~\eqref{eq:background_ansatz}. Near the boundary, solutions to this equation of motion take the form
\begin{multline}
    \cG_{xy}(\w,r) = \cG_{xy}^{(0)}(\w) + \frac{\w^2 r^2}{4} \cG_{xy}^{(0)}(\w)
    \\ + \frac{\w^2 r^4}{48} \log \le(\frac{r}{L}\ri)\le(2 M^2 - 6 b^2 - 3 \w^2 \ri) \cG_{xy}^{(0)}(\w) + r^4 \cG_{xy}^{(4)}(\w) + \cO\le(r^6 \log r\ri) \; ,
\end{multline}
with coefficients \(\cG_{xy}^{(0)}(\w)\) and \(\cG_{xy}^{(4)}(\w)\) fixed by the boundary conditions.

Expanding the action~\eqref{eq:holographic_bulk_action} to quadratic order in \(\cG_{xy}\) and using the equation of motion~\eqref{eq:Gxy_eom}, we find that the on-shell action for fluctuations in this channel is
\begin{equation} \label{eq:Gxy_action}
    S^\star = \frac{L^3}{8 \pi \gn} \int \diff^3 x  \int \frac{\diff \w}{2\pi} \cG_{xy}^{(0)} (-\w) \cG_{xy}^{(4)}(\w) + \dots \; ,
\end{equation}
where the dots denote real contact terms. Applying the Lorentzian correlator prescription of refs.~\cite{Son:2002sd,Herzog:2002pc}, the retarded two-point function of \(T^{xy}\) at zero momentum is then
\begin{equation}
    \vev{T^{xy} T^{xy}}_\mathrm{R}(\w,0) = - \frac{L^3}{4 \pi \gn} \frac{\cG_{xy}^{(4)}(\w)}{\cG_{xy}^{(0)}(\w)} + \dots \;,
\end{equation}
where the dots denote real contact terms descending from the dots in equation~\eqref{eq:Gxy_action}. Since the contact terms are real, they make no contribution to the Kubo formula~\eqref{eq:shear_viscosity_kubo} and can be neglected.

A useful way to rewrite the two-point function expression is
\begin{equation}
    \vev{T^{xy} T^{xy}}_\mathrm{R}(\w,0) = - \frac{1}{16 \pi \gn} \lim_{r \to 0} \frac{r}{L} \sqrt{f} g h\frac{\cG_{xy}(-\w;r) \p_r \cG_{xy}(\w;r)}{|\cG_{xy}^{(0)}(\w)|^2} + \dots \; ,
\end{equation}
where we have made use of reality of \(\cG_{xy}(t,r)\), which implies that \(\cG_{xy}(-\w;r) = \cG_{xy}(\w;r)^*\).
The imaginary part of the Green's function may then be written as
\begin{equation}
    \Im \, \vev{T^{xy} T^{xy}}_\mathrm{R} (\w,0) = \frac{1}{32 \pi \gn |\cG_{xy}^{(0)}(\w)|^2} \cF(\w),
\end{equation}
where
\begin{equation} \label{eq:Gxy_conserved_flux}
    \cF(\w) = i \frac{r}{L} \sqrt{f} g h \le[ \cG_{xy}(-\w;r) \p_r \cG_{xy}(\w;r) - \cG_{xy}(\w;r) \p_r \cG_{xy}(-\w;r)\ri].
\end{equation}
The equation of motion~\eqref{eq:Gxy_action} implies that the right-hand side of this expression is independent of \(r\), so we may evaluate \(\cF(\w)\) at any convenient value of \(r\). We will evaluate it at the horizon at \(r = r_0\).

Near the horizon, the two independent solutions to the equation of motion take the form \(\cG_{xy}(\w;r) = c(\w) (r_0 - r)^{\pm i \w /2\pi T}\). To obtain the retarded two-point function we should choose ingoing boundary conditions, corresponding to the minus sign in the exponent. Then, substituting this solution into equation~\eqref{eq:Gxy_conserved_flux} and using equation~\eqref{eq:hawking_temperature} we find \(\cF(\w) = - 8 \gn s \w |c(\w)|^2 \), and therefore
\begin{equation}
    \Im \, \vev{T^{xy} T^{xy}}_\mathrm{R} (\w,0) = - \frac{\w s}{4 \pi} \frac{|c(\w)|^2}{|\cG_{xy}^{(0)}(\w)|^2} .
\end{equation}
Substituting this into the Kubo formula we find \(\h_{xy,xy} = (s/4\pi) |c(0)|^2/|\cG_{xy}^{(0)}(0)|^2.\)
Finally, we observe that at \(\w=0\) equation~\eqref{eq:Gxy_eom} is solved by constant \(\cG_{xy}\), implying that  \(c(0) = \cG_{xy}^{(0)}(0)\), and therefore
\begin{equation}
    \h_{xy,xy} = \frac{s}{4\pi},
\end{equation}
as advertised.

\subsubsection{\(\h_{xz,xz}\), \(\h_{zx,zx}\), and \(\h_{xz,zx}\)}

We now compute the shear viscosities \(\h_{xz,xz}\), \(\h_{zx,zx}\), and \(\h_{xz,zx}\).\footnote{By rotational symmetry in the \((x,y)\) plane, these will be equal to \(\h_{yz,yz}\), \(\h_{zy,zy}\), and \(\h_{yz,zy}\), respectively.}
These components of the shear viscosity may be computed from the linearised fluctuations \(\cG_{xz}\) and \(\cB_{yz}\), which satisfy the coupled equations of motion
\begin{subequations} \label{eq:Gxz_Byz_eom}
\begin{align}
    \frac{r}{L\sqrt{f}h^2} \p_r \le[\frac{r}{L} \sqrt{f} h^2 \, \p_r \cG_{xz} \ri]
    + \frac{\w^2}{f g} \cG_{xz}
    - \frac{2 r^2 B'}{L^2 h^2} \, \p_r \le( r h \cB_{yz}\ri)
    - \frac{2 r^2 B'^2}{L^2 h^2}\cG_{xz} 
    &
    \nonumber \\   
    - \frac{2 B}{L^2 h^4} \le[h^2 \le(1 + 2 \s \f^2 \ri) + 2 \l  B^2 \ri] \le( B \cG_{xz} + h \cB_{yz}\ri) &= 0,
    \\[1em]
    \frac{1}{ L\sqrt{f}h^2} \p_r \le[\frac{r}{L} \sqrt{f} h^2 \, \p_r \le(r \cB_{yz} \ri) \ri]
    + \frac{\w^2}{fg} \cB_{yz}
    + \frac{r B'}{L^2 h^2} \p_r  \le(h \cG_{xz}\ri)
    - \frac{r B' g'}{L g h} \cG_{xz}
    + C \cB_{yz}
    &= 0,
\end{align}
\end{subequations}
where the coefficient \(C\) is given by
\begin{align}
    C = \frac{r^2}{L^2 h^2} \Biggl[ \frac{\le(f g h^2\ri)' \le(g h^2\ri)'}{3 f g^2 h^2} -& \frac{h' \le(g h\ri)'}{g}
    - \frac{2\le( 2 B'^2 + h^2 \f'^2 \ri)}{3}
    \nonumber \\ &
    - \frac{ \le(3 h^2 + 2 B^2 \ri)}{3 r^2h^2} \le(h^2 (1 + 2 \s \f^2) + 2 \l B^2 \ri)
    \Biggr].
\end{align}
Near the boundary, solutions to the equations of motion take the form
\begin{align}
    \cG_{xz}(\w,r) &= \cG_{xz}^{(0)}(\w) + r^2 \le[ \le(\frac{\w^2}{4} - b^2 \ri) \cG_{xz}^{(0)}(\w) - b \cB_{yz}^{(0)}(\w)\ri]
    \nonumber \\ &\phantom{=}
     + \frac{r^4}{ 48} \log\le(\frac{r}{L}\ri) \le[ \w^2 (2M^2 - 3 \w^2) \cG_{xz}^{(0)} + 24 b \cB_{yz}^{(0)}\ri] + r^4 \cG_{xz}^{(4)} + \cO\le(r^6 \log r\ri)   ,
     \label{eq:Gxz_near_boundary}
      \\
    \cB_{yz}(\w,r) &=  \cB_{yz}^{(0)}(\w) + \frac{r^2}{4} \log\le( \frac{r}{L}\ri) \le[\le(4 \s M^2 - 4 b^2(1-\l) - 2 \w^2 \ri)\cB_{yz}^{(0)}(\w) + \w^2 b \cG_{xz}^{(0)}(\w)\ri] 
    \nonumber \\ &\phantom{=}
    + r^2 \cB_{yz}^{(2)}(\w) + \cO\le(r^6 \log r\ri) ,
    \nonumber
\end{align}
with four coefficients \(\cG_{xz}^{(0)}(\w)\), \( \cG_{xz}^{(4)}\), \(\cB_{yz}^{(0)}(\w)\), and \(\cB_{yz}^{(2)}(\w)\) determined by the boundary conditions. Expanding the action~\eqref{eq:holographic_bulk_action} to quadratic order in \(\cG_{xz}\) and \(\cB_{yz}\), and using the equations of motion~\eqref{eq:Gxz_Byz_eom}, we find that the on-shell action for fluctuations is
\begin{equation} \label{eq:Gxz_Byz_action}
    S^\star = \frac{L^3}{16 \pi \gn} \int \diff^3 x \int \frac{\diff \w}{2\pi} \le[2 \cG_{xz}^{(0)}(-\w) \cG_{xz}^{(4)}(\w) + \le(2 \cB_{yz}^{(0)}(-\w) + b \cG_{xz}^{(0)} (-\w)\ri) \cB_{yz}^{(2)}(\w)\ri] + \dots \; ,
\end{equation}
where the dots once again denote real contact terms that do not contribute to the shear viscosity. From this expression we read off expressions for the retarded two-point functions of \(\tilde{T}^{xz}\) and \(\cO^{yz}\) using the Lorentzian correlator prescription of refs.~\cite{Son:2002sd,Herzog:2002pc}. Using equation~\eqref{eq:TT_two_point_function} we then arrive at expressions for the two-point functions of \(T^{xz}\) and \(T^{zx}\),
\begin{align}
    \vev{T^{xz} T^{xz}}_\mathrm{R}(\w,0) 
    &= \frac{L^3}{4 \pi \gn} \le( \frac{\p \cG^{(4)}_{xz}{(\w)}}{\p \cG^{(0)}_{xz}{(\w)}} + \frac{b}{2} \frac{\p\cB^{(2)}_{yz}(\w)}{\p\cG^{(0)}_{xz}(\w)}\ri) + \dots \;,
    \nonumber \\
    \vev{T^{zx} T^{zx}}_\mathrm{R}(\w,0) 
    &= \frac{L^3}{4 \pi \gn} \le( \frac{\p\cG^{(4)}_{xz}{(\w)}}{\p\cG^{(0)}_{xz}{(\w)}} + \frac{b}{2} \frac{\p\cB^{(2)}_{yz}(\w)}{\p\cG^{(0)}_{xz}(\w)} - 2 b \frac{\p \cG^{(4)}_{xz}(\w)}{ \p \cB^{(0)}_{yz}(\w)}\ri) + \dots \; ,
    \label{eq:Txz_correlators}
    \\
    \vev{T^{xz} T^{zx}}_\mathrm{R}(\w,0)
    &=\frac{L^3}{4 \pi \gn} \le( \frac{\p\cG^{(4)}_{xz}{(\w)}}{\p\cG^{(0)}_{xz}{(\w)}} - \frac{b}{2} \frac{\p\cB^{(2)}_{yz}(\w)}{\p\cG^{(0)}_{xz}(\w)}\ri) + \dots \;,
    \nonumber 
\end{align}
where the dots denote terms descending from the dots in equation~\eqref{eq:Gxz_Byz_action}.

To obtain the Green's functions and shear viscosities numerically we use the method of ref.~\cite{Kaminski:2009dh}. We begin by constructing two linearly independent sets of solutions to the equations of motion~\eqref{eq:Gxz_Byz_eom} that satisfy the ingoing boundary conditions
\begin{equation}
    \cG_{xz} = (r_0 - r)^{-i \w/2\pi T}
    \quad \text{and} \quad
    \cB_{yz} = \pm (r_0 - r)^{-i \w/2\pi T}
\end{equation}
at the horizon. By taking appropriate linear combinations of these solutions we can construct two new solutions with either \(\cG_{xz}^{(0)} = 0\) or \(\cB_{yz}^{(0)} = 0\). Performing a fit to the near-boundary behaviour~\eqref{eq:Gxz_near_boundary}, we can then determine \(\cG_{xz}^{(4)}\) and \(\cB_{yz}^{(2)}\) for these solutions, allowing us to evaluate the derivatives with respect to the sources appearing in equation~\eqref{eq:Txz_correlators}.

\begin{figure}
    \includegraphics[width=\textwidth]{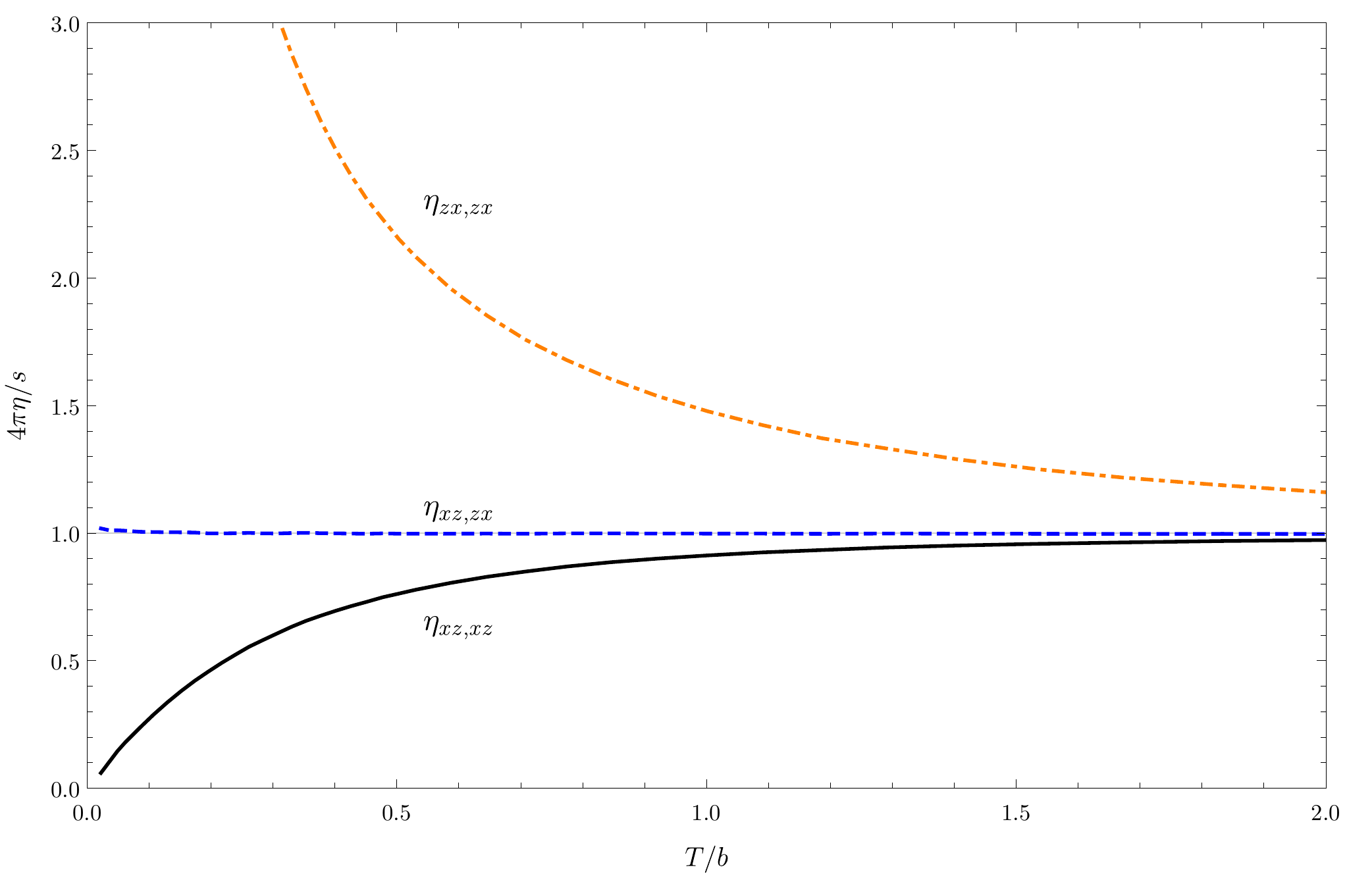}
    \caption{Numerical results for the shear viscosity in units of \(s/4\pi\) at \(M=0\) as a function of \(T/b\).
    The shear viscosity computed from the two-point function of \(T^{xz}\) takes the isotropic value \(\h_{xz,xz}=s/4\pi\) at large \(T/b\). If \(T/b\) is decreased, \(\h_{xz,xz}/s\) also decreases, becoming very small at small \(T/b\).
    The shear viscosity  \(\h_{zx,zx}\), computed from the two-point function of \(T^{zx}\), displays the opposite behaviour, becoming very large at small \(T/b\).
    The shear viscosity \(\h_{xz,zx}\), computed from the mixed two-point function of \(T^{xz}\) with \(T^{zx}\), appears numerically very close to \(\h_{xz,zx}=s/4\pi\) for all \(T/b\).}
    \label{fig:shear_viscosity}
\end{figure}

For simplicity we will restrict to the case \(M=0\), deep in the nodal line phase. Figure~\ref{fig:shear_viscosity} shows our numerical results for the three shear viscosities \(\h_{xz,xz}\), \(\h_{zx,zx}\), and \(\h_{xz,zx}\), each multiplied by \(4\pi/s\), at \(M=0\) and as a function of \(T/b\). The solid black curve shows \(\h_{xz,xz}\), which at large \(T/b\) is given approximately by the universal result for isotropic holographic systems, \(\h_{xz,xz} \simeq s/4\pi\). Moving to small \(T/b\) we find that \(\h_{xz,xz}/s\) decreases monotonically with decreasing \(T/b\), becoming very small as \(T/b \to 0\). Unfortunately our numerics for \(\h/s\) become unstable at small \(T/b\), so it is not possible to determine whether \(\h_{xz,xz}/s \to 0\) exactly as \(T/b \to 0\).

The dot-dashed orange curve shows \(\h_{zx,zx}\), which displays the opposite behaviour, \textit{increasing} with decreasing \(T/b\), becoming very large (and possibly diverging) as we send \(T/b \to 0\). Finally, the dashed blue curve shows \(\h_{xz,zx}\). We find that this component of the shear viscosity is numerically very close to \(s/4\pi\) for all \(T/b\). The figure shows \(\h_{xz,zx}\) deviating slightly from this result at very small \(T/b\), but we cannot rule out the possibility that this deviation is caused by the instability of our numerics at low temperatures.

If \(\h_{xz,zx}\) really does equal \(s/4\pi\), than one would expect to be able to prove this somehow. To do so, one would presumably need to find a formula for \(\h_{xz,zx}\) in terms of quantities evaluated at the horizon at \(r=r_0\). Following ref.~\cite{Blake:2013bqa}, one might try to do so by writing the equations of motion~\eqref{eq:Gxz_Byz_eom} in the form of a matrix equation for the vector of fluctuations \((\cG_{xz}, \cB_{yz})\). If one of the eigenvalues of the ``mass matrix'' \(\mathcal{M}\) appearing in this equation vanishes, then there is a linear combination of the fluctuations that is independent of \(r\) in the limit \(\w\to0\), and we would expect it to be this combination that determined \(\h_{xz,zx}\). However, we find \(\det \mathcal{M} \neq 0\), so no such combination exists.

%% file: discussion.tex
\section{Discussion}

We have studied a variety of properties of the holographic NLSM model~\cite{Liu:2018bye,Liu:2018djq,Liu:2020ymx}, modified by an additional coupling that gives us greater control over the IR physics. We have paid particular attention to transport phenomena. The two-form coupling \(b_{xy}\), responsible for the presence of nodal lines in the fermion spectral function, breaks rotational symmetry, with important consequences for transport. One obvious consequence of the broken rotational invariance is that the coefficients describing transport in different directions are typically different. There is also the subtler effect that broken rotational invariance implies that the stress tensor is not symmetric in its indices, leading to a larger number of transport coefficients than in a rotationally invariant system. There are many possible directions for future work on holographic NLSMs. We discuss a few examples below.

Throughout this paper we have worked at zero chemical potential, i.e. with a Fermi energy equal to the energy of the nodal line, and considering also non-zero chemical potentials would clearly be of interest. Related to this is the implementation of an energy tilt of the nodal line. Moreover, experiments with nodal line semimetals typically consider quantum oscillations. To discuss these in a holographic setting an external magnetic field should be added to the problem as well. Of course we should realize that the nodal lines in the present model take the form of circles in momentum space, due to the unbroken \(\mathrm{SO}(2)\) rotational symmetry in the \((x,y)\) plane. Nodal lines in real materials are typically less symmetric, for instance, the nodal line in Ca\(_3\)P\(_2\) has a six-fold discrete rotational symmetry~\cite{doi:10.1063/1.4926545}. A more realistic holographic NLSM model should therefore include terms that break the \(\mathrm{SO}(2)\) to some discrete subgroup, while preserving the discrete symmetries that protect the nodal line.

As discussed in the introduction, in order to obtain more realistic fermion physics, it may be better to work in the framework of semiholography~\cite{Contino:2004vy,Hartnoll:2009ns,Faulkner:2010tq,Gursoy:2011gz}, supplementing the fermion action in equation~\eqref{eq:fermion_action} with additional boundary terms such that the boundary fermion is elementary, rather than composite. In particular, this means that the fermion spectral function \(\r\) would then satisfy the ARPES sum rule \(
    \int_{-\infty}^\infty \diff \w \, \r = 1
\), satisfied by electrons in real materials. A calculation of the fermion contribution to the electrical conductivity of a semi-holographic Weyl semimetal was performed in ref.~\cite{Jacobs:2015fiv}, and one could attempt to generalise this approach to NLSMs.

By introducing the additional coupling \(\l\) into the holographic model~\eqref{eq:holographic_bulk_action}, we were able to obtain a finite, non-zero DC conductivity \(\s_{zz}^\mathrm{DC}\) in the direction orthogonal to the plane of the nodal line at zero temperature, in agreement with the expected physics of NLSMs. However, we always have \(\s_{xx}^\mathrm{DC} = \s_{yy}^\mathrm{DC} = 0\), whereas these conductivities are also expected to be non-zero in a real NLSM. It would be interesting to explore ways to obtain finite, non-zero values for all three DC conductivities. Simple dimensional analysis shows that in a theory that has both hyperscaling and an anisotropic scale invariance, such that time scales as \(t \to \l t\) while the spatial directions scale as \(x^i \to \l^{d_i} x^i\), the longitudinal conductivity in the \(x^i\) direction scales as\footnote{See ref.~\cite{Hartnoll:2009ns} for a clear discussion of the case when \(d_i\) is the same for all \(i\). The extension to different values of \(d_i\) is straightforward.}
\begin{equation} \label{eq:conductivity_scaling}
    \s_{ii} \to \l^{\Delta_i} \s_{ii}, \qquad \mathrm{where} \qquad  \Delta_i = d_i - \sum_{j\neq i} d_j.
\end{equation}
Since the frequency scales inversely to time, \(\w \to \l^{-1} \w\), this dimensional analysis implies that the conductivity depends on frequency as \(\s_{ii} \propto \w^{-\Delta_i}\) at small \(\w\). For our system, the IR fixed point of the topological phase has just such a scale invariance, with \(d_x = d_y = 1/2\) while \(d_z =  1\), so this dimensional analysis reproduces the low-frequency results \(\s_{xx}, \s_{yy} \propto \w\) and \(\s_{zz} \propto \w^0\) that we found holographically.
To obtain finite, non-zero values of \(\s_{ii}^\mathrm{DC} = \le.\s_{ii}\ri|_{\w=0}\) for all three directions, we would require \(\Delta_i = 0\) for all \(i\), which from equation~\eqref{eq:conductivity_scaling} only occurs if \(d_i = 0\) for all \(i\). This situation would correspond to fermion self-energies depending only on frequency, which is not so uncommon in the literature of strongly correlated electrons studied by dynamical mean-field theory approximation \cite{Georges:1996zz}. In the gravity dual this would arise from an $AdS_2$ near horizon region that occurs in extremal black-holes, see e.g. ref.~\cite{Faulkner:2009wj} and would correspond to a situation where the IR of the field theory is governed by a one-dimensional CFT. An alternative resolution involves hyperscaling violation. Hyperscaling violation modifies the frequency dependence of the conductivity~\cite{Gouteraux:2013oca,Gouteraux:2014hca,Karch:2014mba}, so may provide a way to avoid having to set \(d_i =0 \) for all \(i\). 

The bulk fermions discussed in section~\ref{sec:fermions} were treated purely classically. However, important physics arises from quantum effects involving the fermions, see ref.~\cite{Hartnoll:2016apf} for a review. For NLSMs, possibly the most relevant quantum effect to study would be the fermion contribution to the electrical conductivity, which arises from a loop correction to the propagator of the bulk gauge field \(A_m\)~\cite{Faulkner:2010zz,Faulkner:2011tm,Faulkner:2013bna}. 

NLSMs with boundaries exhibit surface states~\cite{Heikkila:2010yk}, as a consequence of the non-trivial topology of the material's band structure, and one could look for similar surface states in the holographic model. A natural approach would be that of ref.~\cite{Ammon:2016mwa}, which found evidence for surface states in a holographic model of a Weyl semimetal with a boundary in the form of an edge current. Alternatively, in a semiholographic approach the surface states could presumably be directly obtained by considering a half-infinite system and solving the appropriate one-dimensional Schr\"odinger problem, for instance, with hard-wall boundary conditions.

In section~\ref{sec:transport} we computed various transport coefficients of the holographic NLSM, using their Kubo formulas. Due to the asymmetry of the stress tensor, arising from the explicitly broken Lorentz invariance, there are a larger number of coefficients governing transport of energy and momentum compared to in a Lorentz invariant system. To better understand the physics of these coefficients, one should work out the full first-order hydrodynamic expansion of the stress tensor, including antisymmetric terms allowed due to non-zero \(b_{\m\n}\). We leave this for future work.\footnote{See refs.~\cite{Hoyos:2013eza,Hoyos:2013qna} for related work on hydrodynamics when boosts, but not rotations, are broken.}

It would also be very interesting to compute entanglement entropy in the holographic NLSM model. For Lorentz invariant renormalisation group (RG) flows, entanglement entropy may be used to define a \(c\)-function, a quantity that decreases along the RG flow from the UV to the IR, and therefore provides a measure of the number of degrees of freedom at a given energy scale~\cite{Casini:2006es,Myers:2010tj,Myers:2010xs,Casini:2012ei,Myers:2012ed}. For the NLSM, Lorentz invariance is explicitly broken by non-zero \(b\). However, there have been proposals, based on evidence from holography, that an entropic \(c\)-function also exists in anisotropic systems~\cite{Chu:2019uoh}, see also refs.~\cite{Ghasemi:2019xrl,Arefeva:2020uec,Hoyos:2020zeg}. One could test this proposal for the holographic NLSM. The entropic \(c\)-function has also been proposed as a probe of topological phase transitions, based on the fact that it displays a maximum near the critical point in a holographic model of a Weyl semimetal~\cite{Baggioli:2020cld}. One could look for such a maximum in the quantum phase transition exhibited by the holographic NLSM.

We thus conclude that many interesting extensions of our present work exist, and we intend to work on some on them in the near future. In particular, we hope to be able to bring the holographic approach for strongly interacting systems closer to the exciting experiments with the recently discovered nodal line semimetals, which indeed appear to show signs of strong correlations due to the reduced screening near the nodal line. However, for a quantitative comparison between theory and experiments these are still early times and much work needs to be done. Nevertheless, we hope that with our present paper we have at least been able to set an additional step towards this ultimate goal.

%% file: holo_rg.tex
\section{Details of holographic renormalisation}
\label{app:holo_rg}

In this appendix we give some details on the derivation of the formulas for various physical quantities quoted in section~\ref{sec:thermodynamics}.

\paragraph{Free energy.} The free energy at temperature \(T\) is given by \(\cF = T I^\star\), where \(I^\star\) is the Euclidean signature on-shell gravitational action with Euclidean time periodic with period \(1/T\). Using the equations of motion~\eqref{eq:eom_second_order} and~\eqref{eq:eom_first_order}, one can show that the bulk part of the Lagrangian in equation~\eqref{eq:holographic_bulk_action} may be written as a total derivative, \(\sqrt{-G} \le(R + \frac{12}{L^2} - \cL \ri)= - \p_r \le[\frac{r}{L} \sqrt{f} h g' \ri]\), so that the bulk contribution to the on-shell action may be written as a boundary term
\begin{align}
    I^\star_\mathrm{bulk} 
    &= - \frac{1}{ 16 \pi \gn} \int_0^{1/T} \diff t_\mathrm{E} \int \diff^3 x \le[\frac{r}{L} \sqrt{f} h g' \ri]_{r=\e}
    \nonumber\\
    &= \frac{V L^3}{16 \pi \gn T} \le[
        \frac{2}{\e^4} + \frac{1}{\e^2} \le(b^2 - \frac{M^2}{3}\ri)
        - m + \frac{2 + 3 \h}{24} M^4 + \frac{1 - 3\l}{12} b^4 - \frac{5\s}{6} M^2 b^2
    \ri],
\end{align}
where in the second line we have substituted the near boundary expansions~\eqref{eq:bg_near_boundary}, keeping only terms that are non-zero in the limit \(\e \to 0\), and \(V = \int \diff^3 x\) is the volume of the system. The Gibbons-Hawking term and the counterterms are straightforward to evaluate, yielding
\begin{align}
    I_\mathrm{GH}^\star &= \frac{V L^3}{16 \pi \gn T} \le[- \frac{8}{\e^4} + \frac{4 M^2}{3 \e^2}
    - \frac{2+3\h}{6} M^4 + \frac{\l - 1}{3} b^4 - \frac{2\s}{3} M^2 b^2\ri],
    \nonumber \\
    I_\mathrm{ct}^\star &= \frac{V L^3}{16 \pi \gn T} \le[
        \frac{6}{\e^4} - \frac{1}{\e^2} (b^2 + M^2) - M \f_2 - b b_2 + \frac{2+3\h}{6}M^4 + \frac{\l+1}{2} + \s M^2 b^2
    \ri].
\end{align}
Summing these three contributions, we find the free energy \(\cF =T  (I_\mathrm{bulk}^\star + I_\mathrm{GH}^\star + I_\mathrm{ct}^\star)\) is given by
\begin{equation}
    \cF = - \frac{V L^3}{16 \pi \gn} \le(m + M \f_2 + b b_2 + \frac{2 + 3 \h}{24} M^4 + \frac{\s}{2} M^2 b^2 + \frac{\l-1}{4} b^4 \ri).
\end{equation}

\paragraph{Scalar one-point function.} The one-point function of the operator sourced by \(M\) is
\begin{equation}
    \vev{\cO} = \frac{\d S^\star}{\d M},
\end{equation}
where \(S^\star\) is the on-shell gravitational action in Lorentzian signature. The change in bulk action under a small change in \(\f\) is
\begin{align}
    \d S_\mathrm{bulk} &= - \frac{1}{8 \pi \gn} \int \diff^5 x \, \sqrt{-G} \, \nabla^m \le( \d \f \, \nabla_m \f\ri) + \mathrm{EOM}
    \nonumber \\
    &= - \frac{1}{8 \pi \gn} \int_{\p \cM} \diff^4 x \, \sqrt{-\g} \, \d \f \, n^m \p_m \f + \mathrm{EOM},
\end{align}
where \(\mathrm{EOM}\) denotes an integral over a term proportional to the equations of motion, which vanishes on shell by definition. The variation in the counterterm action is
\begin{equation}
    \d S_\mathrm{ct} = - \frac{1}{8 \pi \gn L} \int \diff^4 x \, \sqrt{-\g} \, \d \f \le[
        \f + \le( \frac{4 + 6 \h}{3} \f^3 + 2 \s \f B_{\m\n} B^{\m\n} \ri) \log(\e/L)
    \ri].
\end{equation}
Summming the bulk and counterterm contributions to the variation of the action and inserting the near-boundary expansion~\eqref{eq:bg_near_boundary}, we find
\begin{equation}
    \d S^\star = \frac{L^3}{16 \pi \gn} \int \diff^4 x \, \d M \le(4 \f_2 + \frac{2 + 3 \h}{3} M^3 + 2 \s M b^2 \ri),
\end{equation}
so that the scalar one-point function is
\begin{equation}
    \vev{\cO} = \frac{L^3}{16 \pi \gn} \le(4 \f_2 + \frac{2 + 3 \h}{3} M^3 + 2 \s M b^2 \ri).
\end{equation}

\paragraph{Antisymmetric tensor one-point function.} The calculation of the one-point function if the antisymmetric tensor operator proceeds similarly. The one-point function is given by
\begin{equation}
    \vev{\cO^{\m\n}} = \frac{\d S^\star}{\d b_{\m\n}}.
\end{equation}
The variation of the bulk part of the action under in change of \(B\) is
\begin{equation}
    \d S_\mathrm{bulk} = - \frac{1}{16 \pi \gn} \int \diff^4 x \, \sqrt{-\g} \, \d B_{\m\n} \, n_r H^{r\m\n} + \mathrm{EOM},
\end{equation}
while the variation of the counterterms yields
\begin{align}
    \d S_\mathrm{ct} = \frac{1}{16 \pi \gn L} \int_{\p \cM} \diff^4 x \, \sqrt{-\g} \Bigl[&
        \d B_{\m\n} \, B^{\m\n} 
    \nonumber \\ &
    - \le(2 \s |\f|^2 \d B_{\m\n} \, B^{\m\n} + (\l-1) \d B_{\m\n} \, B^{\m\n} B_{\k\t} B^{\k\t} \ri) \log(\e/L)    
    \Bigr]
\end{align}
Summming the bulk and counterterm contributions to the variation of the action and inserting the near-boundary expansion~\eqref{eq:bg_near_boundary}, we find
\begin{equation}
    \d S^\star = \frac{L^3}{8 \pi \gn} \int \diff^4 x \, \d b \le[2 b_2 + (\l-1) b^3 + \s M^2 b \ri],
\end{equation}
so that the non-zero two-form one-point functions are
\begin{equation}
    \vev{\cO^{xy}} = - \vev{\cO^{yx}} = \frac{L^3}{16 \pi \gn} \le[2 b_2 + (\l-1) b^3 + \s M^2 b \ri].
\end{equation}

\paragraph{Stress tensor.} The naive holographic stress tensor, obtained by differentiation of the on-shell action with respect to the boundary metric, is~\cite{deHaro:2000vlm}
\begin{equation}
    \vev{\tilde{T}_{\m\n}} \equiv - \lim_{\e \to 0} \frac{L^2}{\e^2} \frac{2}{\sqrt{-\g}} \frac{\d S}{\d \g^{\m\n}}
    = \lim_{\e \to 0} \frac{L^2}{\e^2}\le[ - \frac{1}{8\pi\gn} (K_{\m\n} - K \g_{\m\n}) - \frac{2}{\sqrt{-\g}} \frac{\d S_\mathrm{ct}}{\d \g^{\m\n}} \ri].
\end{equation}
As discussed in section~\ref{sec:stress_tensor_comments}, this tensor is not conserved in the presence of a non-zero source \(b_{\m\n}\). The conserved stress tensor is instead \(T_{\m\n} \equiv \tilde{T}_{\m\n} + 2 \cO_{\m\r} b_\n{}^\r\). Inserting the near-boundary expansions, we find \(\vev{T_{\m\n}} = \mathrm{diag}(\ve,p,p,p)\), where the energy density and pressure are given by
\begin{align}
    \ve &= \frac{L^3}{16 \pi \gn} \le(3m - M \f_2 - b b_2 - \frac{2 + 3 \h}{24} M^4 - \frac{\s}{2} M^2 b^2  - \frac{\l-1}{4} b^4\ri),
    \nonumber
    \\
    p &=  \frac{L^3}{16 \pi \gn} \le(m + M \f_2 + b b_2 + \frac{2 + 3 \h}{24} M^4 + \frac{\s}{2} M^2 b^2  + \frac{\l-1}{4} b^4 \ri),
\end{align}
respectively.

\section{Holographic derivation of the Ward identity for translations}
\label{app:holo_ward_identity}

In this appendix we demonstrate that our holographic stress tensor satisfies the expected Ward identity for translations~\eqref{eq:translation_ward_identity}, via an explicit computation of the stress tensor for the case when the antisymmetric tensor operator has a position-dependent source. For simplicity we will neglect the bulk scalar field, since its contribution to the Ward identity is well known~\cite{deHaro:2000vlm}, and the expressions in this section will be rather lengthy even in its absence. For the latter reason we will also neglect the self-coupling of the two-form field, i.e. we set \(\l=0\), since this coupling cannot affect the Ward identity.

Throughout this appendix we will work in units in which the \ads\ radius is \(L=1\). We write the metric of asymptotically locally \ads[5] as
\begin{equation} \label{eq:4_plus_1_metric}
    \diff s^2 = \diff r'^2 + \g_{\m\n} \diff x^\m \diff x^\n,
\end{equation}
where \(r'\) is the radial coordinate, with the boundary at \(r' \to \infty\), and \(x^\m\) are the field theory directions. The radial coordinate is related to the one used in the main text by \(r' =- \log r\). The matrix \(\g_{\m\n}\) is the induced metric on constant-\(r'\) slices. For the remainder of this appendix we will drop the prime on \(r'\) for notational simplicity.

In the metric decomposition~\eqref{eq:4_plus_1_metric}, the bulk Einstein equations may be written as (see for example equation (53) of ref.~\cite{Papadimitriou:2004ap})
\begin{align} \label{eq:einstein_gauss_codazzi}
    K^2 - K_{\m\n} K^{\m\n} &=  12 + R[\g] +2 \Q_{rr}
    \nonumber \\
    \nabla^\m K_{\m\n} - \nabla_\n K &=  \Q_{r\n},
    \\
    \g_{\m\s} \p_r (\g^{\r\s} K_{\s\n}) + K K_{\m\n} &= 4 \g_{\m\n} + R_{\m\n} [\g] - \le( \Q_{\m\n} - \frac{1}{3} \Q^m{}_m \g_{\m\n} \ri),
    \nonumber
\end{align}
where \(K_{\m\n} = \frac{1}{2} \p_r \g_{\m\n}\) is the extrinsic curvature of the constant-\(r\) slices, \(K = \g^{\m\n} K_{\m\n}\) is the mean curvature, \(R_{\m\n}[\g]\) is the Ricci tensor computed with \(\g_{\m\n}\), \(\nabla_\m\) is the covariant derivative with respect to \(\g_{\m\n}\), \(\Q_{mn}\) defined in equation~\eqref{eq:bulk_stress_tensor} is proportional to the bulk stress tensor, and Greek indices have been raised with \(\g^{\m\n} \equiv (\g^{-1})_{\m\n}\)..

Defining \(A_\m = B_{\m r}\), not to be confused with the bulk \(U(1)\) gauge field used in the main text, and \(F_{\m\n} = \p_\m A_\n - \p_\n A_\m\), for \(m_B^2 = 1\) Einstein's equations~\eqref{eq:einstein_gauss_codazzi} read
\begin{align}
        \le(\g^{\m\n} \p_r \g_{\m\n} \ri)^2
        - \g^{\m\r} \g^{\n\s} \p_r \g_{\m\n} \p_r \g_{\r\s}
        &=48 + 4 R[\g] +2 \g^{\m\r} \g^{\n\s} \le( \p_r B_{\m\n} + F_{\m\n} \ri) \le( \p_r B_{\r\s} + F_{\r\s} \ri)
    \nonumber \\ &\phantom{=}
        - \frac{3}{2} \g^{\m\r} \g^{\n\s} \g^{\kappa\t} \p_{[\kappa} B_{\m\n]} \p_{[\t} B_{\r\s]}
        - 2 B^{\m\n} B_{\m\n} +  A^\m A_\m,
    \nonumber \\[1em]
        \nabla^\m \p_r \g_{\m\n} - \g^{\r\s} \nabla_\nu \p_r \g_{\r\s} 
        &= \g^{\m\r} \g^{\l\s} \p_{[\n} B_{\m\l]} \le( \p_r B_{\r\s} + F_{\r\s} \ri)
        + 2 A^\m B_{\m\n},
        \label{eq:4_plus_1_einstein}
    \\[1em]
    2\g_{\m\r} \p_r \le(\g^{\r\s} \p_r \g_{\s\n} \ri) +  \g^{\r\s} \p_r \g_{\r\s} \p_r \g_{\m\n}
    &= 16 \g_{\m\n} +4 R_{\m\n}[\g]
      - 4 g^{\r\s} \le( \p_r B_{\m\r} + F_{\m\r} \ri) \le( \p_r B_{\n\s} + F_{\n\s} \ri)
    \nonumber \\ &\phantom{=}
        - 2 \g^{\r\kappa} \g^{\s\t} \p_{[\m} B_{\r\s]} 
        \p_{[\n} B_{\kappa \t]}
        - 4 B_{\m\r} B_{\n}{}^\r
        -4 A_\m A_\n
    \nonumber \\ &\phantom{=}
    + 4 \g_{\m\n} \biggl[
        \frac{1}{3} \g^{\l\r} \g^{\h\s} \le(\p_r B_{\l\h} + F_{\l\h} \ri) \le( \p_r B_{\r\s} + F_{\r\s} \ri)
    \nonumber \\ &\phantom{=+  4 \g_{\m\n} \biggl[}
        + \frac{1}{9} \p^{[\t} B^{\r\s]} \p_{[\t} B_{\r\s]}
        + \frac{1}{6} B^{\r\s} B_{\r\s}
        + \frac{1}{3} A^\r A_\r
    \biggr],
    \nonumber
\end{align}
where \(\p_{[\m} B_{\n\r]} = \p_\m B_{\n\r} + \p_\n B_{\r\m} + \p_{\r} B_{\m\n}\). In this decomposition, the equation of motion for the two-form field becomes
\begin{align}
    \frac{1}{\sqrt{-\g}} \g_{\m\t} \p_\n \le[ \sqrt{-g} \, \g^{\n\r} \g^{\s\t} \le( \p_r B_{\r\s} + F_{\r\s} \ri) \ri] - A_\m &= 0,
    \nonumber
    \\[1em]
    \frac{1}{\sqrt{-\g}} \g_{\m\r} \g_{\n\s} \p_r \le[\sqrt{-\g} \g^{\r\k} \g^{\s\t} \le( \p_r B_{\k\t} + F_{\k\t} \ri) \ri]  \hspace{2cm}&
    \label{eq:4_plus_1_two_form} \\
    + \frac{1}{\sqrt{-\g}} \, \g_{\m\r} \g_{\n\s} \p_\l \le[
        \sqrt{-\g} \, \g^{\l\g} \g^{\r\k} \g^{\s\t} \p_{[\g} B_{\k\t]}
    \ri] - B_{\m\n} &= 0,
    \nonumber
\end{align}
where \(\g = \det \g_{\m\n}\).

Near the asymptotic \ads\ boundary, the fields have the large-\(r\) expansions
\begin{align}
    \g_{\m\n} &= e^{2r} \g^{(0)}_{\m\n} + \g^{(2)}_{\m\n} + r e^{-2r} \tilde{\g}^{(4)}_{\m\n} + e^{-2r} \g^{(4)}_{\m\n} + \dots \; ,
    \nonumber \\
    B_{\m\n} &= e^{r} b_{\m\n} + r e^{-r} \tilde{B}^{(2)}_{\m\n} + e^{-r} B^{(2)}_{\m\n} + \dots \;,
    \label{eq:holo_rg_large_r_expansions}
    \\
    A_\m &= e^{-r} A^{(0)}_\m + r e^{-3r} \tilde{A}^{(2)}_\m + e^{-3r} A^{(2)}_\m \; ,
    \nonumber
\end{align}
where \(\g^{(0)}_{\m\n}\) is the metric in the dual field theory, \(b_{\m\n}\) is the source for the antisymmetric tensor operator, and the dots indicate terms of higher order in a large-\(r\) expansion. If we impose the boundary condition \(\g^{(0)}_{\m\n} = \h_{\m\n}\), where \(\h_{\m\n}\) is the Minkowski metric, then it is a straightforward but tedious exercise to solve the equations of motion~\eqref{eq:4_plus_1_einstein} and~\eqref{eq:4_plus_1_two_form} order-by-order at large \(r\) to find
\begin{subequations} \label{eq:boundary_coefficients_holo_rg}
\begin{align}
    \g^{(2)}_{\m\n} &= b_{\m}{}^\r b_{\n\r} - \frac{1}{4} \h_{\m\n} b_{\r\s} b^{\r\s},
    \nonumber \\
    \tilde{B}^{(2)}_{\m\n} &= 2 \le( b_\m{}^\r b_\n{}^\s  - \frac{1}{4} b_{\m\n} b^{\r\s} \ri) b_{\r\s}
    - \p_\m \p^\r b_{\r\n} - \p_\n \p^\r b_{\m\r} + \frac{1}{2} \Box  b_{\m\n},
    \nonumber \\
    A^{(0)}_\m &= - \p^\n b_{\m\n},
    \\
    \tilde{A}^{(2)}_\m &= 2 \p^\n \le(b_{\m\r} b_{\n\s} b^{\r\s} \ri) - \frac{1}{2} \p^\n \le(b_{\m\n} b_{\r\s} b^{\r\s} \ri) - \frac{1}{2} \Box \p^\n b_{\m\n},
    \nonumber \\
    A^{(2)}_\m &=  \p^\n B^{(2)}_{\m\n} - b_{\m\r} b^{\r\s} b_{\n\s} + \frac{1}{4} b_{\r\s} b^{\r\s} \p^\n b_{\m\n}  - \frac{1}{2} \Box \p^\n b_{\m\n},
    \nonumber
\end{align}
and
\begin{align}
    \tilde{\g}^{(4)}_{\m\n} &=
    \frac{1}{2} \Box \g^{(2)}_{\m\n}
    - \frac{1}{4}  \p_\r \p_\m \g^{(2)}_{\r\n}- \frac{1}{4}  \p_\r \p_\n \g^{(2)}_{\r\m}
    + \frac{1}{12} \h_{\m\n} \p^\r \p^\s \g^{(2)}_{\r\s}
    + \frac{1}{4} \h^{\r\k} \h^{\s\t} \p_{[\m} b_{\r\s]} \p_{[\n} b_{\k\t]}
    \nonumber \\ & \phantom{=}
    + \frac{1}{2} \p^\r b_{\r\m} \p^\s b_{\s\n} 
    - \frac{1}{2} \p_\r b_{\m\s} \p^\r b_\n{}^\s
    - \frac{1}{12} \h_{\m\n} \p^\r b_{\r\l} \p_\s b^{\s\l}
    + \frac{1}{4} \h_{\m\n}\p_\s b_{\r\l} \p^\r b^{\s\l}
    \nonumber \\ &\phantom{=}
    - \frac{1}{36} \h_{\m\n} \p^{[\l} b^{\r\s]} \p_{[\l} b_{\r\s]}
    + \frac{1}{12} \h_{\m\n} b^{\r\s} b_{\s\k} b^{\k\t} b_{\t\r} - \frac{1}{48} \h_{\m\n} \le(b^{\k\t} b_{\k\t} \ri)^2,
\end{align}
\end{subequations}
where \(\Box \equiv \p^\m \p_\m\). In equation~\eqref{eq:boundary_coefficients_holo_rg} and all subsequent expressions, Greek indices are raised and lowered with the Minkowski metric.

The normalisable coefficients \(\g^{(4)}_{\m\n}\) and \(B^{(2)}_{\m\n}\) are not completely fixed by the near-boundary analysis. However, Einstein's equations do fix the trace of \(\g^{(4)}\) to be
\begin{align}
    \g^{(4)\m}{}_\m &= \frac{1}{6} b^{\m\n} B^{(2)}_{\m\n} 
    + \frac{1}{3} b^{\m\n} b_{\n\r} b^{\r\s} b_{\s\m}
     - \frac{1}{12} (b^{\m\n} b_{\m\n})^2
    - \frac{1}{8} \p^\m \p^\n \le( b_{\m\l} b_\n{}^\l \ri)
    + \frac{1}{96} \Box \le( b^{\m\n} b_{\m\n} \ri)
    \nonumber \\ &\phantom{=}
    + \frac{1}{144} \p_{[\l} b_{\m\n]} \p^{[\l} b^{\m\n]}
    - \frac{1}{24} \h^{\m\n} \p^\r b_{\r\m} \p^\s b_{\s\n}
    + \frac{1}{24} \p^\l b^{\m\n} \p_\l b_{\m\n},
\end{align}
and its divergence to  be given by the solution of
\begin{align} \label{eq:gamma4_divergence}
    4 \p^\m &\g^{(4)}_{\m\n} 
    - 4 \p_\n \g^{(4)\m}{}_\m -  \p^\m \tilde{\g}^{(4)}_{\m\n} + \p_\n \tilde{\g}^{(4)\m}{}_\m 
    \nonumber \\
    &=
    2 \p^\m \le( b_\m{}^\r B^{(2)}_{\n\r} - B^{(2)}_{\m\r} b_\n{}^\r \ri)
    - b^{\r\s} \p_\n B^{(2)}_{\r\s} + B^{(2)}_{\r\s} \p_\n b^{\r\s}
        + \p_\m \le( b_{\n\r} \Box b^{\m\r}\ri)
        - \frac{1}{2} \Box b_{\r\s} \p_\n b^{\r\s}
    \nonumber \\ &\phantom{=}
        +2 \p_\m \le(b^{\m\r} b_{\r\s} b^{\s\k} b_{\k\n} \ri)
        - \p_\m\le(b^{\m\r} b_{\n\r}  b^{\k\t} b_{\k\t}  \ri)
        +  b^{\r\s} b_{\r\s} \p_\n \le(b^{\k\t} b_{\k\t} \ri)
        -3  b^{\r\k} b_{\s\k} \p_\n \le(b_{\r\t} b^{\s\t} \ri).
\end{align}

To compute one-point functions in the dual field theory we take functional derivatives of the on-shell action with respect to the boundary values of \(B_{\m\n}\) and \(\g_{\m\n}\). When the equations of motion~\eqref{eq:eom_general} are satisfied, the variation of the action~\eqref{eq:holographic_bulk_action} with the counterterms written in equation~\eqref{eq:counterterms} under a small change in \(B_{\m\n}\) is
\begin{equation}
    \d S^\star = \frac{1}{16 \pi \gn} \int_{\p\cM} \diff^4 x \, \d b^{\m\n} \le(2 B^{(2)}_{\m\n} - \tilde{B}^{(2)}_{\m\n} - 2 \p_\m A_\n^{(0)} - 2 \Box b_{\m\n} \ri),
\end{equation}
for \(\f=\l=0\). Taking a functional derivative with respect to \(b_{\m\n}\) and inserting the solutions for the near-boundary coefficients written in equation~\eqref{eq:boundary_coefficients_holo_rg}, we find that the one-point function of the operator dual to \(B_{\m\n}\) is
\begin{equation}
    \vev{\cO_{\m\n}} = \frac{\d S^\star}{\d b^{\m\n}}
    = \frac{1}{8 \pi \gn} \le( B^{(2)}_{\m\n} +   b_{\m\r} b^{\r\s} b_{\s\n} + \frac{1}{4} b_{\m\n} b^{\r\s} b_{\r\s}  - \frac{5}{4} \Box b_{\m\n} \ri).
    \label{eq:OB_vev_general}
\end{equation}

The naive stress tensor obtained by functional differentiation of the on-shell action with respect to the boundary metric is~\cite{deHaro:2000vlm}
\begin{equation} \label{eq:stress_tensor_formula}
    \vev{ \tilde{T}_{\m\n}} \equiv - \lim_{r \to \infty} e^{2r} \frac{2}{\sqrt{-\g}} \frac{\d S^\star}{\d \g^{\m\n}} = \lim_{r \to \infty} \le(  \frac{e^{2r}}{8\pi\gn} \le(K \g_{\m\n} - K_{\m\n} \ri) -e^{2r} \frac{2}{\sqrt{-\g}} \frac{\d S^\star_\mathrm{ct}}{\d \g^{\m\n}}\ri),
\end{equation}
where \(S_\mathrm{ct}\) is the counterterm action given in equation~\eqref{eq:counterterms}. Performing the functional derivative on the right-hand side in equation~\eqref{eq:stress_tensor_formula}, inserting the near-boundary expansions~\eqref{eq:holo_rg_large_r_expansions} and simplifying the result using the solutions in equation~\eqref{eq:boundary_coefficients_holo_rg}, we find
\begin{align} \label{eq:holo_rg_stress_tensor_tilde}
    \vev{\tilde{T}_{\m\n}} 
        &= \frac{1}{16 \pi\gn} \biggl[ 4 \g^{(4)}_{\m\n} - 4 \g^{(4)\r}{}_\r \h_{\m\n} - \tilde{\g}^{(4)}_{\m\n} + \tilde{\g}^{(4)\r}{}_\r \h_{\m\n} 
        - 2 B^{(2)}_{\m\r} b_\n{}^\r - 2 b_\m{}^\r B^{(2)}_{\n\r} + \h_{\m\n} b^{\r\s} B^{(2)}_{\r\s}
        \nonumber \\ &\phantom{= \frac{1}{16 \pi\gn} \biggl[}
            + \frac{1}{2} \le( \p^\r \p_\m \g^{(2)}_{\r\n} + \p^\r \p_\n \g^{(2)}_{\m\r} - \Box \g^{(2)}_{\m\n}  - \h_{\m\n} \p^\r \p^\s \g^{(2)}_{\r\s} \ri)
        \nonumber \\ &\phantom{= \frac{1}{16 \pi\gn} \biggl[}
            + 2 b_{\m\r} b^{\r\s} b_{\s\k} b^\k{}_\n +  \h_{\m\n} b^{\r\s} b_{\s\k} b^{\k\t} b_{\t\r} - \frac{3}{8} \h_{\m\n} \le( b^{\r\s} b_{\r\s} \ri)^2
        \\  &\phantom{= \frac{1}{16 \pi\gn} \biggl[}
        -2 \p_\m b_{\r\s} \p_\n b^{\r\s} + \h_{\m\n} \p_\l b_{\r\s} \p^\l b^{\r\s} 
        +2 b_{\m\r} \Box b_\n{}^\r +2 b_{\n\r} \Box b_\m{}^\r
        \nonumber \\  &\phantom{= \frac{1}{16 \pi\gn} \biggl[}
        -2  \p_\l \le(b^{\l\r} \p_\n b_{\m\r} + b^{\l\r} \p_\m b_{\n\r}  - b_{\n\r} \p_\m b^{\l\r} - b_{\m\r} \p_\n b^{\l\r} \ri)
        \biggr].
        \nonumber 
\end{align} 

The divergence of \(\vev{\tilde{T}_{\m\n}}\) may now be computed using equation~\eqref{eq:gamma4_divergence},
\begin{align}
    \p^\m \vev{\tilde{T}_{\m\n}} &= \frac{1}{8 \pi \gn} \biggl[
        B^{(2)}_{\r\s} \p_\n b^{\r\s} - 2 \p^\m \le( B^{(2)}_{\m\r} b_\n{}^\r  \ri) 
        + 2 \p_\m \le(b^{\m\r} b_{\r\s} b^{\s\k} b_{\k\n} \ri)
        - \frac{1}{2} \p_\m \le( b^{\m\r} b_{\n\r} b^{\k\t} b_{\k\t} \ri)
    \nonumber \\ &\phantom{= \frac{1}{16 \pi \gn} \biggl[}
        + \frac{1}{4} b^{\k\t} b_{\k\t} b^{\r\s} \p_\n b_{\r\s}
        + b_{\r\k} b^{\k\t}b_{\t\s}  \p_\n  b^{\r\s} 
        + \frac{5}{2} \p_\m \le(b_{\n\r} \Box b^{\m\r} \ri) - \frac{5}{4} \Box b_{\r\s} \p_\n b^{\r\s}
    \biggr].
\end{align}
Replacing \(B^{(2)}_{\m\n}\) with \(\vev{\cO_{\m\n}}\) using equation~\eqref{eq:OB_vev_general}, this becomes
\begin{equation} \label{eq:holo_rg_stress_tensor_diveregence}
    \p^\m \vev{\tilde{T}_{\m\n}} =  \vev{\cO_{\r\s}} \p_\n b^{\r\s} - 2 \p^\m \le( \vev{\cO_{\m\r}} b_\n{}^\r \ri),
\end{equation}
which is precisely the Ward identity~\eqref{eq:Ttilde_ward_identity} derived from general field theory considerations in section~\ref{sec:stress_tensor_comments}. Thus, the putative stress tensor that we have obtained from differentiation of the action with respect to the metric is not a conserved current. Instead, equation~\eqref{eq:holo_rg_stress_tensor_diveregence} shows that the conserved stress tensor is
\begin{equation}
    T_{\m\n} = \tilde{T}_{\m\n} + 2 \cO_{\m\r} b_\n{}^\r,
\end{equation}
which satisfies the expected Ward identity
\begin{equation}
    \p_\m \vev{T^{\m\n}} =  \vev{\cO^{\r\s}} \p^\n b_{\r\s}.
\end{equation}

%% file: two_point_fn_wi.tex
\section{Derivation of the Ward identities for two-point functions}
\label{app:two_point_ward_identities}

In this appendix we derive the Ward identities for two-point functions quoted in equation~\eqref{eq:two_point_function_ward_identities}, that arise due to the translational symmetry of our system. The discussion follows ref.~\cite{Policastro:2002tn}, but with the addition of the antisymmetric tensor operator \(\cO^{\m\n}\) sourced by \(b_{\m\n}\). We will first work out the Ward identities for two-point functions in Euclidean signature, and then perform a Wick rotation.

As in section~\ref{sec:stress_tensor_comments}, we will compute the Ward identities by working on a curved spacetime of metric \(g\) with general two-form source \(b\), setting \(g\) to be flat and \(b\) to be constant at the end of calculations. In Euclidean signature the one-point functions are obtained from functional derivatives of \(W = \log Z\) as
\begin{equation} \label{eq:euclidean_signature_one_point_functions}
    \vev{\tilde{T}^{\m\n}} = \frac{2}{\sqrt{g}}\frac{\d W}{\d g_{\m\n}},
    \qquad
    \vev{\cO^{\m\n}} = \frac{1}{\sqrt{g}} \frac{\d W}{\d b_{\m\n}}.
\end{equation}
The corresponding two-point functions are
\begin{align}
    \vev{\tilde{T}^{\m\n}(x) \tilde{T}^{\r\s}(y)} &= \frac{4}{\sqrt{g(x)g(y)}} \frac{\d^2 W}{\d g_{\m\n}(x) \d g_{\r\s}(y)},
    \nonumber \\
    \vev{\tilde{T}^{\m\n}(x) \cO^{\r\s}(y)} &= \frac{2}{\sqrt{g(x)g(y)}} \frac{\d^2 W}{\d g_{\m\n}(x) \d b_{\r\s}(y)},
    \\
    \vev{\cO^{\m\n}(x) \cO^{\r\s}(y)} &= \frac{1}{\sqrt{g(x)g(y)}} \frac{\d^2 W}{\d b_{\m\n}(x) \d b_{\r\s}(y)}.
    \nonumber
\end{align}
Note that the factors of \(1/\sqrt{g}\) in equation~\eqref{eq:euclidean_signature_one_point_functions} imply that the functional derivative of a one-point function with respect to \(g_{\r\s}\) differs from the corresponding two-point function by a contact term, for instance
\begin{equation} \label{eq:one_point_function_contact_term}
    \frac{2}{\sqrt{g}} \frac{\d}{\d g_{\r\s}(y)} \vev{\cO^{\m\n}(x)} = \vev{\cO^{\m\n}(x) \tilde{T}^{\r\s}(y)} - \vev{\cO^{\m\n}(x)} \frac{g^{\r\s}(x)}{\sqrt{g(x)}} \d^{(4)}(x-y).
\end{equation}

Now consider the one-point function Ward identity~\eqref{eq:translation_ward_identity}. In a curved spacetime it becomes
\begin{equation} \label{eq:one_point_function_WI_curved}
    \nabla_\m \vev{T^{\m\n}} = \vev{\cO^{\k\t}} \nabla^\n b_{\k\t}.
\end{equation}
Taking a functional derivative of this identity with respect to \(g_{\r\s}\), and then setting \(g_{\m\n} = \d_{\m\n}\) and \(b_{\m\n}\) to be constant, we find
\begin{align}
    \p_\m \vev{T^{\m\n}(x) \tilde{T}^{\r\s}(0)} &+ \vev{T^{(\m\r)}} \d^{\n\s}  \p_\m \d^{(4)}(x) + \vev{T^{(\m\s)}} \d^{\n\r} \p_\m \d^{(4)}(x) - \vev{T^{(\r\s)}} \p^\n \d^{(4)}(x)
    \nonumber \\
    &= - 2 \vev{\cO^{\k\t}} \d^{\m\n} b^\l{}_\t \le[ \d_\l^{(\r} \d_\k^{\s)} \p_\m \d^{(4)}(x)  + \d_\m^{(\r} \d_\l^{\s)} \p_\k \d^{(4)}(x) - \d_\m^{(\r} \d_\k^{\s)} \p_\l \d^{(4)}(x)\ri],
    \label{eq:T_T_tilde_WI_position}
\end{align}
where \((\cdot)\) denotes symmetrisation of indices, with normalisation \(T^{(\m\n)} = \frac{1}{2} (T^{\m\n} + T^{\n\m})\). The various contact terms in equation~\eqref{eq:T_T_tilde_WI_position} arise either from equation~\eqref{eq:one_point_function_contact_term} or from the variation of the Christoffel symbols hidden in the covariant derivatives in equation~\eqref{eq:one_point_function_WI_curved}. We now perform a Fourier transform, defining the momentum-space two-point function
\begin{equation}
    \vev{ T^{\m\n}\tilde{T}^{\r\s}}(k_\mathrm{E}) = \int \diff^4 x \, e^{- i k_\mathrm{E} \cdot x} \vev{T^{\m\n}(x) \tilde{T}^{\r\s}(0)},
\end{equation}
where \(k_\mathrm{E} = (\w_\mathrm{E},\vec{k})\) is the Euclidean four-momentum. From equation~\eqref{eq:T_T_tilde_WI_position}, we find that the momentum space two-point function satisfies
\begin{align}
    (k_\mathrm{E})_\m &\le[ \vev{ T^{\m\n}\tilde{T}^{\r\s}}(k_\mathrm{E}) + \d^{\n\s} \vev{T^{(\m\r)}} + \d^{\n\r} \vev{T^{(\m\s)}}  - \d^{\m\n} \vev{T^{(\r\s)}}  \ri]
    \nonumber \\ &
    = - 2 (k_\mathrm{E})_\m \biggl[
        \d^{\m\n} b^{(\r}{}_\l \vev{\cO^{\s) \l}}
        + \d^{\n\r} b^{[\s}{}_\l \vev{\cO^{\m]\l}} + \d^{\n\s} b^{[\r}{}_\l \vev{\cO^{\m]\l}}
    \biggr],
\end{align}
where \([\cdot]\) denotes antisymmetrisation of indices, with normalisation \(T^{[\m\n]} = \frac{1}{2} (T^{\m\n} - T^{\n\m})\). Now we use the relation \(T^{\m\n} = \tilde{T}^{\m\n} + 2 \cO^{\m\r} b^\n{}_\r\) to replace \(\tilde{T}\) with the stress tensor, finding
\begin{align}
    (k_\mathrm{E})_\m  \le[ 
        \vev{T^{\m\n} T^{\r\s}}(k_\mathrm{E}) + \d^{\n\s} \le( \vev{T^{(\m\r)}} - 2 b^{[\m}{}_\l \vev{\cO^{\r]\l}}\ri)
        + \d^{\n\r} \le( \vev{T^{(\m\s)}} - 2  b^{[\m}{}_\l \vev{\cO^{\r]\l}}\ri)
    \ri]
    \nonumber \\
    - k_\mathrm{E}^\n \le( \vev{T^{(\r\s)}} -2 b^{[\r}{}_\l \vev{\cO^{\s]\l}} \ri) = 0.
    \label{eq:TT_euclidean_pre_simplification}
\end{align}
However, it is straightforward to verify that \(T^{(\m\r)} - 2 b^{[\m}{}_\l \cO^{\r]\l} = T^{\m\r}\) identically, so equation~\eqref{eq:TT_euclidean_pre_simplification} simplifies dramatically to
\begin{equation}
    (k_\mathrm{E})_\m \le[ 
        \vev{T^{\m\n} T^{\r\s}}(k_\mathrm{E}) + \d^{\n\s} \vev{T^{\m\r}}
        + \d^{\n\r} \vev{T^{\m\s}}
    \ri]
     + k_\mathrm{E}^\n \vev{T^{\r\s}} = 0.
     \label{eq:TT_euclidean_post_simplification}
\end{equation}

A second Ward identity may be obtained by instead differentiating equation~\eqref{eq:one_point_function_WI_curved} with respect to \(b_{\r\s}\). After setting \(g_{\m\n} = \d_{\m\n}\) and taking \(b\) to be constant, we find
\begin{equation}
    \p_\m \vev{T^{\m\n}(x) \cO^{\r\s}(0)} = \vev{\cO^{\r\s}} \p^\n \d^{(4)}(x).
\end{equation}
Performing a Fourier transformation, we then find that the momentum-space two-point function satisfies
\begin{equation}
    (k_\mathrm{E})_\m \vev{T^{\m\n} \cO^{\r\s}} (k_\mathrm{E}) - k_\mathrm{E}^\n \vev{\cO^{\r\s}} = 0.
\end{equation}

We now perform a Wick rotation to Lorentzian signature, assuming that under the Wick rotation \(\vev{A B}(k_\mathrm{E}) \to - \vev{A B}_\mathrm{L}(k)\) for any operators \(A\) and \(B\), where \(k^\m = (\w,\vec{k})\) is the Lorentzian four-momentum, and \(\vev{\cdot}_\mathrm{L}\) denotes some Lorentzian signature two-point function.\footnote{This minus sign in \(\vev{A B}(k_\mathrm{E}) \to - \vev{A B}_\mathrm{L}(k)\) is conventional, see e.g. ref.~\cite{Kapusta:2006pm}.} For the moment we will be agnostic about precisely \textit{which} Lorentzian signature two-point function (e.g. retarded, advanced, or time-ordered) we obtain through this procedure. Under the Wick rotation, each upper time-like index picks up a factor of \(i\), while each lower time-like index picks up a factor of \(-i\). See for example ref.~\cite{Herzog:2009xv} for an explicit accounting of all of the factors of \(i\). The result is that we find that the Lorentzian signature two-point functions satsify
\begin{align}
    k_\m \le[ \vev{T^{\m\n} T^{\r\s}}_\mathrm{L}(\w,\vec{k}) - \h^{\n\r} \vev{T^{\m\s}} - \h^{\n\s} \vev{T^{\m\r}} + \h^{\m\n} \vev{T^{\r\s}} \ri] &= 0,
    \nonumber \\
    k_\m \vev{T^{\m\n} \cO^{\r\s}}_\mathrm{L}(\w,\vec{k}) + k^\n \vev{\cO^{\r\s}} &= 0.
    \label{eq:lorentzian_correlator_identities}
\end{align}
These are the Ward identities written in equation~\eqref{eq:two_point_function_ward_identities}, but with \(\vev{\cdot}_\mathrm{L}\) in place of \(\vev{\cdot}_\mathrm{R}\).

Now we turn to the question of which Lorentzian correlation functions satisfy the Ward identities~\eqref{eq:lorentzian_correlator_identities}. It turns out that \(\vev{\cdot}_\mathrm{L}\) cannot be the retarded Green's function, since the components \(\vev{T^{0i}T^{\r\s}}_\mathrm{L}(\w,\vec{k})\) have the wrong \(\w \to 0\) limit~\cite{Policastro:2002tn}. However, one may choose \(\vev{\cdot}_\mathrm{L}\) to differ from the retarded Green's function only by a constant contact term~\cite{Policastro:2002tn,Herzog:2003ke}, which cancels out in the difference between two-point functions taken in the Kubo formula~\eqref{eq:thermal_conductivity_kubo_formula}. We will denote by \(\vev{\cdot}_\mathrm{R}\) the set of two-point functions related to the retarded Green's functions only by contact terms. They satisfy the Ward identities~\eqref{eq:two_point_function_ward_identities}.

%% file: coefficients.tex
\section{Thermal conductivity fluctuation equation}
\label{app:coefficients}

In this appendix we give the coefficients appearing in the equation of motion~\eqref{eq:thermal_conductivity_ZEOM} for the fluctuation \(\cZ\) that determines the thermal conductivity. They are
\begin{align}
    c_1 &=
    - \frac{h r}{D} \biggl\{-2 B^3 f g \left(-g h r X^2 \phi  f'+2 f g X^2 \phi  \left(2 r h'+h\right)+4 \lambda  L^2 r \omega ^2 \left(\phi  h'+h \phi
   '\right)\right)
   \nonumber \\ & \phantom{=- \frac{h r}{D} \biggl\{}
   +4 f g h^3 r X \phi  B' \left(L^2 \omega ^2-f g X\right)
   +8 B^2 f g h \lambda  L^2 r \omega ^2 \phi  B'
   +4 B h^3 f g L^2 r (X-1) \omega ^2 \phi'
   \nonumber \\ & \phantom{=- \frac{h r}{D} \biggl\{}
   +B h^3 \left[X \phi  \left(g
   \left(r f' \left(f g X+L^2 \omega ^2\right)+2 f \left(L^2 \omega ^2-f g X\right)\right)+2 f L^2 r \omega ^2 g'\right)\right]\biggr\},
   \\
   c_2 &= - \frac{2}{D} \biggl\{
       4 B^5 f^2 g^2 X^3 \phi 
       +2 B^3 f g h^2 X^2 \phi  \left(f g X-2 L^2 \omega ^2\right)
    \nonumber \\ & \phantom{=- \frac{h r}{D} \biggl\{}
       + f g h^3 r^2 B' \left(X \phi  \left(h X \left(g f'+f g'\right)+h' \left(L^2 \omega ^2-f g X\right)\right)+2 h L^2 (X-1) \omega ^2 \phi '\right)
   \nonumber \\ & \phantom{=- \frac{h r}{D} \biggl\{}
   +B h^2 \phi  \left(4 f g r^2 B'^2 \left(f g X^2+\lambda  L^2 \omega ^2\right)+h^2 L^2 X \omega ^2 \left(L^2 \omega ^2-f g X\right)\right)
   \nonumber \\ & \phantom{=- \frac{h r}{D} \biggl\{}
   +2 B^2 f g h r^2 B' \left(g X^2 \phi  \left(h f'-3 f h'\right)+f h X^2
   \phi  g'-2 \lambda  L^2 \omega ^2 \left(\phi  h'+h \phi '\right)\right)\biggr\},
   \nonumber
\end{align}
where we have defined \(X = 1 + 2 \s \f^2 + 2 \l \frac{B^2}{h^2}\), and
\begin{equation}
    D = 2 B f g h^2 r^2 X \phi  \left[2 B^2 f g X+h^2 \left(f g X-L^2 \omega ^2\right)\right].
\end{equation}